\documentclass{aa}  

\makeatletter
\renewcommand*\aa@pageof{, page \thepage{} of \pageref*{LastPage}}
\makeatother

\usepackage{graphicx}
\usepackage{txfonts}
\usepackage{color}
\usepackage{natbib}
\usepackage{amsmath}
\usepackage{hyperref}
\usepackage{ccaption}
\bibpunct{(}{)}{;}{a}{}{,} 
\usepackage{float}
\usepackage{natbib}
\usepackage{placeins}
\usepackage{txfonts}
\usepackage{forloop}
\usepackage{subcaption}
\usepackage{ulem}
\usepackage{xcolor}

\definecolor{todo}{rgb}{0.8,0.2,0.3}

\definecolor{comm}{rgb}{0,0.7,0}

\definecolor{idea}{rgb}{0,0.5,1}

\definecolor{new}{rgb}{0.4,0.8,0.8}

\definecolor{old}{rgb}{0.7,0.7,0.3}

\definecolor{modified}{HTML}{FF8C00}

\def\omicron{${\scriptstyle\mathcal{O}}$ }

\begin{document}

\title{Catalogue of BRITE-Constellation targets\thanks{Based on data collected by the BRITE-Constellation satellite mission, designed, built, launched, operated and supported by the Austrian Research Promotion Agency (FFG), the University of Vienna, the Technical University of Graz, the University of Innsbruck, the Canadian Space Agency (CSA), the University of Toronto Institute for Aerospace Studies (UTIAS), the Foundation for Polish Science \& Technology (FNiTP MNiSW), and National Science Centre (NCN).}}
       \titlerunning{BRITE Catalogue I}

   \subtitle{I. Fields 1 to 14 (November 2013\,--\,April 2016)}

   \author{K. Zwintz
          \inst{1},
          A. Pigulski\inst{2},
          R. Kuschnig\inst{3},
          G. A. Wade\inst{4},
          G. Doherty\inst{4},
          M. Earl\inst{4},
          C. Lovekin\inst{5},
          M. M\"ullner\inst{1},
          S.~Pich\'e-Perrier\inst{4},
          T. Steindl\inst{1},
        P. G. Beck\inst{3,6,7},
        K. Bicz\inst{2},
        D. M. Bowman\inst{8,9},
          G. Handler\inst{10},
          B. Pablo\inst{11},
          A. Popowicz\inst{12},
          T.~R\'o\.za\'nski\inst{2},
          P. Miko{\l}ajczyk\inst{2,13},
          D. Baade\inst{14},
          O. Koudelka\inst{15},
          A. F. J. Moffat\inst{16},
          C. Neiner\inst{17},
          P. Orlea\'nski\inst{18},
          R.~Smolec\inst{10},
         N.~St.~Louis\inst{16},
         W. W. Weiss\inst{19},
         M. Wenger\inst{15}, \and
         E. Zoc{\l}o\'nska\inst{10,20}
          }
        \authorrunning{K. Zwintz et al.}

   \institute{Universit\"at Innsbruck, Institut f\"ur Astro- und Teilchenphysik, Technikerstra{\ss}e 25, 6020 Innsbruck, Austria\\
              \email{konstanze.zwintz@uibk.ac.at}
         \and
Instytut Astronomiczny, Uniwersytet Wroc{\l}awski, Kopernika 11, 51-622 Wroc{\l}aw, Poland \and
Institut für Physik, Karl-Franzens Universit\"at Graz, Universit\"atsplatz 5/II, NAWI Graz, 8010 Graz, Austria \and 
Department of Physics and Space Science, Royal Military College of Canada, PO Box 17000, Kingston, ON K7K 7B4, Canada \and
Mount Allison University, 69 York St, Sackville, NB, Canada \and
  Instituto de Astrof\'{\i}sica de Canarias, 38200 La Laguna, Tenerife, Spain \and
  Departamento de Astrof\'{\i}sica, Universidad de La Laguna, 38206 La Laguna, Tenerife, Spain \and
School of Mathematics, Statistics and Physics, Newcastle University, Newcastle upon Tyne, NE1 7RU, UK \and
Institute of Astronomy, KU Leuven, Celestijnenlaan 200D, Leuven 3001, Belgium \and
Nicolaus Copernicus Astronomical Center of the Polish Academy of Sciences, Bartycka 18, 00-716 Warsaw, Poland \and
American Association of Variable Star Observers, 185 Alewife Brook Pkwy Suite. 410, Cambridge, MA 01238, USA \and
Silesian University of Technology, Department of Electronics, Electrical Engineering and Microelectronics, Akademicka 16, 44-100 Gliwice, Poland \and
Astronomical Observatory, University of Warsaw, Al. Ujazdowskie 4, 00-478 Warszawa, Poland \and
European Organisation for Astronomical Research in the Southern Hemisphere (ESO), Karl-Schwarzschild-Str.\ 2, 85748 Garching b. M\"unchen, Germany \and 
Technische Universit\"at Graz, Inffeldgasse 12, 8010 Graz, Austria \and
D\'epartement de physique, Universit\'e de Montr\'eal, Campus MIL, 1375 Th\'er\'ese-Lavoie-Roux, Montr\'eal (Qc), Canada \and
LESIA, Paris Observatory, PSL University, CNRS, Sorbonne University, Paris Cité University, 5 place Jules Janssen, 92195 Meudon, France \and
Space Research Center, Polish Academy of Sciences, Bartycka 18A, 00-716 Warsaw, Poland \and
University of Vienna, Department for Astrophysics, T\"urkenschanzstrasse 17, 1180 Vienna, Austria \and
{\L}ukasiewicz Research Network ‒ Institute of Aviation, Al. Krakowska 110/114, 02-256 Warsaw, Poland
 }

   \date{Received ; accepted }

\abstract
   {The BRIght Target Explorer (BRITE) mission collects photometric time series in two passbands aiming to investigate stellar structure and evolution. Since their launches in the years 2013 and 2014, the constellation of five  BRITE nano-satellites has observed a total of more than 700 individual bright stars in 64 fields. Some targets have been observed multiple times. Thus, the total time base of the data sets acquired for those stars can be as long as nine years.}
   {Our aim is to provide a complete description of ready-to-use BRITE data, to show the scientific potential of the BRITE-Constellation data by identifying the most interesting targets, and to demonstrate and encourage how scientists can use these data in their research.}
   {We apply a decorrelation process to the automatically reduced BRITE-Constellation data to correct for instrumental effects. We perform a statistical analysis of the 
   light curves obtained for the 300 stars observed in the first 14 fields during the first $\sim$2.5 years of the mission. We also perform cross-identification with the International Variable Star Index.}
   {We present the data obtained by the BRITE-Constellation mission in the first 14 fields it observed from November 2013 to April 2016. We also describe the properties of the data for these fields and the 300 stars observed in them. Using these data, we detected variability in 64\% of the presented sample of stars. Sixty-four stars or 21.3\,\% of the sample have not yet been identified as variable in the literature and their data have not been analysed in detail. They can therefore provide valuable scientific material for further research. All data are made publicly available through the BRITE Public Data Archive and the Canadian Astronomy Data Centre.} 
{}
   

   \maketitle

%

\section{Introduction}
\label{sec:brite}

The BRITE-Constellation\footnote{http://www.brite-constellation.at} is the first nano-satellite mission to study the structure and evolution of the brightest stars in the sky. Providing photometric time series in two passbands for up to half a year continuously, BRITE-Constellation data are used to study different types of stellar variability such as pulsations, wind phenomena, the feeding of decretion disks around Be stars, binarity, and rotational modulation \citep[e.g., ][]{weiss2021}. A majority of the brightest stars are also some of the most massive and most luminous objects in the Milky Way. Consequently, a large fraction of the sample observed by BRITE-Constellation consists of stars with spectral types O and B. For an overview of the BRITE-Constellation mission, including its history and selected scientific highlights, see \citet{weiss2021}.

The BRITE-Constellation mission originally included six 20-cm cube-shaped nano-satellites that were funded by three countries (Austria, Canada and Poland). Each satellite has a mass of about 7 kg, is stabilised in three axes, has access to almost the entire sky and carries a 3-cm telescope which feeds an uncooled CCD detector \citep{weiss2014}. Each partner country contributed a pair of BRITE satellites that were deployed into Low Earth Orbits, with one satellite equipped with a custom-defined red filter (transmitting in the range 550 -- 700\,nm) and the other with a custom-defined blue filter (390 -- 460\,nm). Five of the six nano-satellites were operational after launch; the sixth, named BRITE-Montr\'eal, did not separate from the upper stage of its rocket and, hence, was never active. The BRITE-Constellation therefore consists of three satellites with a red filter, named BRITE-Toronto (BTr), UniBRITE (UBr) and BRITE-Heweliusz (BHr), and two satellites equipped with a blue filter, named BRITE-Austria (BAb) and BRITE-Lem (BLb). The `r' and `b' at the end of the abbreviations indicate whether the satellite is equipped with a red or blue filter, respectively. The paper by \citet{pablo2016} provides more details on the detectors, pre-launch and in-orbit tests. \citet{popowicz2017} described the pipeline that was used to process the observed images giving the instrumental magnitudes that were originally provided to users.

The nominal goal of the BRITE-Constellation mission was to observe stars brighter than $V$~$\sim$~4\,mag, but the first observa\-tions showed that scientifically useful photometry could be easily obtained for stars down to $V$~$\sim$~6\,mag. The faintest star ever observed by BRITE was the M-type subgiant HD\,96265 with $V=8.03$\,mag. In addition, BRITE collected data for two novae, Nova Carinae 2018 (V906\,Car) and Nova Reticulum 2020 (YZ\,Ret), which were tracked by red-filtered BRITE satellites until their $V$ magnitudes dropped to about 9.8\,mag and 8.7\,mag, respectively. The brightest BRITE target was Canopus ($\alpha$~Car), with $V = -0.72$\,mag. Despite the intentional large and position-dependent blurring of the images, all stars brighter than $V$~$\sim$~1\,mag saturated the detectors when the integration times were on the order of one second. The light curves of saturated stars can still be used if the decorrelation parameters are chosen correctly, as the saturation is analogue. This means that the full-well electron capacity of a pixel is translated into a digital signal smaller than the maximum allowed. The total flux from a star can therefore be measured, despite the overflow of charge to neighbouring pixels.

The field of view of each satellite is approximately 24\,$\times$\,20 degrees, allowing observations of fields with typically 15\,--\,20 bright stars, at least three of which must be brighter than $V=3$\,mag. The observed fields are named according to the celestial constellations covered by them and numbered consecutively. A given field is observed for at least 15 minutes in each $\sim$100-minute orbit for up to six months \citep{weiss2014}. Observations of several fields are repeated every year (for example, the Orion field), which currently results in observations covering nine consecutive seasons for some objects. Table \ref{tab:obs} shows the names and properties of the first 14 fields that are the subject of this publication. 

Since the launch of the first two BRITE satellites in February 2013, a total of 716 stars have been observed to date, contained in 64 fields with publicly released data. Most of the stars observed by the BRITE satellites have also been targeted by the National Aeronautics and Space Administration Transiting Exoplanet Survey Satellite (TESS) mission \citep{Ricker2015}. A combination of BRITE and TESS data provides a fruitful synergy and excellent complementarity.
Due to the different passbands used by BRITE-Constellation and TESS, a combination of the data allows for extracting colour information for the targets.
While the per-observation accuracy of TESS is much better than BRITE, many of the much longer BRITE time series with higher cadence provide a better frequency resolution. For example, BRITE data can be very useful to study pulsations in binary systems based on Doppler shifts if the assignment of frequencies to one of the two components is not clear.

In the present paper, the first in a series of three catalogue papers in which we discuss all BRITE mission data, we provide an overview of the astrophysical content of the BRITE-Constellation data obtained in the first 14 fields\footnote{The second catalogue paper will cover fields 15 to 44 and the third, fields 45 to the last, which will most likely be field 74.}. The main motivation for publishing this catalogue is to raise the awareness of the BRITE-Constellation data and its properties and to enable non-experts to use it for reliable quantitative studies.
The present section introduces the BRITE-Constellation mission. In Sect.\,\ref{sec:data_properties} we discuss the properties of the BRITE data, including statistical information on the first 14 BRITE fields. Section \ref{sec:classification} is devoted to a discussion of the variability of selected objects and general information on the known variability. More details on the BRITE-Constellation data, notes on data for the 300 stars in Fields 1\,--\,14, maps of the individual fields and additional supplementary information are given in the Appendices. The work is summarised in Sect.\,\ref{sec:conclusions}.

\begin{figure*}[!ht]
\centering
\includegraphics[width=0.93\textwidth]{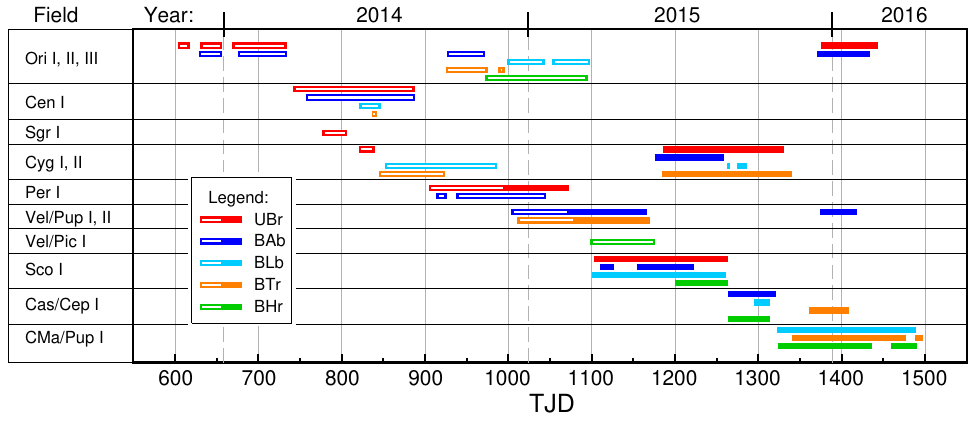}
\caption{Distribution of observations by the five BRITE satellites for Fields 1 to 14. The satellites are colour-coded in the legend. Data obtained in the stare and the chopping observing modes (see Sec. 2.1) are shown with unfilled and filled bars, respectively.}
\label{fig:tdistr}
\end{figure*}

\section{Properties of the BRITE-Constellation data}
\label{sec:data_properties}
In this first part of the BRITE-Constellation catalogue, we present the data obtained in the first 14 BRITE fields, which were observed between November 2013 and April 2016. 
The selection of these particular fields for the first part of the catalogue is based on the version of the BRITE data reduction pipeline that was used for performing photometry on the BRITE CCD images. For these 14 fields, the raw data include a reduced number of decorrelation parameters. The format of these Data Release 2 (DR2), 3 (DR3), and 4 (DR4) data is described by \cite{popowicz2017}.\footnote{Starting with field 15, data were made available to users in the Data Release 5 format \citep{popowicz2017}.} Within these first 14 fields, 300 individual stars were observed; 208 of these more than once. Table \ref{tab:obs} details the observations and Fig.\,\ref{fig:tdistr} illustrates the distribution of the observations over time. Appendix \ref{sec:fieldmaps} shows the positions of the 14 fields in the sky. Throughout the paper we use the truncated Heliocentric Julian Date (TJD), which is defined as follows: TJD [d] $=$ HJD [d] $-$ 2456000.0, where HJD is the Heliocentric Julian Date at mid-exposure. This definition has been used in all figures and tables.

\begin{table*}[htb]
\caption{Properties of the BRITE-Constellation observations of the first 14 fields. }
\label{tab:obs}
\begin{tabular}{rllccccrrc}
\hline\hline\noalign{\smallskip} 
\multicolumn{2}{c}{Field} & \multicolumn{1}{c}{Sat} & \multicolumn{1}{c}{Field centre}&
\multicolumn{1}{c}{Setup(s)} & \multicolumn{1}{c}{Observation} & \multicolumn{1}{c}{Observation}& \multicolumn{1}{c}{Time} & \multicolumn{1}{c}{$N_{\rm stars}$} & \multicolumn{1}{c}{Observing} \\
& & &\multicolumn{1}{c}{(RA$_{\rm 2000}$, DEC$_{\rm 2000}$)} & & \multicolumn{1}{c}{start}   & \multicolumn{1}{c}{end} & \multicolumn{1}{c}{span (d)} & & \multicolumn{1}{c}{mode}\\
\noalign{\smallskip}\hline\noalign{\smallskip} 
01 & Ori-I-2013	&  UBr & (5$^{\rm h}$30$^{\rm m}$, $+$0$\degr$30$\arcmin$) & 7 & 2013.11.07 & 2014.03.17 & 130.7 & 15 & S\\
&& BAb & as above & 3, 4 & 2013.12.01 & 2014.03.17 & 105.7 & 15 & S\\
02 & Cen-I-2014 & UBr &	(14$^{\rm h}$45$^{\rm m}$, $-$51$\degr$20$\arcmin$) & 4 & 2014.03.25 & 2014.08.17 & 145.3 & 30 & S\\
&& BAb & as above & 4 & 2014.04.09 & 2014.08.19 & 132.0 & 29 & S\\
&& BLb & as above & 1 & 2014.06.12 & 2014.07.08 & 26.6 & 15 & S\\
&& BTr & as above & 1 & 2014.06.27 & 2014.07.03 & 6.0 & 30 & S\\
03 & Sgr-I-2014 & UBr &	(18$^{\rm h}$00$^{\rm m}$, $-$30$\degr$20$\arcmin$) & 1 & 2014.04.29 & 2014.05.28	& 29.9 & 19 & S\\
04 & Cyg-I-2014	& UBr &	(20$^{\rm h}$40$^{\rm m}$, $+$40$\degr$10$\arcmin$) & 1 & 2014.06.12 & 2014.07.01	& 19.0 & 25 & S\\
&& BTr & as above & 1, 2& 2014.07.06 & 2014.09.23 & 79.8 & 32 & S\\
&& BLb & as above & 1, 2& 2014.07.12 & 2014.11.24 & 135.0 & 22 & S\\
05 & Per-I-2014 & UBr &	(3$^{\rm h}$27$^{\rm m}$, $+$37$\degr$06$\arcmin$)  & 2, 3, 6, 7 & 2014.09.04 & 2015.02.18 & 167.9 & 33 & S/Ch\\
&& BAb & as above & 1 -- 5 & 2014.09.13 & 2015.01.23 & 132.1 & 21 & S\\
06 & Ori-II-2014 & BTr & (5$^{\rm h}$12$^{\rm m}$, $-$0$\degr$30$\arcmin$) & 1, 2, 3 & 2014.09.24 & 2014.12.04 & 70.9 & 34 & S\\
&& BAb & as above  & 2, 3, 4 & 2014.09.25 & 2014.11.10 & 45.7 & 23 & S\\
&& BHr & as above & 2, 5, 7 & 2014.11.10 & 2015.03.14 & 123.3 & 26 & S\\
&& BLb & as above & 3, 6 & 2014.12.07 & 2015.03.16 & 99.8 & 25 & S\\
07 & VelPup-I-2014 & BAb & (8$^{\rm h}$40$^{\rm m}$, $-$47$\degr$30$\arcmin$) & 1 -- 7 & 2014.12.10 & 2015.05.23 & 163.8 & 32 & S/Ch\\
&& BTr & as above & 1 -- 5 & 2014.12.19 & 2015.05.27 & 159.3 & 36 & S/Ch\\
08 & VelPic-I-2015 & BHr & (7$^{\rm h}$14$^{\rm m}$, $-$50$\degr$40$\arcmin$) & 3, 4, 5 & 2015.03.16	& 2015.06.02 & 78.3 & 20 & S\\
09 & Sco-I-2015 & UBr & (15$^{\rm h}$58$^{\rm m}$, $-$30$\degr$00$\arcmin$) & 1, 2, 3, 4 & 2015.02.20 & 2015.08.29 & 162.1 & 19 & Ch\\
&& BLb & as above & 1 -- 6 & 2015.03.17 & 2015.08.26 & 162.0 & 20 & Ch\\
&& BAb & as above & 2 & 2015.03.28 & 2015.07.19 & 112.9 & 8 & Ch\\
&& BHr & as above & 3 & 2015.06.26 & 2015.08.28 & 63.8 & 19 & Ch\\
10 & Cyg-II-2015 & BAb & (20$^{\rm h}$40$^{\rm m}$, $+$38$\degr$30$\arcmin$) & 1 -- 5 & 2015.06.01 & 2015.08.24 & 83.4 & 20 & Ch\\
&& BTr & as above & 1 -- 5 & 2015.06.11 & 2015.11.14 & 156.0 & 24 & Ch\\
&& UBr & as above & 2, 3, 4 & 2015.06.12 & 2015.11.04 & 145.9 & 18 & Ch\\
&& BLb & as above & 1, 2 & 2015.08.27 & 2015.09.21 & 25.3 & 25 & Ch\\
11 & CasCep-I-2015 & BAb & (0$^{\rm h}$40$^{\rm m}$, $+$59$\degr$00$\arcmin$) & 1 & 2015.08.27 & 2015.10.26 & 59.1 & 12 & Ch\\
&& BHr & (23$^{\rm h}$25$^{\rm m}$, $+$62$\degr$04$\arcmin$) & 1, 2, 3 & 2015.08.29 & 2015.10.17 & 49.8 & 18 & Ch\\
&& BLb & as above & 2, 3, 4 & 2015.09.29 & 2015.10.17 & 18.2 & 15 & Ch\\
&& BTr & as above & 1, 2 & 2015.12.04 & 2016.01.20 & 47.7 & 7 & Ch\\
12 & CMaPup-I-2015 & BLb & (7$^{\rm h}$12$^{\rm m}$, $-$24$\degr$50$\arcmin$) & 3, 4, 5 & 2015.10.26 & 2016.04.10 & 166.5 & 26 & Ch\\
&& BTr & as above & 1, 2, 3 & 2015.11.15 & 2016.04.18 & 156.6 & 21 & Ch\\
&& BHr & as above & 3, 4 & 2015.10.27 & 2016.04.14 & 169.3 & 17 & Ch\\
13 & Ori-III-2015 & BAb & (5$^{\rm h}$19$^{\rm m}$, $-$0$\degr$45$\arcmin$) & 1, 2 & 2015.12.14 & 2016.02.15 & 63.3 & 17 & Ch\\
&& UBr & as above & 1, 2 & 2015.12.18 & 2016.02.24 & 68.8 & 16 & Ch\\
14 & VelPup-II-2015	& BAb & (8$^{\rm h}$40$^{\rm m}$, $-$47$\degr$30$\arcmin$) & 1, 2 & 2015.12.17 & 2016.01.31 & 44.5 & 12 & Ch\\
\noalign{\smallskip}\hline
\end{tabular}
\tablefoot{Columns are: field identification (Field), BRITE satellite (Sat), equatorial coordinates of the centre of the observed field for epoch 2000.0 (RA$_{\rm 2000}$, DEC$_{\rm 2000}$), setup number(s) (Setup(s)), observation start and end dates (Observation start, Observation end) in the format [yyyy.mm.dd], observation time span in days (Time span), the number of stars in the field ($N_{\rm stars}$), and the observing mode used (S: stare mode, Ch: chopping mode).}
\end{table*}

\begin{figure*}[htb]
\centering
\includegraphics[width=0.93\textwidth]{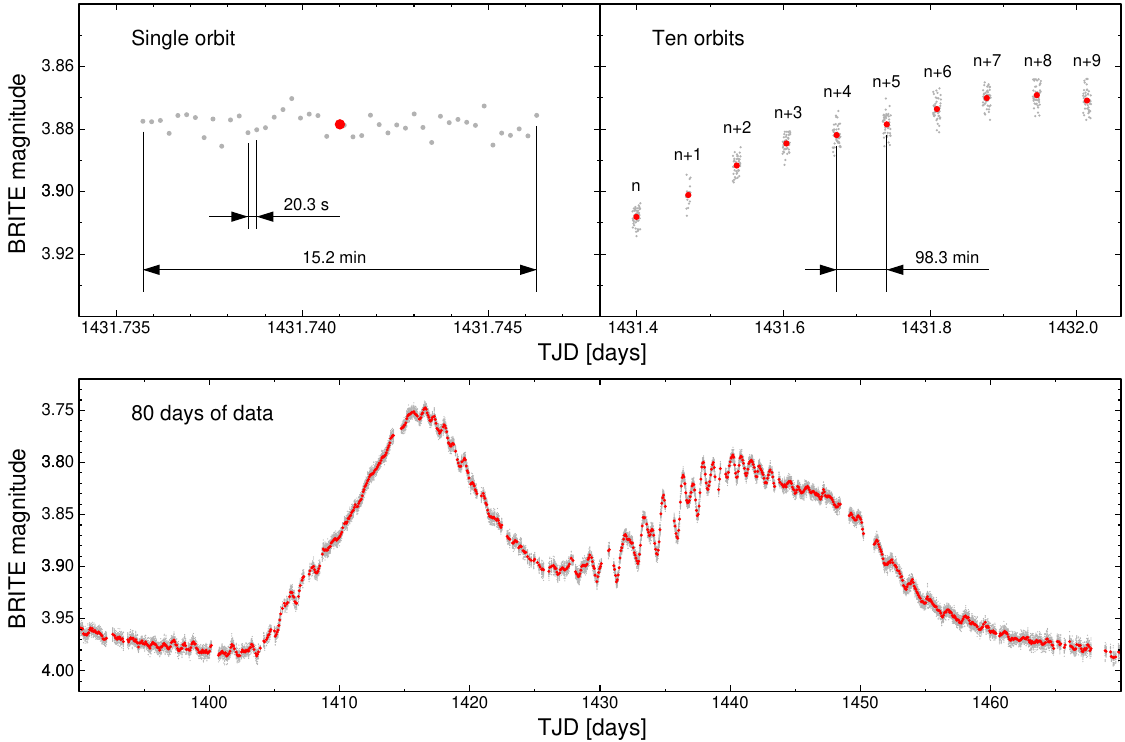}
\caption{Typical photometric sequence observed by BTr during a single orbit (top left), during ten consecutive orbits (top right) and for 80 days (lower panel). Grey dots stand for individual measurements, red dots indicate orbit averages. The presented photometry is for the Be star HD\,56139 ($\omega$~CMa) observed by BTr in Field 12 (CMa/Pup I).}
\label{fig:sampling}
\end{figure*}

The typical observing cadence of BRITE observations amounts to about 20 seconds and includes an exposure which is one to eight seconds long\footnote{In very rare cases, exposures shorter than one second were used.}, depending on the filter and brightness of the most important targets in the field. The cadence itself is independent of the exposure time. More relevant in the context of the BRITE data are the lengths of observations over a single orbit. During each $\sim$100-minute orbit, observations were made for up to 42 minutes; the median value for a single-orbit observation length was 13.5 minutes. However, during some observations two different fields were observed by a single satellite. An example of the distribution of observations during a single orbit, ten orbits, and $\sim$3 months is shown in Fig.\,\ref{fig:sampling}.

Data for the first 14 BRITE fields are available through the BRITE Public Data Archive (PDA) and the Canadian Astronomy Data Centre (CADC). The BRITE data were first processed through a reduction pipeline described by \cite{popowicz2017}. As they suffer from various instrumental effects \citep{2018pas8.conf..106P}, a procedure called `decorrelation' had to be applied (Sect.\,\ref{sec:decorrelations}). This procedure corrects the BRITE data for the strongest instrumental effects. The BRITE team has decorrelated the data for the 14 fields in question and these data are now available through the BRITE PDA and CADC. The decorrelated data form the basis for the results presented here. Detailed information about the BRITE data in the archives can be found in Appendix \ref{sec:archives}.

The following is a brief overview of the BRITE observing modes, the necessary decorrelation process, and the properties of the data discussed here. 

\subsection{Observing modes: stare mode versus chopping mode}
BRITE-Constellation data in Fields 1\,--\,14 were obtained using two different observing modes. Initially, the BRITE satellites maintained a fixed orientation in the sky and the stars in fixed positions on the CCD. This observing mode is called the stare mode. It was used for the first eight fields (see Table \ref{tab:obs}). After two years from the onset of the BRITE mission, it became clear that the number of hot pixels on the detectors was increasing, seriously affecting the photometry. Therefore, a new observing mode, the chopping mode, was tested at the end of 2014. 

In this mode, the satellite changes pointing slightly between consecutive exposures. The shifts are small, about 9$\arcmin$ on the sky or about 20 pixels. This mode forced a change of the subraster\footnote{A subraster is a small piece of a CCD image surrounding a star selected for observations with BRITE (see also Appendix \ref{sec:archives})}. size to slightly wider than in the stare mode \citep{pablo2016}. The BRITE reduction pipeline was adapted to this observing mode, as described by \cite{popowicz2017}. In this mode, consecutive images from alternating positions are subtracted to provide differential images, almost free of hot pixels, which are then used to extract photometric time series. Some residual hot pixels may still remain, as the dark current in the used CCD sensor exhibits random-telegraph-signal noise \citep[more details are given in][]{popowicz2020}. The first field for which the chopping mode was applied was field 5 (Perseus I), and regular observations in this mode have been carried out from mid-2015 to the present. Figure \ref{fig:tdistr} and Table \ref{tab:obs} show which observing mode was used for each field. More details on the two observing modes and the corresponding versions of the BRITE data reduction pipeline are given by \citet{popowicz2017}. 
Some attempts to obtain point-spread-function (PSF) photometry were made \citep{popowicz2018}, but this method has finally not been utilised in the official data sets.

\subsection{Description of decorrelation process}\label{sec:decorrelations}
The raw BRITE data suffer from instrumental effects, the most important of which is the dependence of the stellar PSF on changes in the temperature of the optics \citep{pablo2016}. Measuring through a fixed aperture leads to the raw magnitudes being strongly dependent on both the temperature of the optics and the position of the star on the detector. The temperature of the optics is not measured and only the temperature of the detector is known. These two temperatures may differ. Therefore, many parameters are introduced into the decorrelation process, the number increasing as the reduction pipeline evolved. In DR2, there are four parameters, in DR3 six, in DR4 seven and finally, in DR5, nine parameters. All parameters are available for each observational point in the raw photometry, except for the orbital phase, which is calculated by the decorrelation pipeline. For a description of the meaning of the decorrelation parameters, see Appendix B of \cite{popowicz2017}.

The goal of the decorrelation process is to eliminate the final dependence of magnitudes on the known instrumental parameters. The problem we face here is that the raw data also contain the intrinsic variability, which is what we are most interested in. The decorrelation therefore aims to separate the intrinsic variability from the variability caused by instrumental effects. For this reason, the choice of an appropriate variability model, which is fitted to the data and subtracted at each iteration step, plays an important role in the decorrelation process. It is sufficient if the variability model takes into account only the strongest variability so that the time series can be assumed to be dominated by instrumental effects. The decorrelations themselves include one- and two-dimensional dependencies of residual magnitudes on decorrelation parameters. Details of the entire procedure are described by \cite{2018pas8.conf..106P}.

All BRITE data from fields 1\,--\,14 were decorrelated by several members of the BRITE team. At least three team members examined each data set presented here. The decorrelation procedure is semi-automatic, but the final result may vary depending on how the extreme data in each parameter space were discarded and which intrinsic variability model was adopted. Therefore, we would like to emphasize that the decorrelated data were not entirely automatically reduced, but the procedure required human supervision. The decorrelations of data for periodic variables are the most reliable, as the variability model can be well established for them. Problems may occur for stars showing long-period variability or an intrinsic trend. In such cases, the separation of intrinsic variability from instrumental effects may not be unambiguous and the decorrelated data for such stars should be used with caution.

\subsection{Data characteristics for the observed fields}
\begin{figure*}[!ht]
\centering
\includegraphics[width=\textwidth]{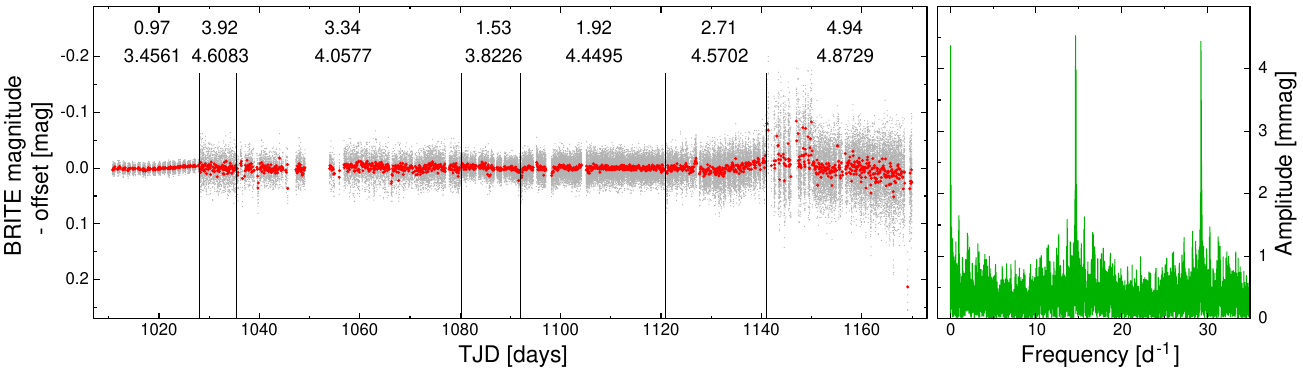}
\caption{Light curve (left) of the merged BTr field 7 data of HD\,74180 (b Vel) and its Fourier spectrum (right). The red points in the light curve mark the mean instrumental magnitudes per orbit. Vertical lines separate individual setups that were processed separately during decorrelations. The two numbers above each setup show the values of $\sigma_{\rm orbit}^{\rm med}$ (see Sect.\,\ref{sec:data_statistics}) in mmag (top number) and  subtracted offsets in mag (bottom number). The light curve illustrates some instrumental effects such as different amount of scatter, gaps in data, offsets and long term trends. These effects result in the presence of very low frequencies in the Fourier spectrum reproduced close to the satellite's orbital frequency of 14.66~d$^{-1}$ and its multiples.}
\label{fig:lc_problems}
\end{figure*}

There are two main issues that affect the quality of the BRITE data, not only for the first 14 fields presented in this paper, but throughout the mission. Firstly, this is the progressive radiation damage to the CCD detectors, and secondly is the ability of the satellites to achieve stable pointing during observations. There are cases (described below) where, after optimising the initial setup\footnote{Data from a given satellite pointing are split into parts, called setups (see Appendix \ref{sect:setups} for additional information).}, the area of the CCD on which a star's image was projected was hit by cosmic ray particle(s) and started to build up hot pixels, which after some time resulted in a sudden enhancement of  scatter to the reduced data. Such an example is shown in Fig.\,\ref{fig:lc_problems}. The data -- in this case collected by BTr -- were at the lower noise limit of the instrument before TJD $\sim$ 1028. Then, the noise increased rapidly when the area at and near the star's profile was ‘polluted’ by hot pixels. For some time the scatter remained at this higher level until TJD $\sim$ 1130. At this time, the scatter increased again by a significant amount persisting until the end of the observations in this field. These jumps in scatter occurred more frequently before chopping mode was introduced in 2015, which greatly mitigated such effects.

In the following paragraphs the data characteristics for the first 14 fields of BRITE-Constellation observations are described in detail.

{\it Field 1:} The Orion I (Ori-I-2013) field was the first field in which a regular campaign following the commissioning phase of UBr and BAb was carried out. UBr was able to achieve stable pointing at the beginning of November 2013 and maintained it for a number of orbits. Consistent stability and hence data collection by both UBr and BAb began on 12 December 2013 and continued until the end of the planned observing time in mid-March 2014. All observations were performed in stare mode. Exposure times for both BAb and UBr were mostly set to one second, 
providing a high signal-to-noise ratio for most of the 15 stars selected in this field. Due to their brightness, $\alpha$\,Ori (Betelgeuse) and $\beta$\,Ori (Rigel) caused saturation in the CCD images. The former star was saturated only in the red-filter UBr images, whereas the latter, both in the red-filter UBr and in the blue-filter BAb images.

{\it Field 2:} The Centaurus I (Cen-I-2014) campaign was scheduled for full coverage by UBr and BAb in stare mode, starting in late March and early April 2014, respectively. UBr was set up to observe 30 stars; BAb initially observed 29 stars. However, data transfer to European ground stations was severely hampered by interference, so that the number of stars in the BAb setup was reduced to 14 for the full time base. For the remaining stars, only the first 40 d and the last 18 d were covered. On 12 June 2014, BLb collected the first scientific data for 15 stars after commissioning, and BTr followed on 27 June 2014 with 30 objects. For BLb and BTr, the Cen I field observations, also only in the stare mode, were scheduled mainly as a test and hence only a few days were allocated to them.  

{\it Field 3:} The Sagittarius I (Sgr-I-2014) campaign was intended to be a test of whether BRITE satellites, in this case UBr, could observe two fields (properly separated) during each orbit. The primary target field was Cen I, the secondary field Sgr I. Trials began on 20 April 2014 and were stopped after about four weeks. While UBr was still able to produce good-quality data for Cen I, it failed to do so for the stars in Sgr I. The satellite struggled to achieve fine pointing during many orbits, and even when it did, the guiding was poor. Hence, the data quality for all 19 selected stars in this field is rather poor.

{\it Field 4:}
The Cygnus I (Cyg-I-2014) field was assigned as the first scientific campaign field for BLb and BTr. This field with 25 stars was planned to be used to continue test observations of two fields during a single orbit. This time the fields were Cen I and Cyg I. Unfortunately, the pointing stability of  BTr in Cyg I was insufficient. Consequently, this test was stopped after only 19 d and BTr was assigned to observe another field.

{\it Field 5:}
The Perseus I (Per-I-2014) campaign started in early September 2014 with both UBr and BAb in stare mode. Due to the progressive damage of detectors resulting in an increasing number of hot pixels in the CCDs, a new observing mode, the chopping mode, was introduced, as outlined above \citep[see also ][]{pablo2016,popowicz2017}. It was first tested on UBr starting on 5 December 2014, about halfway through the Per~I run. The positions of stars in these early tests were changed along the columns rather than the rows of the detector, as later became the default observing scheme. The chopping mode did reduce the scatter of the data. BAb data were collected in the stare mode. There are several long gaps in the data from this satellite, when it did not achieve fine pointing during numerous orbits at the beginning and also at the end of the run.

{\it Field 6:} Observations of the Orion II (Ori-II-2014) field began at the end of September 2014, when the field became observable. This was the first repeat observation of a field, but this time with more satellites observing more stars. Observations of the Orion II field were started by BAb and BTr, they were joined shortly after launch on 10 November 2014 by BHr, and finally BLb also joined on 12 December 2014 (Fig.\,\ref{fig:tdistr}). All satellites operated in stare mode. BAb also observed the Per I field regularly on each orbit again to test the capability of observing two fields per orbit. This test was stopped after 45 d due to a lack of fine pointing for about 50\% of the orbits. Images of $\alpha$\,Ori in the red filter (BTr and BHr) and $\beta$\,Ori in the red and blue filters were saturated.

{\it Field 7:} The Vela/Puppis I (VelPup-I-2014) field was observed by BAb and BTr starting in December 2014 in stare mode, which was changed to chopping mode in February 2015 for testing purposes. For the first four weeks, BAb combined observations with the Per I field, which showed better performance compared to previous tests. The chopping mode generally gave better noise levels, with a few specific exceptions described in the notes on individual stars.

{\it Field 8:} The Vela/Pictor I (VelPic-I-2015) field was the first test to verify that stars located relatively far (up to 31$\degr$ in this case) from the Galactic plane could be observed with a BRITE satellite. BHr operating in stare mode was assigned to observe this field. The pointing stability was poor at the beginning, but improved over time and overall good data were collected for 20 stars over almost 80 d.

{\it Field 9:} The Scorpius I (Sco-I-2015) field was the first observing run in which all satellites operated in chopping mode.  UBr and BLb were assigned to this campaign for the entire observing run and BAb and BHr for shorter periods. The BAb data generally do not have good time coverage due to frequent failures to achieve fine pointing in this field.

{\it Field 10:} The Cygnus II (Cyg-II-2015) field was the second campaign for which all satellites were operating in chopping mode. UBr and BTr were assigned to this field for the entire observing period, while BAb and BLb provided shorter coverage. UBr and BAb had Sco I as an alternate field for scheduled observations. The performance of BAb was generally quite poor in terms of data production and quality. 

{\it Field 11:} The Cassiopeia/Cepheus I (CasCep-I-2015) campaign was started with BAb, BHr and BLb. All three satellites had problems achieving fine pointing on many orbits. After about two months, observations by these three satellites were suspended. A few weeks later, observations of the Cas/Cep I field were resumed by BTr, which observed 7 stars in this field for 47 days in combination with the CMa/Pup I field, achieving good data quality.

{\it Field 12:} The Canis Major/Puppis I (CMaPup-I-2015) campaign was conducted by BLb, BTr and BHr, which observed this field for the entire scheduled time.

{\it Field 13:} The Orion III (Ori-III-2015) campaign, the third devoted to this field, was assigned to UBr and BAb. Both satellites observed this field for more than 60 d. The BAb satellite had significant problems achieving fine pointing.

{\it Field 14:}
The Vela/Puppis II (VelPup-II-2015) was a short observing run with BAb, which was conducted in combination with the campaign in the Ori III field.

\subsection{Noise characteristics}
\label{subsec:noise}

\begin{figure}
\includegraphics[width=0.48\textwidth]{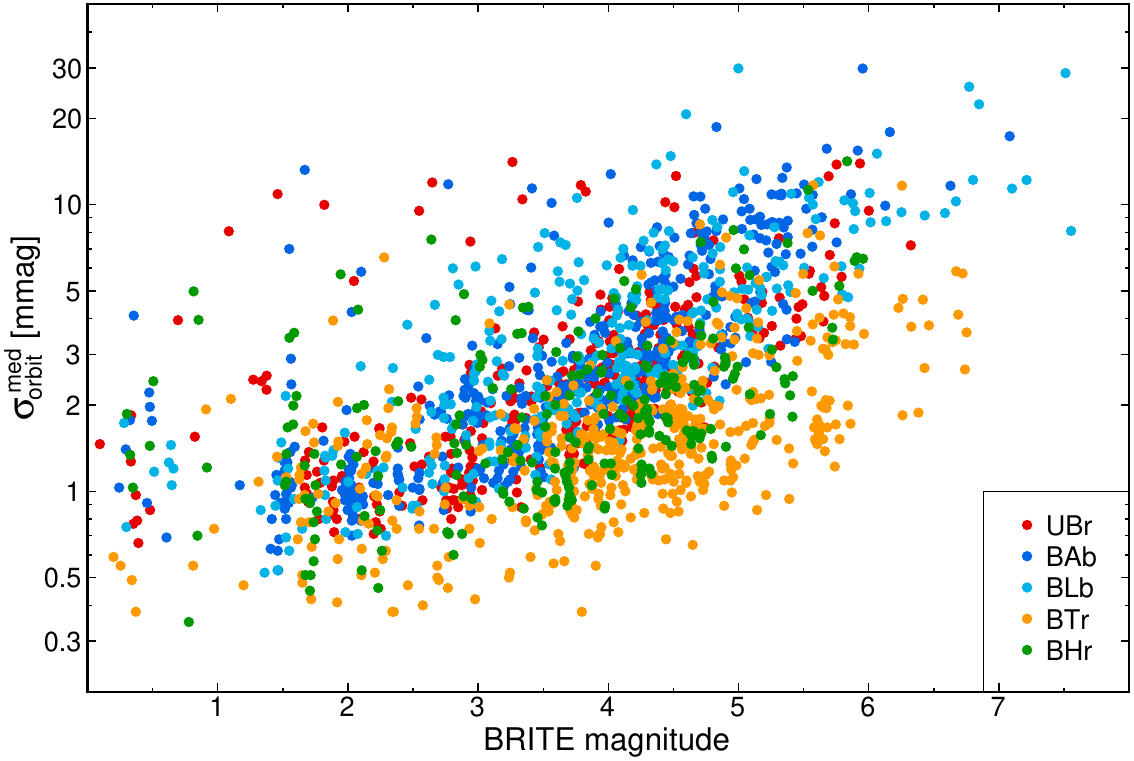}
\caption{Values of $\sigma_{\rm orbit}^{\rm med}$ plotted as a function of instrumental BRITE magnitude. Data for different BRITE satellites are shown with different colours.}
\label{fig:rms_all}
\end{figure}

In order to characterise the typical scatter in the BRITE data for stars observed in Fields 1\,--\,14, we first calculated the standard deviation of the mean for samples consisting of single satellite's orbit observations. Then, we derived the median values of these standard deviations, $\sigma_{\rm orbit}^{\rm med}$, for all 1840 setups with more than 10 orbits. The parameter $\sigma_{\rm orbit}^{\rm med}$ is similar to the root-mean-square noise per orbit.
The values of $\sigma_{\rm orbit}^{\rm med}$ are shown in Fig.\,\ref{fig:rms_all} for all setups and all five BRITE satellites as a function of mean BRITE magnitudes, calculated independently for each setup. These magnitudes are instrumental, that is, they were calculated from raw fluxes, as explained by \cite{popowicz2017}. They represent the total flux rate measured through the optimal aperture defined by the reduction pipeline. The zero point of the flux rate to magnitude conversion was chosen arbitrarily.

As can be seen in Fig.\,\ref{fig:rms_all}, the scatter in the BRITE data increases towards fainter magnitudes, as expected. For many setups the value of $\sigma_{\rm orbit}^{\rm med}$, which is a good representation of an uncertainty of data obtained during a single orbit, is smaller than 1~mmag and can be as low as $\sim$0.4~mmag for a few setups. There are also differences in scatter between different satellites which can be seen in Figs.~\ref{fig:rms_UBr} to \ref{fig:rms_BHr} in Appendix \ref{sec:quality}. 
While the values of $\sigma_{\rm orbit}^{\rm med}$ are influenced by many factors, the good performance of some satellites in the first 14 fields, especially BTr, was due to a good fine pointing and a relatively small number of defects in the detectors at the beginning of the BRITE mission.

\begin{figure*}
\sidecaption
\includegraphics[width=12cm]{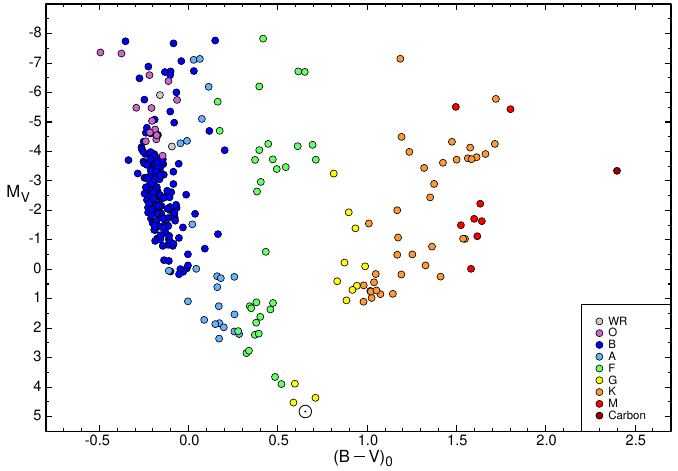}
\caption{$V$-filter absolute magnitudes ($M_V$) plotted as a function of dereddened $(B-V)_0$ colours for 300 stars observed in BRITE Fields 1\,--\,14. Stars of different spectral types are plotted with different colours, explained in the legend. The Sun symbol ($\odot$) denotes the location of the Sun.}
\label{fig:HRD}
\end{figure*}

\subsection{Data statistics}
\label{sec:data_statistics}
BRITE-Constellation has observed 300 individual targets in the first 14 fields. The $(B-V)_0$ colour versus absolute $V$ magnitude diagram for these stars is shown in Fig.\,\ref{fig:HRD}. To calculate absolute $V$ magnitudes, $M_V$, the apparent $V$ magnitudes were taken from the catalogue of \cite{2015A&A...580A..23P} or, if lacking in this catalogue, from the Simbad\footnote{http://simbad.u-strasbg.fr/simbad/} database. They were transformed to absolute magnitudes using photo-geometric distances derived by
\cite{2021AJ....161..147B} and based on the Gaia EDR3
\citep{GaiaEDR3} parallaxes. If the Gaia distances were lacking or unreliable (this was the case for about 30\% of the stars in our sample), we used Hipparcos parallaxes \citep{2007A&A...474..653V}. The total absorption, $A_V$, was estimated from the Gaia-2MASS three-dimensional map of absorption \citep{2019A&A...625A.135L} using their online tool\footnote{https://astro.acri-st.fr/gaia\_dev/}. The observed $(B-V)$ colour indices were taken primarily from \cite{2007A&A...474..653V}. If not available in this catalogue, they were taken from the Simbad database. Calculating $E(B-V)$ colour excesses, we adopted a total-to-selective absorption ratio of $A_V/E(B-V) = 3.1$. Dereddened colours shown in Fig.\,\ref{fig:HRD} were calculated as $(B-V)_0 = (B-V) - E(B-V)$. As can be seen from this figure, all stars from our sample are more luminous than the Sun.

\begin{figure}
\centering
\includegraphics[width=\columnwidth]{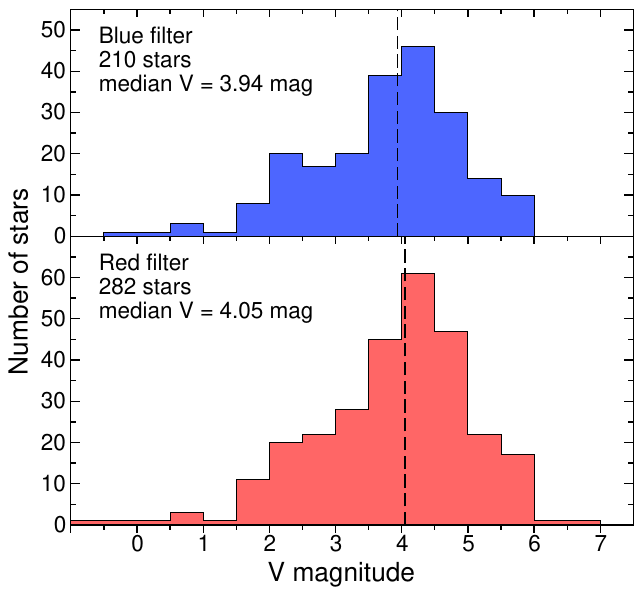}
\caption{Histograms of $V$ magnitudes for stars in our sample of stars, observed in blue (top) and red (bottom) BRITE passbands. Median values of $V$ magnitudes for subsamples of stars observed in each of the passbands are marked with vertical dashed lines.
}
\label{fig:MagDist}
\end{figure}

\begin{figure}
\centering
\includegraphics[width=\columnwidth]{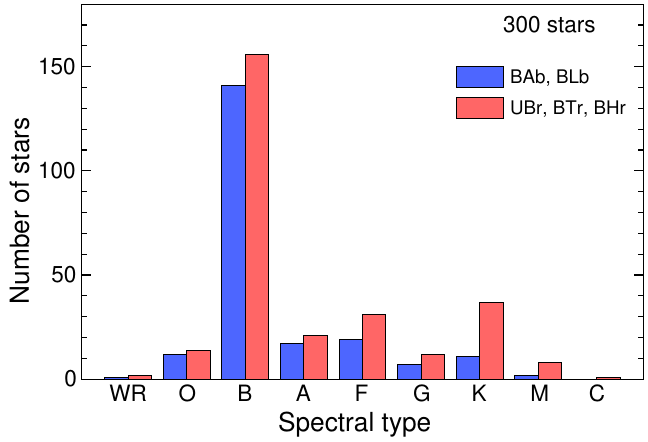}
\caption{Histogram of the spectral types of stars observed with BRITEs in Fields 1\,--\,14. Wolf-Rayet stars are put left of the O-type stars in a separate class `WR', the carbon star V460\,Cyg is marked with `C' and added to the right of the M-type stars. Blue-coloured bars (left sides) illustrate the numbers for the blue-filter observations conducted by BAb or BLb, while red-coloured bars (right sides) show the numbers for the red-filter observations carried out by UBr, BTr, or BHr.}
\label{fig:SpecTypes}
\end{figure}

BRITE-Constellation prioritises the brightest stars in the sky, as can be seen in the histograms of $V$ magnitudes of stars observed in blue and red BRITE passbands (Fig.\,\ref{fig:MagDist}). 
The median $V$ magnitude of stars for our sample amounts to 4.10\,mag. In Fig.\,\ref{fig:SpecTypes}, we show the distribution of spectral types (references are provided in the footnotes to Table \ref{stars300}) observed in Fields 1\,--\,14. It can be seen that, as mentioned in Sect.\,\ref{sec:brite}, the sample of BRITE stars is dominated by stars with the earliest spectral types, O and B. These stars account for 62\% of the sample. 

\section{Variability}\label{sec:classification}
\subsection{Known variability and VSX}\label{sect:vsx}
For the sample of 300 BRITE stars discussed in this paper, we checked whether or not any variability has been found in the past or can be seen in the BRITE data. The most complete source of information on the known stellar variability is currently The International Variable Star Index (VSX\footnote{https://www.aavso.org/vsx/index.php}, \citealt{2016yCat....102027W}) catalogue maintained by The American Association of Variable Star Observers. The VSX catalogue was initially populated with the data from the General Catalogue of Variable Stars, GCVS\footnote{http://www.sai.msu.su/gcvs/gcvs/} \citep{2017ARep...61...80S}, but has been and is being updated based on both published papers and users' analyses. The VSX classification scheme was inherited from the GCVS, but has been extended and adapted to the current state of knowledge on stellar variability. 

The variability types for known variable stars at the time of writing this text are listed in the sixth column of Table\,\ref{stars300}. The meanings of the individual classifications can be found on the VSX web page. If there is a `---' in the sixth column, it means that the star has an entry in the VSX catalogue, for example because it was suspected to be variable, but no variability type was assigned. If the sixth column contains abbreviation `n.e.' meaning  `no entry', this means that the star has no entry in the VSX catalogue. There are 64 such stars in our sample.

\subsection{Variability from BRITE data}\label{sect:var-brite}
The variability of many of the stars in our sample has already been studied using BRITE data. A list of papers based on BRITE data can be found on the BRITE Wiki page\footnote{http://brite-wiki.astro.uni.wroc.pl/bwiki/doku.php?id=bscience}. This information is also given in the eighth column of Table\,\ref{stars300}. The authors of these papers were very often able to identify the nature of the variability of the stars studied and therefore we can classify their variability. There is another source of knowledge about variability based on BRITE data. The decorrelation process of the BRITE data and the statistical analysis of their quality (Sect.\,\ref{sec:data_properties}), in which a visual inspection of the light curves was performed and frequency spectra were calculated, allow for general comments on the variability. For example, for some stars eclipses are visible, which allows the type of variability to be easily defined. In most cases, however, this inspection only allowed us to conclude that a star was variable or that variability was not detected.

The variability classification of the stars in our sample, based on the BRITE data, takes into account the two above-mentioned sources of information on variability: published papers and the inspection of light curves and frequency spectra. This classification is presented in the seventh column of Table\,\ref{stars300}. It follows the VSX catalogue classification system. For stars for which we did not detect variability based on the BRITE data from Fields 1\,--\,14, a `---' appears in the seventh column. Stars for which we have detected variability but have not investigated the nature of this variability in detail are marked as `VAR'. 

In order to present in a simple way the results of our BRITE-based classifications for the sample analysed, we divided all stars into seven broad categories: (i) pulsating stars, (ii) binary stars, (iii) Be stars, (iv) stars with stochastic variability, (v) stars with rotational modulation, (vi) stars with not yet identified type of variability, and (vii) non-variable stars. The stars were assigned to one of these seven categories based on the classification process described above. If a star presented more than one type of variability, the dominant variability is given first.

\begin{figure*}[!ht]
\centering  
\includegraphics[width=\textwidth]{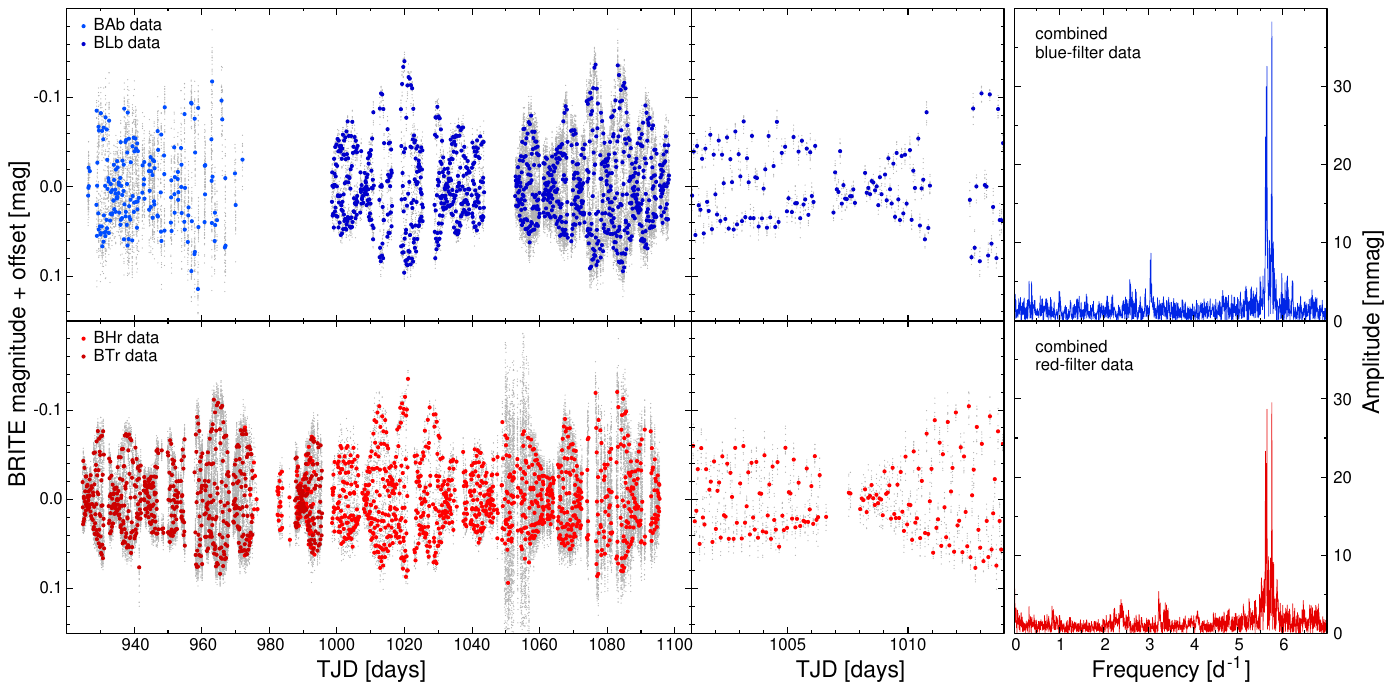}
\caption{BRITE-Constellation observations of the $\beta$ Cephei pulsator HD\,29248 ($\nu$~Eri) in field 6. Top panels show observations gathered by the blue-filter satellites, BAb and BLb, bottom panels, observations with the red-filter satellites, BHr and BTr. Left panels show the full light curve, the middle panels illustrate a 13-day long subset of the light curve, and the right panels display the Fourier spectra of the combined blue- and red-filter data. Grey dots are the decorrelated data points, while the red and blue points mark the mean instrumental magnitudes per orbit. }
\label{fig:lc_puls}
\end{figure*}

The category of `pulsating stars' includes 46 stars classified as DCEP (classical Cepheids; 4 stars), DCEPS (a subcategory of classical Cepheids; 3 stars), BCEP ($\beta$ Cephei stars; 17 stars), SPB (Slowly Pulsating B stars; 12 stars), DSCT ($\delta$ Scuti stars; 8 stars), roAp (rapidly oscillating Ap stars; 1 star) and GDOR ($\gamma$ Doradus stars; 2 stars) in the VSX classification scheme. An example of a light curve of a $\beta$ Cephei pulsator is shown in Fig.\,\ref{fig:lc_puls}. Given the observed sample of stars (Fig.\,\ref{fig:HRD}), the pulsators are mostly located on the upper main sequence. In addition to these 46 pulsating stars, a further six (3 BCEP and 3 DSCT) also show pulsational variability, although they were classified as binaries because eclipsing or proximity effects dominate their light curves. A few stars included into the Be category also show g-mode variability what classifies them as SPB.

\begin{figure*}[!ht]
\centering
\includegraphics[width=\textwidth]{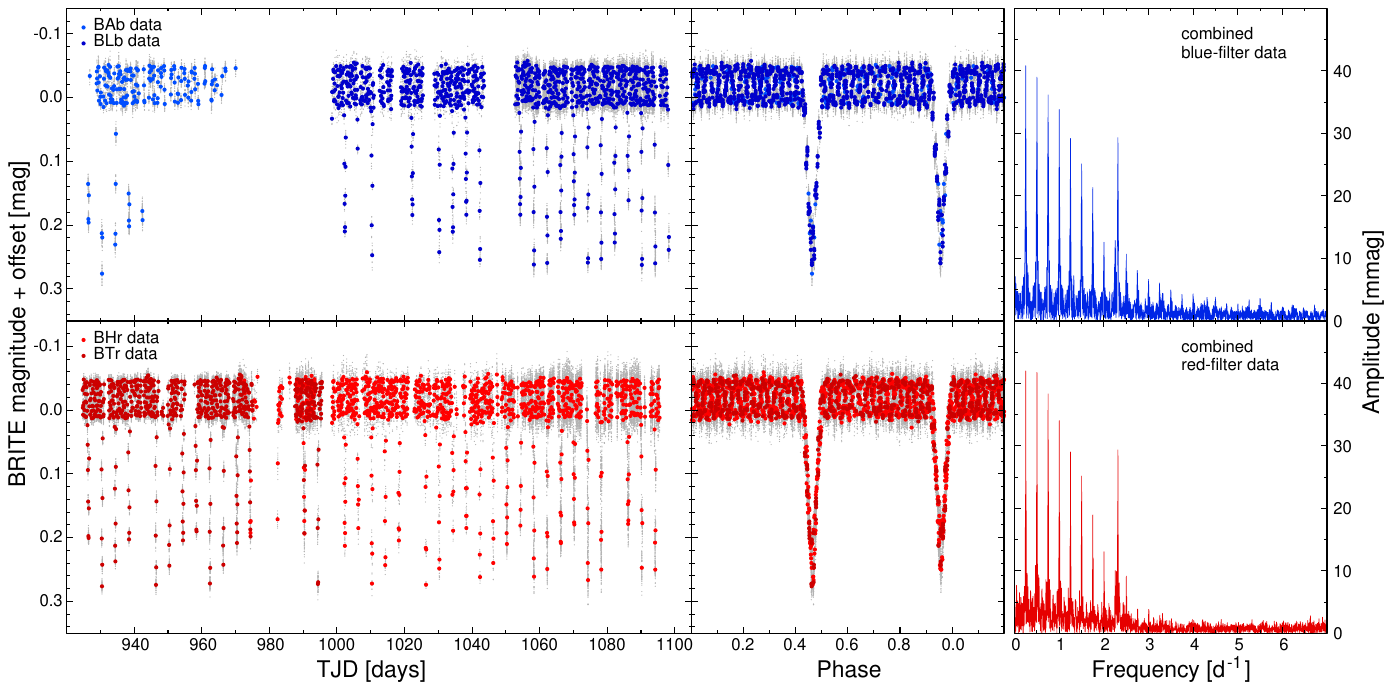}
\caption{BRITE-Constellation observations of the eclipsing binary HD\,35411 ($\eta$\,Ori) in field 6. 
Top panels show observations gathered by the blue-filter satellites, BAb and BLb, bottom panels, observations with the red-filter satellites, BHr and BTr. Left panels show the full light curve, the middle panels illustrate the light curve phased with the orbital period of 7.989454\,d, and the right panels display the Fourier spectra of the combined blue- and red-filter data. Grey dots are the decorrelated data points, while the red and blue points mark the mean instrumental magnitudes per orbit.}
\label{fig:lc_binary}
\end{figure*}

The category `binary stars' refers to stars in which binarity was detected via variability, which included eclipsing binaries and stars showing proximity effects. We did not distinguish different types of eclipsing binaries assigning to all eclipsing stars a single designation, E. In total, we included 27 stars into this category, although two more eclipsing stars were included in the category of pulsating stars, as pulsations dominate their light curves. The binary star category includes 15 stars classified as E and 12 stars in which proximity effects dominate: ELL (ellipsoidal variables; 5 stars), HB (heartbeat stars; 5 stars), and R (close binaries showing reflection effect; 2 stars). In 11 stars from the binary star category, there is also an additional variability characteristic for other categories.
As an example for the group of binaries, the blue (BAb and BLb) and red (BHr and BTr) data for HD 35411 ($\eta$\,Ori) are shown in Fig.\,\ref{fig:lc_binary}.

Be stars show variability on different time scales that are included in the VSX catalogue as short-period variability (LERI, $\lambda$~Eridani stars) and large-amplitude long-term variability (GCAS, $\gamma$~Cassiopeiae stars); less specific overall variability in Be stars is designated simply BE in VSX if the data were not sufficient to characterise the variability more precisely. Our sample includes 16 Be stars; Fig.\,\ref{fig:sampling} shows BRITE data for HD 56139 ($\omega$ CMa) as an example. Many Be stars also show pulsations in g-modes, which resulted in SPB-type variability being attributed to them as well. Of the 16 Be stars in our sample, four showed clear SPB-type pulsations.

We included 13 stars into the category of stars showing stochastic variability. The category comprises variable blue supergiants, ACYG ($\alpha$~Cygni, 11 stars), Wolf-Rayet stars, WR (1 star), and luminous blue variables, SDOR (S\,Doradus, 1 star). In addition, in four other stars the stochastic variability (3 stars classified as ACYG and one classified as WR) was associated with the dominating rotational or ellipsoidal modulation. Since stochastic solar-type oscillations do not have a separate variability class in VSX, we assigned a VAR class to stars that exhibit this type of variability, for example those studied by \cite{kallinger2019}. 

\begin{figure*}[!ht]
\includegraphics[width=\textwidth]{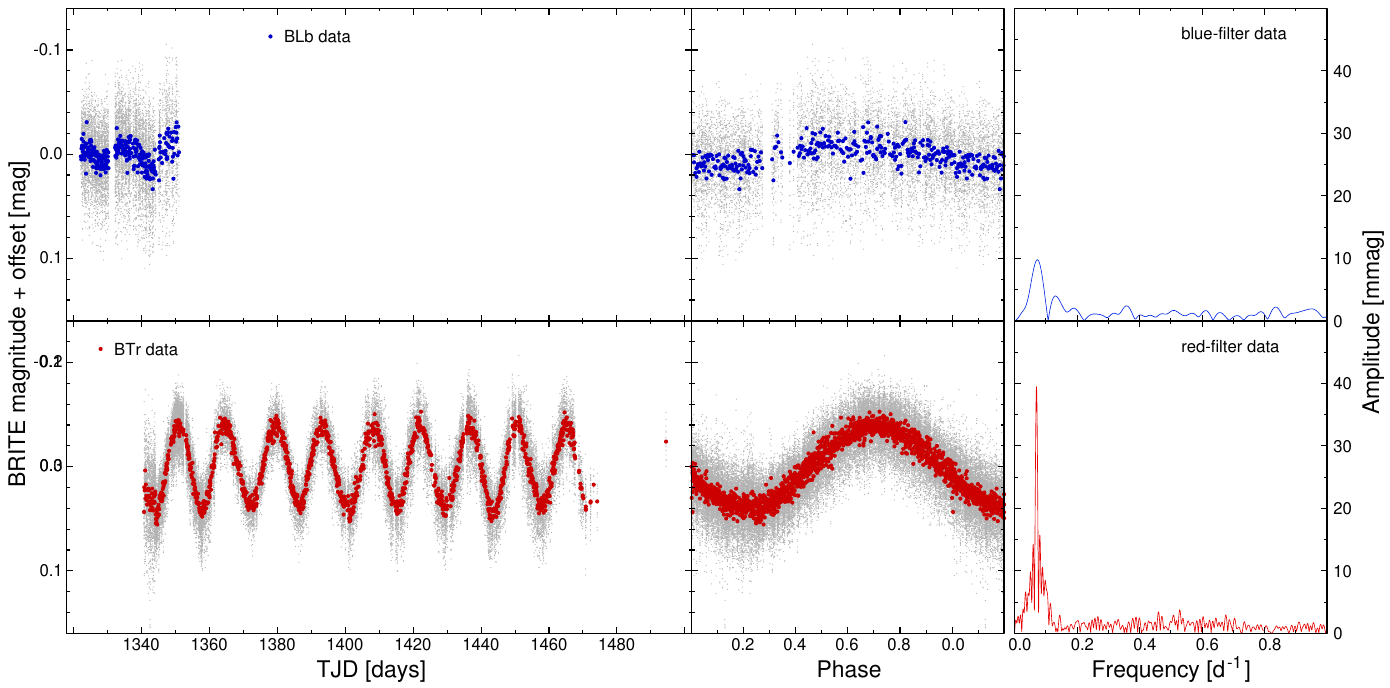}
\caption{The same as in Fig.\,\ref{fig:lc_binary} but for HD\,50123 observed in field 12. The star shows variability due to rotational modulation.}
\label{fig:lc_rotation}
\end{figure*}

Next, three stars showed a likely rotational modulation in their BRITE data, in two cases associated with another type of variability (WR and ACYG), obtained the VSX class ROT (after rotational variability). Figure \ref{fig:lc_rotation} shows HD\,50123 as an example for a light curve with rotational modulation.

All other stars showing variability, but of as yet unidentified type, were assigned to the VSX type VAR. One star was assigned the VAR type in addition to the dominating eclipse light curve. The VAR class has also been assigned to stars showing long-term variability, mostly red giants. The variability they show is usually difficult to characterise if observations are short in comparison to the time scale of the variability as is often the case for BRITE data sets. In total, there are 86 stars in our sample assigned to the VAR class.

\begin{figure}[!ht]
\centering  
\includegraphics[width=\columnwidth]{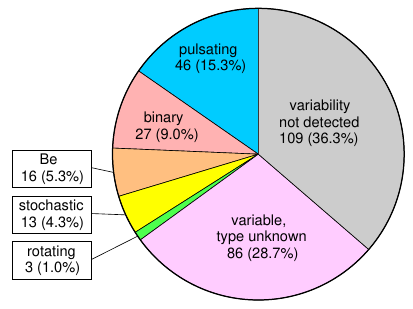}
\caption{Pie chart showing how many of the 300 stars observed in the BRITE fields 1\,--\,14 fall into each of the seven categories defined in Sect.\,\ref{sect:var-brite}.}
\label{fig:piechart}
\end{figure}

The numbers above give a total of 191 stars (64\% of the sample) included in one of the six categories for which variability was found in the BRITE data. For the remaining 109 stars (36\% of the sample) we did not find variability. These stars are marked by `---' in the seventh column of Table\,\ref{stars300}. Many of them are known to be variable stars, however, which means that either the BRITE data are too short in duration, which resulted in a poor detection threshold, or the data are of insufficient quality to detect the reported variability. The proportion of the number of stars of each category in the total sample of 300 stars observed by BRITEs in Fields 1\,--\,14 is illustrated in Fig.\,\ref{fig:piechart}.

For 125 stars (or 41.7\% of the total sample), there is agreement between the variability classes available from VSX and those inferred from the BRITE data analysis. We considered both classifications to be in agreement also when we assigned the class VAR based on the BRITE data and there was a more detailed classification in VSX. For 64 stars (21.3\% of the sample), the VSX does not have an entry that we could compare to. In addition, there are 24 stars that have a VSX entry `---' or `CST' (constant stars that have previously been identified as variable, but do not actually show any variability) for which no variability was found in the BRITE data. 

For 39 stars (13.0\% of the sample), the BRITE data do not reveal any variability, whereas in VSX they have been assigned different types of variability. The reasons why these stars were not found as variables in the BRITE data may be different. The most likely reason is an insufficient quality of the BRITE data or very short coverage. A good example is $\beta$~Cep, the prototype of a class of variable stars for which the BRITE data have a time base that is too short and a data quality that is too poor to reveal its variability. 

For another 42 stars (14.0\% of the sample), the classification based on BRITE data can be considered more reliable in comparison to previous literature results. These are also stars for which no variability type was assigned in VSX, but they were found variable in the BRITE data. Finally, for 6 stars (2.0\% of the sample) the classifications from VSX and BRITE are different (see Table \ref{stars300}. These are $\varepsilon$~Cas, $\psi$~Per, $\eta$~Tau, $\eta$~Ori, a~Car, and $\kappa$~Cen).

\section{Summary and conclusions}\label{sec:conclusions}
In this first part of the catalogue of stars observed by BRITE-Constellation, we discussed the data obtained in the first 14 fields between November 2013 and April 2016 by the five operational BRITE nano-satellites. The $V$ magnitudes of the 300 stars in this sample range from $-$0.72 to 6.91, showing that the BRITE satellites were able to obtain photometry for stars much fainter than the originally assumed limit of $V =$~4.0 mag. All 14 fields but Field 8 (VelPic-I-2015) were located in the Galactic plane, as this ensures the availability of a significant number of bright stars in the field of view which is required for attitude control. Many of the stars have been observed multiple times in the successive years, giving a longer, sometimes up to 9-year long time base. Here we discuss only the first $\sim$2.5 years of data. In twelve of the 14 fields, photometric time series were obtained by more than one BRITE satellite. Only Fields 3 (Sgr-I-2014) and 8 (VelPic-I-2015) were observed by a single satellite. 

The raw BRITE data suffer from instrumental effects due to changing temperature of the optics and changing position of a star on a detector. Over the past few years, dedicated software has been developed to remove instrumental effects to a large extent. We describe the features of the decorrelated data, which we make available to the public through the BRITE-Constellation PDA and CADC.

One of the aims of this work is to encourage researchers to use BRITE photometry in their scientific work. The decorrelated data that are presented here are ready for use. From the point of view of a user, the most interesting is probably the group of 86 stars known to be variable from BRITE data, which has not been analysed in detail and has not yet been used in scientific work. For another 115 stars (38\% of the sample), BRITE data have been used in publications. However, new data have been secured for many of them since the first analysis. For these stars, it is still possible to obtain new results and, in particular, to find low-amplitude variability not detectable in a shorter data set (see \cite{weiss2021} as an example). Another advantage of the BRITE data is the temporal coverage, which for some stars spans nine consecutive seasons. Such a long coverage greatly improves the resolution in frequency, which can be important, for example, for the studies of stars with dense pulsation spectra, such as SPB-type stars.

\begin{acknowledgements}
APi was supported by the National Science Centre, Poland, grant No.\,2022/45/B/ST9/03862. GAW acknowledges scientific funding from the Canadian Space Agency in support of decorrelation and data release efforts. CL and GAW acknowledge Discovery Grant support by the Natural Science and Engineering Research Council (NSERC) of Canada. GH acknowledges financial support by the Polish National Science Centre (NCN) under grants 2015/18/A/ST9/00578 and 2021/43/B/ST9/02972. PGB acknowledges the financial support by NAWI Graz and support by the Spanish Ministry of Science and Innovation with the \textit{Ram{\'o}n\,y\,Cajal} fellowship number RYC-2021-033137-I and the number MRR4032204.  DMB gratefully acknowledges a senior postdoctoral fellowship from the Research Foundation Flanders (FWO) with grant agreement no.\,1286521N, and the Engineering and Physical Sciences Research Council (EPSRC) of UK Research and Innovation (UKRI) in the form of a Frontier Research grant under the UK government’s Horizon Europe funding guarantee (grant number [EP/Y031059/1]). RS acknowledges financial support by the Polish National Science Centre (NCN) under the Sonata Bis grant 2018/30/E/ST9/00598.
\end{acknowledgements}

\bibliographystyle{aa}
\bibliography{references}

\begin{appendix}
\section{BRITE-Constellation targets in the first 14 fields}

Table \ref{stars300} provides an overview of the 300 stars observed by BRITE-Constellation included in the first 14 fields. The columns used are HD identifier (HD), other commonly used identifier (Other designation), $V$ magnitude ($V$) and spectral classification with references (MK spectral type). The BRITE field numbers in which the star was observed are then given (BRITE field(s)). Numbers in brackets indicate fields from which data will be discussed in subsequent papers. In the two columns entitled `Class of variability', we list the variability type assigned in the VSX database (VSX, see Sect.\,\ref{sect:vsx}) and the classification based on the BRITE-Constellation data (this paper; see also Sect.\,\ref{sect:var-brite}). We follow the classification scheme used in the VSX database.\footnote{A description of the VSX variability types is available at https://www.aavso.org/vsx/index.php?view=about.vartypes.}
The last column in Table \ref{stars300} provides the publications that used BRITE-Constellation data for presented stars and notes on the objects. In a few cases, two HD numbers are given in one line, which may be due to one of two reasons: either these are visual double stars that were not resolved in the BRITE photometry (e.g.~for HD\,36861/2), or a single star was given two HD numbers in the original catalogue (e.g.~for HD\,195068/9).

\begin{table*}[htb]
\caption{General information on the 300 BRITE-Constellation target stars included in Fields 1 to 14.}
\label{stars300}
\centering\scriptsize
\begin{tabular}{rccccccl}
\noalign{\smallskip}\hline\hline\noalign{\smallskip}
HD & Other & $V$ & MK spectral & BRITE & \multicolumn{2}{c}{Variability class}  & Notes, references\\
& designation & (mag) & type & field(s) & VSX & this paper & to BRITE papers \\
\noalign{\smallskip}\hline\noalign{\smallskip}
432 & $\beta$ Cas & 2.27 & F2\,IV [1] & 11 (19,30,39,57) & DSCTC & DSCT & (28,61,75) \\
2905 & $\kappa$ Cas & 4.16 & B1\,Ia [2] & 11 (19,30,39,57) & ACYG & ACYG & (21,46) \\
3360 & $\zeta$ Cas & 3.66 & B2\,IV [3] & 11 (19,30,39) & --- & SPB & \\
3712 & $\alpha$ Cas & 2.23 & K0\,IIIa [4] & 11 (19) & ROT: & --- &  \\
3901 & $\xi$ Cas & 4.81 & B2\,V [1] &  11 & --- & --- &  \\
4614 & $\eta$ Cas & 3.44 & G0\,V\,+\,K7\,V [4,5] & 11 (19) & CST &  --- &  \\
5394 & $\gamma$ Cas & 2.47 & B0\,IV:e [6] & 11 (19,30,39,57) & GCAS+X+LERI: & GCAS+SPB & (26,64,70) \\
6811 & $\phi$ And & 4.25 & B7\,III [7]&  11 & BE: & --- &  \\
6961 & $\theta$ Cas & 4.33 & A7\,IV [8] & 11 (19,30,39) & --- & --- &  \\
8538 & $\delta$ Cas & 2.68 & A5\,IV [9] & 11 (19,30,39) & EA: & --- &  \\
\noalign{\smallskip}
11415 & $\varepsilon$ Cas & 3.37 & B3\,Vp shell [10] & 11 (19,30,39) & SXARI: & BE & \\
16908 & 35 Ari & 4.66 & B3\,V [10] & 5 & n.e. & --- & \\
17573 & 41 Ari & 3.63 & B8\,Vn [7] & 5 & --- & --- & \\
17584 & 16 Per & 4.23 & F2\,III [1] & 5 & DSCT: & DSCT & \\
17709 & 17 Per & 4.53 & K7\,III [3] & 5 & LB & --- & \\
18296 & 21 (LT) Per & 5.11 & B9p\,Si [8] & 5 & ACV & VAR & \\
19058 & $\rho$ Per & 3.39 & M4\,II [4] & 5 & SRB & VAR & (57) \\
19356 & $\beta$ Per & 2.12 & B8\,V\,+\,A/F [11] & 5 & EA/SD & E & \\
19373 & $\iota$ Per & 4.05 & G0\,V [4] & 5 & n.e. & --- & \\
19476 & $\kappa$ Per & 3.80 & K0\,III [3] & 5 & --- & --- & \\
\noalign{\smallskip}
20365 & 29 Per & 5.15 & B3\,V [10] & 5 & n.e. & --- & $\alpha$~Per, (4)\\
20418 & 31 Per & 5.03 & B5\,IV [10] & 5 & n.e. & --- & $\alpha$~Per, (4)\\
20468 & --- & 4.82 & K2\,II\,CN0.5 [12] & 5 & n.e. & --- & \\
20809 & --- & 5.29 & B4\,IV [10] & 5 & SPB: & VAR & $\alpha$~Per, (4)\\
20902 & $\alpha$ Per & 1.79 & F5\,Ib [2] & 5 & --- & --- & $\alpha$~Per\\
21428 & 34 Per & 4.67 & B5\,V [10] & 5 & n.e. & --- & $\alpha$~Per, (4)\\
21552 & $\sigma$ Per & 4.36 & K3\,III [13] & 5 & LB & VAR & (57) \\
22192 & $\psi$ Per & 4.23 & B5\,Ve [10] & 5 & GCAS & SPB & $\alpha$~Per, (4)\\
22780 & --- & 5.57 & B7\,Vn [10] & 5 & GCAS: & --- & \\
22928 & $\delta$ Per & 3.01 & B5\,III [10] & 5 (20) & GCAS: & VAR & \\
\noalign{\smallskip}
23180 & $o$ Per & 3.83 & B1\,III [3] & 5 & ELL & VAR & (21,46) \\
23230 & $\nu$ Per & 3.77 & F5\,II [2] & 5 (20) & --- & --- & \\
23302 & 17 Tau & 3.70 & B6\,III [1] & 5 & SPB & --- & Pleiades\\
23338 & 19 Tau & 4.30 & B6\,V [1] & 5 & SPB & --- & Pleiades\\
23408 & 20 Tau & 3.87 & B7\,III [1] & 5 & ACV & VAR & Pleiades\\
23480 & 23 (V971) Tau & 4.18 & B6\,IVnn & 5 & SPB | LERI & --- & Pleiades\\
23630 & $\eta$ Tau & 2.87 & B7 III [3] & 5 & ROT+SPB & BE & Pleiades\\
23850 & 27 Tau & 3.63 & B8\,III [1] & 5 & SPB & SPB & Pleiades, (49)\\
24398 & $\zeta$ Per & 2.85 & B1\,Ib [2] & 5 (20) & --- & ACYG & (21,46) \\
24640 & --- & 5.49 & B1.5\,V [10] & 5 & BCEP: & --- & \\
\noalign{\smallskip}
24760 & $\varepsilon$ Per & 2.89 & B0.5\,III [10] & 5 (20) & BCEP & BCEP+SPB & (20,42) \\
24912 & $\xi$ Per & 4.04 & O7.5\,III((f))(n) [14] & 5 (20) & $\ast$ & VAR & \\
25823 & 41 (GS) Tau & 5.20 & B9\,Vp\,Si [8] & 5 & ACV & VAR & \\
25940 & 48 (MX) Per & 4.04 & B3\,Ve [10] & 5 (20) & GCAS & VAR & \\
25998 & 50 (V582) Per & 5.51 & F7\,V [15] & 5 & RS: & --- & \\
26322 & 44 (IM) Tau & 5.41 & F3\,V [15] & 5 & DSCT & DSCT & \\
26630 & $\mu$ Per & 4.14 & G0\,Ib [2] & 5 (20) & --- & --- & \\
27396 & 53 (V469) Per & 4.85 & B4\,IV [10] & 5 (20) & SPB & SPB & (27,51) \\
29248 & $\nu$ Eri & 3.93 & B2\,III [10] & 6,13 (22,32,40,49,58,59,65) & BCEP+SPB & BCEP+SPB & (6,7,11,19,23) \\
30211 & $\mu$ Eri & 4.02 & B4\,IV [10] & 6,13 (22,32,40,49,59,65) & EA+SPB & E+SPB & \\
\noalign{\smallskip}
30652 & $\pi^3$\,Ori & 3.19 & F6\,V [3] & 6,(40,65) & --- & --- & \\
30836 & $\pi^4$\,Ori & 3.69 & B2\,III [1] & 6,13 (22,32,40,49,58,59,65) & VAR & VAR& \\
31109 & $\omega$ Eri & 4.39 & A9\,IVn [16] & 6 & GDOR+DSCT & VAR & \\
31139 & 5 Ori & 5.33 & M1\,III [17] & 6 & --- & VAR & \\
31237 & $\pi^5$\,Ori & 3.72 & B2\,III [10] & 1,6,13 (22,32,40,49,58,59,65) & ELL & ELL+SPB & (67)\\
31767 & $\pi^6$\,Ori & 4.47 & K2\,II [2] & 6 (58) & --- & --- & \\
33111 & $\beta$ Eri & 2.79 & A3\,III [18] & 1,6 (32,40,48,58) & --- & VAR & \\
33328 & $\lambda$ Eri & 4.27 & B2\,III(e)p [19] & 6,13 (22,32) & LERI+GCAS & VAR & \\
33904 & $\mu$ Lep & 3.31 & B9p HgMn [8] & 6 & ACV & --- & \\
34085 & $\beta$\,Ori & 0.12 & B8\,Ia [2]& 1,6,13 (22,32,40,49,59,65) & ACYG & VAR & \\
\noalign{\smallskip}
34503 & $\tau$\,Ori & 3.60 & B5\,III [1] & 1,6,13 (22,49,59,65) & n.e. & HB & (44) \\
34816 & $\lambda$ Lep & 4.29 & B0.5\,IV [6] & 6 (22) & n.e. & VAR & \\
35039 & 22 Ori & 4.74 & B2\,IV-V [10] & 6 (40,49,59,65) & BCEP: & --- & \\
35369 & 29 Ori & 4.14 & G8\,III [13] & 6 & n.e. & --- & \\
35411 & $\eta$\,Ori & 3.36 & B0.7\,V\,+\,B1.5:\,V [20] & 1,6,13 (22,32,40,49,58,59,65) & EA+BCEP: & E+ELL & (17) \\
35439 & $\psi^1$\,Ori & 4.95 & B1\,V:ep [6] & 6,13 (22,32,40,49,58,59,65) & GCAS & BE+SPB & (26,39) \\
35468 & $\gamma$\,Ori & 1.64 & B2\,III [1] & 1,6,13 (22,32,40,49,59,65) & --- & --- & \\
35715 & $\psi^2$\,Ori & 4.59 & B1\,V [10] & 1,6 (22,32,40,49,59,65) & E/D & ELL+BCEP & (17) \\
36267 & 32 Ori & 4.20 & B5\,V [10] & 6 (32,40,65) & n.e. & VAR & \\
36486/5 & $\delta^1$/$\delta^2$\,Ori & 2.23/6.83 & O9.5\,II-III\,+\,B2\,IV-V [10] & 1,6,13 (22,32,40,48,49,58,59,65) & EA+VAR: & E+VAR & (65,79) \\
\noalign{\smallskip}
36512 & $\upsilon$\,Ori & 4.62 & B0\,V [10] & 6 & BCEPS: & --- & \\
36822 & $\phi^1$\,Ori & 4.41 & B0.5\,IV-V [10] & 6 (40,58) & n.e. & --- & \\
36861/2 & $\lambda^1$/$\lambda^2$\,Ori & 3.54/5.45 & O8\,IIIf\,+\,B0.5\,V [21] & 1,6,13 (22,32,40,48,58,65) & --- & --- & \\
36960/59 & --- & 5.67/4.78 & B0.5\,V\,+\,B1\,V [10] & 6 & VAR: & --- & \\
37018 & 42 Ori & 4.59 & B1\,V [10] & 6 (49,65) & CST & --- & \\
37022/41 & $\theta^1$/$\theta^2$\,Ori & 5.13/5.09 & O6pe\,+\,O9.5\,Vpe [10] & 6,13 & --- & VAR & \\
37043 & $\iota$\,Ori & 2.77 & O8.5\,III\,+\,B0 [21] & 1,6,13 (22,32,49,58,59,65) & HB & HB & (8,11,22) \\
37128 & $\varepsilon$\,Ori & 1.70 & B0\,Ia [2] & 1,6,13 (22,32,40,48,49,58,59,65) & ACYG & ACYG & (52) \\
\noalign{\smallskip}\hline
\end{tabular}
\end{table*}

\setcounter{table}{0}
\begin{table*}[!ht]
\caption{continued.}
\centering\scriptsize
\begin{tabular}{rccccccl}
\noalign{\smallskip}\hline\hline\noalign{\smallskip}
HD & Other & $V$ & MK spectral & BRITE & \multicolumn{2}{c}{Variability class}  & Notes, references\\
& designation & (mag) & type & field(s) & VSX & this paper & to BRITE papers \\
\noalign{\smallskip}\hline\noalign{\smallskip}
37468 & $\sigma$\,Ori & 3.81 & O9.5\,V [10] & 1,6,13 (22,32,40,49,58,59,65) & VAR: & VAR & (76) \\
37490 & $\omega$\,Ori & 4.57 & B2\,III(e) [10] & 6,13 (22,32,40,58) & GCAS & VAR & (26) \\
\noalign{\smallskip}
37742/3 & $\zeta$\,Ori A/B & 1.88/3.70 & O9.5\,Ib\,+\,O9.5\,[2] & 1,6,13 (22,32,40,48,49,58,59,65) & VAR & ACYG & (5,9,15) \\
38771 & $\kappa$\,Ori & 2.06 & B0.5\,Ia [2] & 1,6,13 (22,32,49,59,65) & ACYG & ACYG & \\
39060 & $\beta$ Pic & 3.85 & A5\,IV shell [19] & 8 (23,33,50,61) & DSCT+EP: & DSCT& (28,37,48,60) \\
39801 & $\alpha$\,Ori & 0.50 & M2\,Iab [2] & 1,6,13 (22,32,40,49,59,65) & SRC & VAR & \\
42933 & $\delta$ Pic & 4.81 & B0.5\,IV [23] & 8 (23,33,50,61) & EB/D: & E+BCEP & (17) \\
44402 & $\zeta$ CMa & 3.02 & B2.5\,IV [23] & 12 & BCEP: & VAR & \\
44743 & $\beta$ CMa & 1.98 & B1\,II-III [10] & 12 & BCEP & BCEP & (19) \\
45348 & $\alpha$ Car & $-$0.72 & F0\,Iab [24] & 8 (23) & n.e. & --- & \\
45871 & IY CMa & 5.74 & B5\,V(e) [19] & 12 & E: & VAR & \\
46328 & $\xi^1$ CMa & 4.33 & B0.5\,IV [6] & 12 & BCEP & BCEP & (19,47,63) \\
\noalign{\smallskip}
47306 & N Car & 4.40 & A0\,II [25] & 8 (23) & n.e. & --- & \\
47670 & $\nu$ Pup & 3.17 & B8\,III [26] & 8 (23,33,41,50,61) & LERI & SPB+BE & (53) \\
48917 & 10 (FT) CMa & 5.20 & B2\,IIIe [23] & 12 & GCAS & BE & (26) \\
49131 & HP CMa & 5.80 & B1.5\,Vne [27] & 12 & GCAS+LERI & --- & \\
50013 & $\kappa$ CMa & 3.96 & B1.5\,IVne [23] & 12 (41) & GCAS & BE & (26) \\
50123 & HZ CMa & 5.70 & B6\,IVe\,+\,A [19] & 12 & ELL & ELL & \\
50337 & V415 Car & 4.42 & G6\,II\,+\,A0\,V [28,29] & 8 (33,50,61,66) & EA/GS & --- & \\
50707 & 15 (EY) CMa & 4.83 & B1\,IV [6] & 12 & BCEP & BCEP & (19) \\
50877 & $o^1$ CMa & 3.87 & K3 Iab [2] & 12 & LC & VAR & \\
50896 & EZ CMa & 6.91 & WN5-B [30] & 12 & WR & ROT+WR & WR 6, (13,38,56,72) \\
\noalign{\smallskip}
51309 & $\iota$ CMa & 4.37 & B3\,II [2] & 12 & BCEP & BCEP& \\
52089 & $\varepsilon$ CMa & 1.50 & B2\,II [2] & 12 (41) & n.e. & VAR & (14) \\
52670 & LS CMa & 5.63 & B2\,V [23] & 12 & EA+SPB & E & (59) \\
52877 & $\sigma$ CMa & 3.47 & K7\,Ib [3] & 12 (41) & LC & VAR & (57) \\
53138 & $o^2$ CMa & 3.02 & B3\,Ia [2] & 12 (41) & ACYG & VAR & \\
53244 & $\gamma$ CMa & 4.12 & B8\,II [10] & 12 & --- & --- & \\
54309 & FV CMa & 5.71 & B2\,IVe [23] & 12 & GCAS & --- & \\
54605 & $\delta$ CMa & 1.84 & F8\,Ia [2] & 12 (41) & VAR & --- & \\
55892 & 71 (QW) Pup & 4.49 & F0\,IV [31] & 8 (23,33,50,61,66) & GDOR & GDOR & (28,69) \\
56014 & 27 (EW) CMa & 4.66 & B3\,IVe [32] & 12 (41) & GCAS & BE+SPB & (26)\\
\noalign{\smallskip}
56022 & 72 (OU) Pup & 4.87 & Ap\,SiSr [18] & 8 (23,61) & ACV & --- & \\
56139 & $\omega$ CMa & 3.85 & B3\,IVe [33] & 12 (41) & GCAS & BE & (26) \\
56455 & PR Pup & 5.71 & Ap\,Si [34] & 8 & ACV & VAR & \\
56855 & $\pi$ Pup & 2.70 & K4\,III\,+\,B5 [35,36] & 8 (33,41,50,61,66) & SRD: & --- & \\
57060 & 29 (UW) CMa & 4.98 & O8.5\,If [21] & 12 (41) & EB/KE: & E+VAR & (43) \\
57061 & $\tau$ CMa & 4.40 & O9\,III [1] & 12 (41) & EB & E & in NGC\,2362\\
58155 & NO CMa & 5.43 & B3\,IV(e) [33] & 12 & BE: & VAR & \\
58286 & --- & 5.39 & B2\,V [23] & 12 & n.e. & --- & \\
58343 & FW CMa & 5.33 & B2.5\,V(e) [10] & 12 & GCAS & VAR & (26) \\
58350 & $\eta$ CMa & 2.45 & B5\,Ia [2] & 12 (41) & ACYG & VAR & \\
\noalign{\smallskip}
61068 & PT Pup & 5.74 & B2\,III [6] & 12 & BCEP & BCEP+SPB & (19) \\
61715 & MY Pup & 5.68 & F7\,Ib/II [37] & 8 (23,33,50,61) & DCEPS & DCEPS & (31,41) \\
62623 & 3 Pup & 3.96 & A2\,Iab [23] & 12 (41) & ACYG & ACYG & \\
62747 & V390 Pup & 5.60 & B1.5\,III [23] & 12 & EA & E & \\
63462 & $o$ Pup & 4.50 & B1\,Ve [24] & 12 & LERI: & VAR & \\
63744 & Q Pup & 4.71 & K0\,III [38] & 7 & n.e. & --- & \\
63922 & P Pup & 4.11 & B0.5\,III [24] & 7,8 (23,50,61) & n.e. & --- & \\
63949 & QS Pup & 5.81 & B1.5\,IV [23] & 8 (23,33) & BCEP & --- & \\
64440 & a Pup & 3.73 & K1.5\,II\,+\,A [37] & 7,8 (34,41,42,50,61,66) & n.e. & VAR & \\
64740 & --- & 4.63 & B1.5\,IVp [23] & 7,8,14 (23,25,66) & --- & VAR & \\
\noalign{\smallskip}
64760 & J Pup & 4.24 & B0.5\,Ib [23] & 7,8,14 (23,25,33,50,61) & VAR & VAR & \\
65575 & $\chi$ Car & 3.47 & B3\,IVp Si [23] & 7,8,14 (23,25,33,34,42,50,51,61,66) & BCEP & VAR & \\
65818 & V Pup & 4.41 & B1\,Vp\,+\,B2: [23] & 7,8,14 (23,25,33,50,61,66) & EB/SD & E & (17) \\
66811 & $\zeta$ Pup & 2.25 & O4\,I(n)fp [39] & 7,8,14 (23,25,34,41,42,50,51,60,61,66) & ROT: & ROT+ACYG & (33,71) \\
67523 & $\rho$ Pup & 2.81 & F6\,II [40] & 12 (41) & DSCT & DSCT & \\
68273/43 & $\gamma^2$/$\gamma^1$ Vel & 4.20/1.83 & WC8\,+\,O7\,+\,B1\,IV [41,23] & 7,8,14 (23,25,34,42,51,60,66) & WR & WR & WR\,11, (12,13,38) \\
68553 & h$^1$ (NS) Pup & 4.45 & K4\,III [42] & 7 & LC & VAR & (57) \\
69142 & h$^2$ Pup & 4.44 & K1\,II/III [37] & 7 & n.e. & --- & \\
71129 & $\varepsilon$ Car & 2.01 & K3:\,III\,+\,B2:\,V [28] & 7 (25,34,42,51,60) & E: & VAR & \\
72127 & --- & 4.99 & B2\,IV [23] & 7 & VAR: & --- & \\
\noalign{\smallskip}
73634 & e Vel & 4.14 & A9\,I-II [24] & 7 (25,66) & n.e. & --- & \\
74006 & $\beta$ Pyx & 3.97 & G5 II/III [42] & 7 (34,42) & n.e. & VAR & (57) \\
74180 & b Vel & 3.84 & F2\,Ia [40] & 7 (25,66) & --- & --- & \\
74195 & $o$ Vel & 3.62 & B3\,III [24] & 7,14 (25,34,42,51,66) & SPB & SPB & \\
74375 & d (V343) Car & 4.33 & B1\,III [24] & 7 (24,25,34,36,42,43,51) & BCEP: & VAR & \\
74560 & HY Vel & 4.86 & B4\,IV [43] & 7 & SPB & VAR & in IC\,2391\\
74575 & $\alpha$ Pyx & 3.68 & B1.5\,III [23] & 7 & BCEP & VAR & \\
74772 & d Vel & 4.07 & G5\,III [37] & 7 & n.e. & --- & \\
74956 & $\delta$ Vel & 1.96 & A1\,V [18] & 7 (25,34,42,51,60,66) & EA & E & (59) \\
75063 & a Vel & 3.91 & A1\,III [18] & 7 & n.e. & VAR & \\
\noalign{\smallskip}
75311 & f (V344) Car & 4.49 & B3\,Vne [24] & 7,14 (24,25,36,43,53,62,67) & GCAS & BE & \\
75821 & f (KX) Vel & 5.12 & B0\,III [6] & 7 & EA & --- & \\
76728 & c Car & 3.84 & B8\,II [24] & 7,14 (24,25,34,42,51) & n.e. & VAR & \\
77002 & b$^1$ (V376) Car & 4.89 & B3\,IV [24] & 7 & BCEPS & --- & \\
78004 & c Vel & 3.75 & K2\,III [40] & 7 & n.e. & --- & \\
\noalign{\smallskip}\hline
\end{tabular}
\end{table*}

\setcounter{table}{0}
\begin{table*}[!ht]
\caption{continued.}
\centering\scriptsize
\begin{tabular}{rccccccl}
\noalign{\smallskip}\hline\hline\noalign{\smallskip}
HD & Other & $V$ & MK spectral & BRITE & \multicolumn{2}{c}{Variability class}  & Notes, references\\
& designation & (mag) & type & field(s) & VSX & this paper & to BRITE papers \\
\noalign{\smallskip}\hline\noalign{\smallskip}
78647 & $\lambda$ Vel & 2.21 & K4\,Ib [44] & 7 (25,34,42) & LC & VAR & \\
79351 & a (V357) Car & 3.44 & B2\,IV [24] & 7,14 (24,25,34,36,42,43,51,53) & BE: & SPB & (18) \\
79940 & k Vel & 4.62 & F5\,III [24] & 7 & n.e. & --- & \\
80230 & g Car & 4.34 & M1\,III [28] & 7 (24) & --- & VAR & (57) \\
80404 & $\iota$ Car & 2.25 & F0\,Iab [24] & 7,14 (24,25,34,42,51) & --- & VAR & (30) \\
\noalign{\smallskip}
81188 & $\kappa$ Vel & 2.50 & B2\,IV [24] & 7,14 (24,25,34,36,42,51,53,60) & n.e. & SPB & (18) \\
82434 & $\psi$ Vel & 3.60 & F2\,IV [24] & 7 & n.e. & --- & \\
82668 & N Vel & 3.13 & K5\,III [35] & 7 (24,25,34,36,42,43) & SR & VAR & (57) \\
83183 & h Car & 4.08 & B5\,II [23] & 7 (24,36,53) & --- & --- & \\
83446 & M Vel & 4.35 & A5\,V [18] & 7 (36,43) & n.e. & DSCT & (28) \\
86440 & $\phi$ Vel & 3.54 & B5\,II [24] & 7 (24,36,43,53) & n.e. & --- & \\
118716 & $\varepsilon$ Cen & 2.30 & B1\,III [23] & 2 (15,35) & BCEP & BCEP & (20) \\
120307 & $\nu$ Cen & 3.41 & B2\,IV [23] & 2 (35) & R & R & (17,77) \\
120324 & $\mu$ Cen & 3.04 & B2\,IV-Ve [19] & 2 (35) & GCAS & BE & (3,11) \\
121263 & $\zeta$ Cen & 2.55 & B2.5\,IV [23] & 2 (15,35) & VAR & HB & (11) \\
\noalign{\smallskip}
121743 & $\phi$ Cen & 3.83 & B2\,IV [23] & 2 (35) & --- & VAR  & \\
121790 & $\upsilon^1$ Cen & 3.87 & B2\,IV-V [23] & 2 (35) & L & --- & (24) \\
122451 & $\beta$ Cen & 0.61 & B1\,III [23] & 2 (15,35,62) & BCEP & BCEP+SPB & (2,11,68) \\
122980 & $\chi$ Cen & 4.36 & B2 V [23] & 2 (35) & BCEPS & --- & (24) \\
125238 & $\iota$ Lup & 3.55 & B2.5\,IV [23] & 2 (35) & BCEP: & VAR & \\
125823 & a (V761) Cen & 4.42 & B7\,IIIp [23] & 2 (35) & SXARI & ROT & (14,35,66) \\
126341 & $\tau^1$ Lup & 4.56 & B2\,IV [23] & 2 (35) & BCEP & BCEP+SPB & (25) \\
126354 & $\tau^2$ Lup & 4.35 & F4\,IV\,+\,A7: [45] & 2 & n.e. & VAR & \\
127381 & $\sigma$ Lup & 4.42 & B2\,III [23] & 2 (35) & SXARI & VAR & \\
127972/3 & $\eta$ Cen & 2.31 & B2\,Vnne [46] & 2 (35) & GCAS+LERI & BE & (3,11) \\
\noalign{\smallskip}
128345 & $\rho$ Lup & 4.05 & B5 V [23] & 2 (35) & SPB: & VAR & \\
128620/1 & $\alpha$ Cen & $-$0.01 & G2\,V\,+\,K1\,V [28] & 2 (15,35,62) & BY: & VAR & (73) \\
128898 & $\alpha$ Cir & 3.19 & Ap\,SrEuCr: [28] & 2 (15,35,62) & roAp+ACV & roAp+ACV & (1,11,74) \\
129056 & $\alpha$ Lup & 2.30 & B1.5\,III [23] & 2 (35) & BCEP & BCEP+SPB & (19,23) \\
129116 & b Cen & 4.00 & B3\,V [23] & 2 & n.e. & --- & (24) \\
130807 & $o$ Lup & 4.32 & B5\,IV [23] & 2 (35) & VAR: & SPB+ROT & (55) \\
132058 & $\beta$ Lup & 2.68 & B2\,III [23] & 2 (35) & n.e. & SPB & (25) \\
132200 & $\kappa$ Cen & 3.13 & B2\,IV [23] & 2 (35) & BCEP & SPB & (18) \\
133242/3 & $\pi$ Lup & 4.72 & B5\,V\,+\,B5\,IV & 2 (35) & n.e. & --- & (24) \\
134481/2 & $\kappa^1$/$\kappa^2$ Lup & 3.87 & B9.5\,Vn\,+\,A3/5\,V [28] & 2 (35) & n.e. & --- & \\
\noalign{\smallskip}
134505 & $\zeta$ Lup & 3.41 & G8\,III [44] & 2 (35) & n.e. & --- & \\
135379 & $\beta$ Cir & 4.07 & A3\,IV [24]& 2 & n.e. & VAR & \\
135734 & $\mu^1$/$\mu^2$ Lup & 4.27 & B7\,V\,+\,B8\,Ve [27,19] & 2 (35) & n.e. & --- & \\
136298 & $\delta$ Lup & 3.22 & B1.5\,IV [23] & 2,9 (35,44) & BCEP & BCEP+SPB & (25) \\
136415/6 & $\gamma$ Cir & 4.51 & B5\,IV\,+\,F8\,V [23,47] & 2 (18) & GCAS & VAR & \\
136504 & $\varepsilon$ Lup & 3.37 & B2\,IV-V [23] & 2,9 (35) & HB+SPB & HB+SPB & (11,14,62) \\
136664 & $\phi^2$ Lup & 4.54 & B4\,V [23] & 9 (44) & n.e. & --- & \\
138690 & $\gamma$ Lup & 2.78 & B2\,IV [23] & 2,9 (35,44) & R & R & (77) \\
139127 & $\omega$ Lup & 4.33 & K3\,III [37] & 2 & n.e. & --- & \\
139365 & $\tau$ Lib & 3.66 & B2.5\,V [23] & 9 (44) & n.e. & HB+E & (44) \\
\noalign{\smallskip}
141556 & $\chi$ Lup & 3.95 & B9.5\,III [25] & 9 & n.e. & --- & \\
142669 & $\rho$ Sco & 3.88 & B2\,IV-V [23] & 9 (44) & --- & ELL & \\
143018 & $\pi$ Sco & 2.89 & B1\,V\,+\,B2 [23] & 9 (44) & ELL & E+BCEP & (17) \\
143118 & $\eta$ Lup & 3.41 & B2.5\,IV [23] & 9 (44) & CST: & VAR & \\
143275 & $\delta$ Sco & 2.32 & B0.5\,IV [23] & 9 (44) & GCAS & VAR & \\
144217/8 & $\beta^1$/$\beta^2$ Sco & 2.62/4.69 & B0.5\,V\,+\,B2\,V [1] & 9 (44) & --- & VAR & \\
144294 & $\theta$ Lup & 4.23 & B2.5\,Vn [23] & 9 (44) & VAR: & --- & (24) \\
144470 & $\omega^1$ Sco & 3.96 & B1\,V [23] & 9 (44) & n.e. & --- & \\
145482 & 13 c$^2$ Sco & 4.57 & B3\,Vn [24] & 9 & R+PULS & VAR & (24) \\
145502/1 & $\nu$ Sco & 4.01 & B2\,V\,+\,B9\,Vp\,Si [48]& 9 (44) & ACV & VAR & (24) \\
\noalign{\smallskip}
147165 & $\sigma$ Sco & 2.89 & B1\,III [23] & 9 (44) & BCEP & BCEP+SPB & (40) \\
148478/9 & $\alpha$ Sco & 0.96 & M0.5\,Iab\,+\,B3\,V: [27] & 9 (44) & SRC & VAR & \\
148688 & V1058 Sco & 5.39 & B1\,Ia [24] & 9 (44) & ACYG & ACYG & \\
148703 & N Sco & 4.23 & B2\,III [23] & 9 (44) & EA & E & (45)\\
149038 & $\mu$ Nor & 4.94 & B0\,Ia [23] & 9 (44) & ACYG: & ACYG & (21) \\
149404 & V918 Sco & 5.47 & O8.5\,Iab(f)p [39] & 9 (44) & ELL & ELL+ACYG & (54) \\
149438 & $\tau$ Sco & 2.81 & B0\,V [23] & 9 (44) & n.e. & --- & (14) \\
151680 & $\varepsilon$ Sco & 2.29 & K2.5\,III [35] & 9 (44) & --- & VAR & (57) \\
151804 & V973 Sco & 5.22 & O8\,Ifp [23] & 9 (44) & ACYG & ROT+ACYG & (50) \\
151890 & $\mu^1$ Sco & 2.98 & B1.5\,IV\,+\,B [42] & 9 (44) & EB/SD & E & \\
\noalign{\smallskip}
151985 & $\mu^2$ Sco & 3.54 & B2\,IV [23] & 9 (44) & n.e. & --- & \\
157056 & $\theta$ Oph & 3.27 & B2\,IV [23] & 3 (16,27,37,45,54,63) & BCEP+SPB & BCEP+SPB & (58)\\
157792 & 44 b Oph & 4.17 & A3m [45] & 3 (16,27,37,45,54,63) & --- & --- & \\
157919 & 45 d Oph & 4.29 & F5\,IV [38] & 3 (16,27,37,45,54,63) & n.e. & --- & \\
158408 & $\upsilon$ Sco & 2.69 & B2\,IV [23] & 3 (16,27,37,45,54,63) & n.e. & VAR & \\
158926 & $\lambda$ Sco & 1.63 & B1.5\,IV\,+\,B [42] & 3 (16,27,37,45,54,63) & BCEP+EA & BCEP+E & \\
159433 & Q Sco & 4.29 & K0\,III [49] & 3 & --- & --- & \\
159532 & $\theta$ Sco & 1.86 & F1\,II [31] & 3 (16,27,37,45,54,63) & --- & VAR & \\
160578 & $\kappa$ Sco & 2.41 & B1.5\,III [23] & 3 (16,27,37,45,54,63) & BCEP & BCEP & \\
161471 & $\iota^1$ Sco & 3.03 & F2\,Ia [2] & 3 (16,27,37,45,54,63) & n.e. & --- & \\
\noalign{\smallskip}
161592 & 3 (X) Sgr & 4.54 & F3\,II [31] & 3 (16,27,45) & DCEP & DCEP & (31,41) \\
161892 & G Sco & 3.21 & K2\,III [50] & 3 (16,27,37,54) & n.e. & VAR & (57) \\
\noalign{\smallskip}\hline
\end{tabular}
\end{table*}

\setcounter{table}{0}
\begin{table*}[!ht]
\caption{continued.}
\centering\scriptsize
\begin{tabular}{rccccccl}
\noalign{\smallskip}\hline\hline\noalign{\smallskip}
HD & Other & $V$ & MK spectral & BRITE & \multicolumn{2}{c}{Variability class}  & Notes, references\\
& designation & (mag) & type & field(s) & VSX & this paper & to BRITE papers \\
\noalign{\smallskip}\hline\noalign{\smallskip}
164975 & $\gamma^1$ (W) Sgr & 4.69 & F8p\,+\,A0 V [31,51] & 3 (16,27,45,54) & DCEP & DCEP & (31,41)\\
165135 & $\gamma^2$ Sgr & 2.99 & K1\,III [50] & 3 (16,27,37,54) & --- & VAR & (57) \\
166937 & $\mu$ Sgr & 3.86 & B8\,Iap [6] & 3 (16,27,37,45,54,63) & EA+ACYG & VAR & \\
167618 & $\eta$ Sgr & 3.11 & M3.5\,III [35] & 3 (16,27,45,54) & LB: & VAR & (57) \\
168454 & $\delta$ Sgr & 2.67 & K3\,III [42] & 3 (16,27,37,45,54) & VAR: & --- & \\
169022 & $\varepsilon$ Sgr & 1.85 & B9.5\,III [42] & 3 (16,27,37,45,54,63) & n.e. & --- & \\
169916 & $\lambda$ Sgr & 2.81 & K0\,IV [50] & 3 (16,27,37,45,54) & n.e. & --- & \\
173300 & $\phi$ Sgr & 3.17 & B8\,III [26] & 3 (27,45) & n.e. & --- & \\
\noalign{\smallskip}
186882 & $\delta$ Cyg & 2.87 & B9.5\,III [1] & 4,10 (38) & CST & --- & (76) \\
187849 & 19 (V1509) Cyg & 5.12 & M2\,III [52] & 4 (38,46) & LB & VAR & \\
188892 & 22 Cyg & 4.94 & B5\,IV [10] & 4,10 (38,46) & --- & --- & \\
188947 & $\eta$ Cyg & 3.88 & K0\,III [13] & 4 (17,64) & --- & VAR & (57) \\
189178 & --- & 5.45 & B5\,V [10] & 10 (38) & n.e. & --- & \\
189687 & 25 (V1746) Cyg & 5.19 & B2.5\,V(e) [19] & 4,10 (38,46,64) & GCAS+BCEP & BE & \\
189849 & 15 (NT) Vul & 4.66 & A4\,IIIm: [8] & 10 (17) & ACV & --- & (34) \\
191610 & 28 (V1624) Cyg & 4.93 & B3\,IVe [19] & 4,10,(17,38,46,64) & SXARI+BE & BE+SPB & (36) \\
192577/8 & 31 (V695) Cyg & 3.80 & K2\,II\,+\,B3\,V [40] & 4,10 (38,46) & EA/GS/D & VAR & \\
192640 & 29 (V1644) Cyg & 4.97 & A2\,V [8] & 4,10 (17,38,46,64) & DSCT & DSCT & \\
\noalign{\smallskip}
192685 & QR Vul & 4.78 & B3\,V [10] & 10 (17,46) & GCAS & VAR & \\
192806 & 23 Vul & 4.52 & K3\,III [13] & 4 & n.e. & VAR & (57) \\
192909/10 & $o^2$ (V1488) Cyg & 3.98 & K3\,Ib-II\,+\,B/A [40] & 4 (38) & EA/GS/D & VAR & \\
193092 & --- & 5.24 & K4\,II [13] & 4 (64) & VAR & --- & \\
193237 & 34 (P) Cyg & 4.81 & B1ep [10] & 4,10 (17,38,46,64) & SDOR & SDOR & (13,78) \\
194093 & $\gamma$ Cyg & 2.20 & F5\,II [2] & 4,10 (38) & --- & VAR & \\
194317 & 39 Cyg & 4.43 & K3\,III [2] & 4 (38) & n.e. & VAR & (57,73) \\
194335 & V2119 Cyg & 5.90 & B2\,IIIe [19] & 10 & BE & --- & \\
195068/9 & 43 (V2121) Cyg & 5.74 & F2\,V [16] & 10 (29,38,46,64) & GDOR & GDOR & (29,32) \\
195295 & 41 Cyg & 4.01 & F5\,II [2] & 4,10 (46) & --- & --- & \\
\noalign{\smallskip}
195556 & $\omega^1$ (V2014) Cyg & 4.94 & B2.5\,IV [10] & 10 (38) & LERI: & VAR & \\
196093/4 & 47 (V2125) Cyg & 4.66 & K3\,Ib\,+\,B7 [53] & 4 & LC & VAR & \\
197345 & $\alpha$ Cyg & 1.25 & A2\,Ia [2] & 4,10 (17,38,46,64) & ACYG & VAR & \\
197912 & 52 Cyg & 4.22 & K0\,III [13] & 4 & n.e. & --- & \\
197989 & $\varepsilon$ Cyg & 2.46 & K0\,III [13] & 4,10 (46) & n.e. & --- & (57) \\
198183 & $\lambda$ Cyg & 4.53 & B5\,V [1] & 4,10 (46,64) & GCAS & VAR & \\
198478 & 55 (V1661) Cyg & 4.84 & B3\,Ia [2] & 4,10 (29,64) & ACYG & ACYG & (21,46) \\
198639 & 56 Cyg & 5.05 & A6\,V [16] & 10 (29,64) & CST: & --- & \\
198726 & T Vul & 5.77 & F5\,Ib [54] & 4,10 & DCEP & DCEP & (31,41) \\
198809 & 31 Vul & 4.59 & G5\,III [54] & 4 & --- & --- & \\
\noalign{\smallskip}
199081 & 57 Cyg & 4.78 & B5\,V [10] & 4,10 (29,38,64) & --- & VAR & \\
199629 & $\nu$ Cyg & 3.94 & A1\,Vn [8] & 4,10 (38,64) & n.e. & --- & \\
200120 & 59 (V832) Cyg & 4.74 & B1.5\,Vnne [10] & 4,10 (46,64) & GCAS+R & BE & \\
200310 & 60 (V1931) Cyg & 5.37 & B1\,Vn [10] & 10 (29,46,64) & LERI & BE & (26) \\
200905 & $\xi$ Cyg & 3.72 & K5\,Ib [2] & 4 (38,46,64) & LC & VAR & \\
201078 & DT Cyg & 5.82 & F8\,Ib-II [15] & 4,10 (46) & DCEPS & DCEPS & (31,41) \\
201251 & 63 Cyg & 4.55 & K4\,II [2] & 4 & --- & --- & \\
201433 & V389 Cyg & 5.69 & B9\,V [8] & 10 (17) & SPB & SPB+E & (10,16) \\
202109 & $\zeta$ Cyg & 3.20 & G8\,II [2] & 4,10 (46) & n.e. & --- & \\
202444 & $\tau$ Cyg & 3.72 & F0\,IV [55] & 4,10 (38) & DSCT & --- & \\
\noalign{\smallskip}
202850 & $\sigma$ Cyg & 4.23 & B9\,Ia [2] & 4,10 (38,46) & ACYG & ACYG & (46)\\
202904 & $\upsilon$ Cyg & 4.43 & B2\,Ve [10] & 4,10 (46) & GCAS & BE & \\
203064 & 68 (V1809) Cyg & 5.00 & O7.5\,III((f))n [22] & 4,10 (29,64) & ELL & --- & \\
203156 & V1334 Cyg & 5.83 & F1\,II [15] & 4,10 & DCEPS & DCEPS & (31,41) \\
203280 & $\alpha$ Cep & 2.44 & A7\,IV/V [1] & 11 & DSCT & --- & \\
205021 & $\beta$ Cep & 3.23 & B2\,III [6] & 11 (39) & BCEP & --- & \\
205435 & $\rho$ Cyg & 4.02 & G8\,III [13] & 10 (38) & RS: & --- & \\
206570 & V460 Cyg (DS Peg) & 6.07 & C6,3 [56] & 4 & SRB & --- & \\
207260 & $\nu$ Cep & 4.29 & A2\,Iae [25] & 11 & ACYG & VAR & \\
209790 & $\xi$ Cep & 4.29 & F7\,V [48] & 11 & n.e. & --- & \\
\noalign{\smallskip}\hline
\end{tabular}
\end{table*}

\setcounter{table}{0}
\begin{table*}[!ht]
\caption{continued.}
\centering\scriptsize
\begin{tabular}{rccccccl}
\noalign{\smallskip}\hline\hline\noalign{\smallskip}
HD & Other & $V$ & MK spectral & BRITE & \multicolumn{2}{c}{Variability class}  & Notes, references\\
& designation & (mag) & type & field(s) & VSX & this paper & to BRITE papers \\
\noalign{\smallskip}\hline\noalign{\smallskip}
209975 & 19 Cep & 5.11 & O9.5\,Ib [2] & 11 (57) & n.e. & ACYG & \\
210745 & $\zeta$ Cep & 3.35 & K1\,Ib [2] & 11 (29,39) & LC & --- & \\
211336 & $\varepsilon$ Cep & 4.19 & F0\,IV [1] & 11 (29,39,57) & DSCTC & DSCT & (28) \\
213306/7 & $\delta$ Cep & 3.75 & F5\,Ib\,+\,B8\,V [57,48] & 11 (29,39,57) & DCEP  & DCEP & (31,41) \\
216228 & $\iota$ Cep & 3.52 & K1\,III [13] & 11 & n.e. & --- & \\
218376 & 1 Cas & 4.84 & B0.5\,IV [6] & 11 (19,39,57) & n.e. & --- & \\
220652 & 4 Cas & 4.98 & M2\,III [58] & 11 (39) & LB & VAR & \\
221253 & AR Cas & 4.88 & B3\,IV [10] & 11 (19,39,57) & EA/DM & E & \\
224572 & $\sigma$ Cas & 5.00 & B1\,V [10] & 11 (19,39) & n.e. & --- & \\
225289 & V567 Cas & 5.79 & B8p\,HgMn [59] & 11 & ACV & --- & \\
\noalign{\smallskip}\hline
\end{tabular}
\tablebib{
MK spectral types: [1] \cite{1953ApJ...117..313J}, [2] \cite{1950ApJ...112..362M}, [3] \cite{1973ARA&A..11...29M}, [4] \cite{1980ApJS...42..541K}, [5] \cite{1994AJ....108.1437H}, [6] \cite{1955ApJS....2...41M}, [7] \cite{1972AJ.....77..750C}, [8] \cite{1969AJ.....74..375C}, [9] \cite{1989ApJS...70..623G}, [10] \cite{1968ApJS...17..371L}, [11] \cite{1957ApJ...125..689S}, [12] \cite{1987PASP...99..629K}, [13] \cite{1952ApJ...116..122R}, [14] \cite{1972AJ.....77..312W}, [15] \cite{1976PASP...88...95C}, [16] \cite{1995ApJS...99..135A}, [17] \cite{1985AbaOB..59...83B}, [18] \cite{1972PASP...84..584L}, [19] \cite{1982ApJS...50...55S}, [20] \cite{2018A+A...615A.161M}, [21] \cite{1971ApJ...170..325C}, [22] \cite{2011ApJS..193...24S}, [23] \cite{1969ApJ...157..313H}, [24] \cite{1957MNRAS.117..449D}, [25] \cite{1987ApJS...65..581G}, [26] \cite{1994AJ....107.1556G}, [27] \cite{1984ApJS...55..657C}, [28] \cite{1975mcts.book.....H}, [29] \cite{1987IBVS.3002....1A}, [30] \cite{1966ApJ...143..770H}, [31] \cite{1975AJ.....80..637M}, [32] \cite{1978PASP...90..453D}, [33] \cite{1965PASP...77..376J}, [34] \cite{1959PASP...71...48J}, [35] \cite{1966ApJ...146..587L}, [36] \cite{1998ApJS..119...83P}, [37] \cite{1978mcts.book.....H}, [38] \cite{1957MNRAS.117..534E}, [39] \cite{2014ApJS..211...10S}, [40] \cite{1951ApJ...113..304B}, [41] \cite{1968MNRAS.138..109S}, [42] \cite{1982mcts.book.....H}, [43] \cite{1961MNRAS.122..325M}, [44] \cite{1989ApJS...71..245K}, [45] \cite{1973PASP...85..328M}, [46] \cite{1969MNRAS.144....1B}, [47] \cite{1971MNRAS.152...37B}, [48] \cite{1963ApJ...138..118S}, [49] \cite{1950ApJ...111..221W}, [50] \cite{2006AJ....132..161G}, [51] \cite{1991ApJ...372..597E}, [52] \cite{1974AJ.....79..866B}, [53] \cite{1981AJ.....86..271H}, [54] \cite{1947ApJ...106...20N}, [55] \cite{1955ApJ...121..653S}, [56] \cite{1941ApJ....94..501K}, [57] \cite{1960ApJ...131..330K}, [58] \cite{1967PDAO...13...47Y}, [59] \cite{1968PASP...80..453C}.
\\BRITE papers: (1) \citet{weiss2016}, (2) \cite{Pigulski2016}, (3) \cite{Baade2016}, (4) \cite{2016pas..conf...98M}, (5) \cite{2016sf2a.conf..229B}, (6) \cite{handler2017}, (7) \cite{daszynska2017}, (8) \cite{pablo2017}, (9) \cite{Buysschaert2017}, (10) \cite{kallinger2017}, (11) \cite{2017EPJWC.16001001H}, (12) \cite{Richardson2017}, (13) \cite{2017sbcs.conf...55R}, (14) \cite{2017sbcs.conf...94W}, (15) \cite{2017sbcs.conf..101B}, (16) \cite{2017sbcs.conf..113K}, (17) \cite{2017sbcs.conf..120P}, (18) \cite{2017sbcs.conf..138D}, (19) \cite{2017sbcs.conf..151H}, (20) \cite{2017sbcs.conf..158Z}, (21) \cite{2017sbcs.conf..163R}, (22) \cite{2017sbcs.conf..167P}, (23) \cite{2017sbcs.conf..173W}, (24) \cite{2017sbcs.conf..180P}, (25) \cite{2017sbcs.conf..186C}, (26) \cite{2017sbcs.conf..196B}, (27) \cite{2017sbcs.conf..217N}, (28) \cite{2017sbcs.conf..228Z}, (29) \cite{2017sbcs.conf..236G}, (30) \cite{2017sbcs.conf..240K}, (31) \cite{2017sbcs.conf..265S}, (32) \cite{zwintz2017}, (33) \cite{ramiaramanantsoa2018b}, (34) \cite{2018BlgAJ..28...27S}, (35) \cite{2018CoSka..48..170K}, (36) \cite{baade2018}, (37) \cite{mollous2018}, (38) \cite{2018pas8.conf...37M}, (39) \cite{2018pas8.conf...69B}, (40) \cite{2018pas8.conf...77P}, (41) \cite{2018pas8.conf...88S}, (42) \cite{2018pas8.conf...94Z}, (43) \cite{2018pas8.conf..101P}, (44) \cite{2018pas8.conf..115P}, (45) \cite{2018pas8.conf..118R}, (46) \cite{2018pas8.conf..134R}, (47) \cite{2018pas8.conf..154B}, (48) \cite{2018pas8.conf..161Z}, (49) \cite{2018pas8.conf..170K}, (50) \cite{ramiaramanantsoa2018a}, (51) \cite{2018pas7.conf..162N}, (52) \cite{2018A&A...617A.121K}, (53) \cite{2018A&A...620A.145B}, (54) \citet{rauw2019}, (55) \citet{buysschaert2019}, (56) \cite{schmutz2019}, (57) \cite{kallinger2019}, (58) \cite{Walczak2019}, (59) \cite{2019CoSka..49..252R}, (60) \cite{zwintz2019}, (61) \cite{2019ASPC..518...59Z}, (62) \cite{2019MNRAS.488...64P}, (63) \cite{2020MNRAS.492.2762W}, (64) \cite{2020A+A...635A.140B}, (65) \cite{2020CoSka..50..585O}, (66) \cite{krticka2020}, (67)  \cite{Jerzykiewicz2020}, (68) \cite{2020svos.conf...99L}, (69) \cite{2020svos.conf..119Z}, (70) \cite{2020svos.conf..121B}, (71) \cite{2020svos.conf..147R}, (72) \cite{2020svos.conf..423S}, (73) \cite{2020svos.conf..457H}, (74) \cite{weiss2020}, (75) \cite{2020A+A...643A.110Z}, (76) \cite{2021PASP..133a4401Y}, (77) \cite{2021MNRAS.503.5554J}, (78) \cite{elliott2022}, (79) \cite{2023A&A...672A..31O}.
}
\end{table*}

\FloatBarrier

\section{Notes on individual stars}
\label{sec:notes_on_stars}
In this appendix, we provide short notes on the BRITE-Con\-stel\-la\-tion data from Fields 1\,--\,14 for each of the stars discussed in this paper, including interesting features appearing in the Fourier spectrum (FS). The objects are sorted by HD number. Information on additional observations in other BRITE fields and about variability, either known previously or found from BRITE data is included in Table~\ref{stars300}. 

HD\,432 ($\beta$~Cas, Field 11): Known $\delta$~Scuti-type star. Clear variability is seen with a dominant signal at frequency 9.897~d$^{-1}$ in the FS of the BAb, BLb, BHr and BTr data. The BRITE data are discussed in detail by \citet{2020A+A...643A.110Z}.

HD\,2905 ($\kappa$~Cas, Field 11): The star is a B1\,Ia-type supergiant observed by BAb, BHr and BLb, showing stochastic oscillations at low frequencies in the FS typical for evolved O and B stars \citep[see, for example,][]{Bowman2019b}.

HD\,3360 ($\zeta$ Cas, Field 11): This is a $\beta$ Cephei star for which the FS of the BLb data shows some weak variability with a frequency of 7.6715\,d$^{-1}$. The FSa of the BAb and BTr data show multi-frequency variability at low amplitudes.

HD\,3712 ($\alpha$~Cas, Field 11): No significant variability was detected in the BAb, BHr, BLb and BTr data.

HD\,3901 ($\xi$~Cas, Field 11): No significant variability was detected in the BAb data.

HD\,4614 ($\eta$~Cas, Field 11): No significant variability was detected in the BAb and BHr data.

HD\,5394 ($\gamma$~Cas, Field 11): This well-known Be star is also the prototype of a subclass of Be stars showing large-amplitude long-term variations. Clear variability detected in the BAb, BHr, BLb and BTr data. The BRITE data for this star are discussed in detail by \citet{2020A+A...635A.140B}.

HD\,6811 ($\phi$~And, Field 11): No variability was detected in the BAb data.

HD\,6961 ($\theta$~Cas, Field 11): No variability was detected in the BAb data.

HD\,8538 ($\delta$~Cas, Field 11): No variability was detected in the BAb, BLb or BTr data.

HD\,11415 ($\varepsilon$~Cas, Field 11): Known Be star. Some low-frequency variability is present in the FS of the BAb data.

HD\,16908 (35 Ari, Field 5): No variability was detected in the BAb data. The data set is short and scarce.

HD\,17573 (41 Ari, Field 5): No variability was detected in the BAb data.

HD\,17584 (16 Per, Field 5): Observed with UBr and BAb. The FS of the UBr data shows a frequency at $\sim$4.9~d$^{-1}$ or its alias at $\sim$9.3~d$^{-1}$ that can be attributed to $\delta$ Scuti-type pulsation. The BAb data set is short and scarce.

HD\,17709 (17 Per, Field 5): No variability was detected in the UBr data. The data set is short and scarce.

HD\,18296 (21 Per, Field 5): Observed with UBr and BAb. The FS of the UBr data reveals a dominant frequency of 0.343~d$^{-1}$ (corresponding to a period of $\sim$2.92~d). No variability was detected in the BAb data. Both data sets are short.

HD\,19058 ($\rho$ Per, Field 5): Observed with UBr. The light curve shows clear long-term variability.

HD\,19356 ($\beta$ Per, Algol, Field 5): Known eclipsing binary. Its variability with the 2.87-d period is clearly detectable in the UBr and BAb data.

HD\,19373 ($\iota$ Per, Field 5): No variability was detected in the UBr data.

HD\,19476 ($\kappa$ Per, Field 5): No variability was detected in the UBr data.

HD\,20365 (20 Per, Field 5): No variability was detected in the UBr data.

HD\,20418 (31 Per, Field 5): No variability was detected in the UBr data.

HD\,20468 (Field 5): No variability was detected in the UBr data.

HD\,20809 (Field 5): Observed with UBr. The FS of the data reveals a signal at frequency 2.25~d$^{-1}$. The available data have a short time coverage.

HD\,20902 ($\alpha$ Per, Field 5): No variability was detected in the UBr and BAb data.

HD\,21428 (34 Per, Field 5): No variability was detected in the UBr data.

HD\,21552 ($\sigma$ Per, Field 5): Observed with UBr. The available data set is long and the FS of the data shows some moderate variability at low frequencies. This K3-type star was included in the study of \citet{kallinger2019}.

HD\,22192 ($\psi$ Per, Field 5): Known Be star, observed by UBr and BAb. The FS of the data reveals variability with a frequency of 0.986~d$^{-1}$.

HD\,22780 (Field 5):  No variability was detected in the UBr data.

HD\,22928 ($\delta$ Per, Field 5): Observed by UBr and BAb. The FS of the UBr data shows clear variability with a frequency of $\sim$1.27~d$^{-1}$, but additional peaks can also be identified. The BAb data are of lower quality, but their analysis confirms the frequency peaks identified from the UBr observations.

HD\,23180 (\omicron Per, Field 5): Observed with UBr and BAb. The FSa of the available data reveal clear variability with a frequency of 0.453~d$^{-1}$.

HD\,23230 ($\nu$ Per, Field 5): No variability was detected in the BAb data.

HD\,23302 (17 Tau, Field 5): No variability was detected in the BAb data. The FS of the UBr data shows only some weak variability at low frequencies.

HD\,23338 (19 Tau, Field 5): No variability was detected in the UBr or BAb data.

HD\,23408 (20 Tau, Field 5): Observed with UBr and BAb. Both data sets have long time bases. The FS reveals a significant peak at frequency of 0.965~d$^{-1}$. The light curve phased with the corresponding period does not show coherent variation, however.

HD\,23480 (23 Tau, Field 5): No variability was detected in the UBr or BAb data.

HD\,23630 ($\eta$ Tau, Field 5): Known Be star, observed with UBr and BAb. The FSa show a single significant frequency at 0.03~d$^{-1}$.

HD\,23850 (27 Tau, Field 5): Observed with UBr and BAb. The FS of the UBr data reveals three strong peaks at frequencies of $\sim$0.41, $\sim$0.82 and $\sim$1.3~d$^{-1}$. The FS of the BAb data confirms the dominant frequency identified from the UBr observations.

HD\,24398 ($\zeta$ Per, Field 5): Observed with UBr and BAb. The UBr light curve shows irregular variability resulting in several peaks in the low-frequency domain of the FS, the highest of which occurs at 0.04~d$^{-1}$. The BAb data confirm the character of the variability, which fits the ACYG classification.

HD\,24640 (Field 5): No variability was detected in the BAb data.

HD\,24760 ($\varepsilon$ Per, Field 5): Observed with UBr and BAb. The FSa of the available data show multi-periodic hybrid $\beta$~Ce\-phei/SPB variability with the dominant peak at a frequency of 5.898~d$^{-1}$.

HD\,24912 ($\xi$ Per, Field 5): Observed with UBr and BAb. The FS of the UBr data shows clear variability with a frequency of 0.49~d$^{-1}$. The analysis of the BAb data confirmed the frequency detected based on the UBr observations.

HD\,25823 (41 Tau, Field 5): Observed with UBr. The available data set is short, but its FS reveals variability with a frequency of 0.13~d$^{-1}$.

HD\,25940 (48 Per, Field 5): Observed with UBr and BAb. The UBr data set is long and its FS shows variability with a frequency of $\sim$2.26~d$^{-1}$, which is confirmed with the shorter and scarcer BAb data set.

HD\,25998 (50 Per, Field 5): No variability was detected in the UBr data.

HD\,26322 (44 Tau, Field 5): Known $\delta$ Scuti star, observed with UBr. The $\delta$ Scuti variability is clearly present with the strongest peak in the FS at a frequency of 6.89~d$^{-1}$.

HD\,26630 ($\mu$ Per, Field 5): No variability was detected in the UBr data.

HD\,27396 (53 Per, Field 5): Observed with UBr. The FS of the available data shows multi-mode SPB-type variability with the strongest peak at a frequency of 0.46~d$^{-1}$.

HD\,29248 ($\nu$ Eri, Fields 6 \& 13): Known hybrid $\beta$~Ce\-phei/SPB star. {\it Field 6:} Observed with BAb, BTr, BHr and BLb. The available data clearly show the known $\beta$ Cephei variability. {\it Field 13:} Observed with BAb and UBr. The available data also show the known pulsations. BRITE data for this star were discussed in detail by \citet{handler2017} in combination with ground-based observations.

HD\,30211 ($\mu$ Eri, Fields 6 \& 13): Known SPB star and eclipsing binary. {\it Field 6:} Observed with BTr, BHr, BLb and BAb. The strongest frequency in the FS is found at $\sim$0.61~d$^{-1}$ for BTr, BHr, and BAb data and 0.69~d$^{-1}$ for the BLb data, but the FSa show also the presence of many additional peaks. {\it Field 13:} Observed with UBr and BAb. The available data confirm the variability seen in the data from Field 6.

HD\,30652 ($\pi^3$\,Ori, Field 6): No variability was detected in the BTr data.

HD\,30836 ($\pi^4$\,Ori, Fields 6 \& 13): Known spectroscopic binary. {\it Field 6:} Observed with BTr, BHr, BAb and BLb.  The available data show clear variability with a period of $\sim$4.1 days. {\it Field 13:} Observed with BAb and UBr. The UBr data confirm the variability found from Field 6 data. The BAb data have a short time base and are scarce.

HD\,31109 ($\omega$ Eri, Field 6). Known spectroscopic binary. Observed with BHr and BLb. The FSa of the available data show a dominant peak at a frequency of $\sim$3.64~d$^{-1}$, but the star is multi-periodic. The BLb light curve is shorter and noisier than that secured by BHr, but its analysis confirms the main frequency found from the BHr data.

HD\,31139 (5 Ori, Field 6): Observed with BHr and BTr. Some long-term variability may be present in the BHr data, but the data are inconclusive. The BTr light curve is only $\sim$8 d long and does not reveal any variation.

HD\,31237 ($\pi^5$\,Ori, Fields 1, 6 \& 13): Known ellipsoidal binary variable with an SPB component. {\it Field 1:} Observed with BAb and UBr. The available data clearly show the ellipsoidal variability with a period of $1.85$~d. {\it Field 6:} Observed with BHr, BTr, BLb and BAb. The period of $1.85$~d is again clearly visible. {\it Field 13:} Observed with UBr and BAb. The UBr data confirm the periodicity found from Field 1 and 6 data. The BAb data set is short and scarce. BRITE data are discussed in detail by \citet{Jerzykiewicz2020}.

HD\,31767 ($\pi^6$\,Ori, Field 6): No clear variability was detected in the BTr data.

HD\,33111 ($\beta$~Eri, Fields 1 \& 6): {\it Field 1:} Observed with BAb and UBr.  Variability was detected in both data sets, although the strongest peaks in the FSa were found at different frequencies, 1.055~d$^{-1}$ for the BAb data and 10.44~d$^{-1}$ for the UBr data. {\it Field 6:} No variability was detected in the BTr data.

HD\,33328 ($\lambda$ Eri, Fields 6 \& 13): Known Be star. {\it Field 6:} Observed with BAb, BHr, BTr and BLb. Variability detected in all data sets, although the strongest frequency peak is at $\sim$2.85~d$^{-1}$ in the FS of the BHr data and at half of this frequency in the FSa of the BAb, BTr and BLb data. {\it Field 13:} No variability was detected in the BAb data. The available data have a short time base and are scarce.

HD\,33904 ($\mu$ Lep, Field 6): No variability was detected in the BTr and BAb data. The BAb data set is scarce and of poor quality.

HD\,34085 ($\beta$\,Ori, Rigel, Fields 1, 6 \& 13): {\it Field 1:} Observed with BAb and UBr. The acquired CCD subraster of this star is severely overexposed resulting in low quality data. {\it Field 6:} Observed with BHr, BTr, BLb and BAb. The available data show some long-period variability. Due to the brightness of the star, the images were saturated and the origin of the variability may be partly instrumental. {\it Field 13:} Observed with UBr and BAb. The UBr data again show some long-period variability. The BAb data set is shorter and scarcer, but displays a similar behaviour to the UBr data set.

HD\,34503 ($\tau$\,Ori, Fields 1, 6 \& 13): {\it Field 1:} No variability was detected in the BAb or UBr data. {\it Field 6:} No variability was detected in the BTr, BHr, BLb or BAb data. {\it Field 13:} No variability was detected in the BAb data. According to \cite{2018pas8.conf..115P}, the star shows a weak heartbeat signal.

HD\,34816 ($\lambda$ Lep, Field 6): Early B type star observed with BHr, BLb and BTr. The FS of the BHr data shows the strongest frequency peak at 5.79~d$^{-1}$. The BLb and BTr data are of lower quality and do not show any variability.

HD\,35039 (22 Ori, Field 6): No variability was detected in the BTr data.

HD\,35369 (29 Ori, Field 6): No variability was detected in the BTr data.

HD\,35411 ($\eta$\,Ori, Fields 1, 6 \& 13): Eclipsing binary star with an orbital period of 8~d. {\it Field 1:} Observed with BAb and UBr. Both data sets show eclipses and regular variability with a period of $0.432$~d. {\it Field 6:} Observed with BTr, BHr, BLb and BAb. The data confirm the binary period and the additional regular variability. {\it Field 13:} Observed with UBr and BAb. Both data sets confirm the binary period and the additional variability.

HD\,35439 ($\psi^1$\,Ori, Field 6): Observed with BTr, BHr and BLb. The FSa of the available data show several distinct peaks in the $g$-mode range and additional variability typical for Be stars. The first broad assessment of the complex variability observed by BRITE-Constellation was given by \citet{2018pas8.conf...69B}.

HD\,35468 ($\gamma$\,Ori, Fields 1, 6 \& 13):{\it Field 1:} No variability was detected in the BAb and UBr data. {\it Field 6:} No variability was detected in the BTr, BHr, BLb and BAb data. {\it Field 13:} No variability was detected in the UBr and BAb data.

HD\,35715 ($\psi^2$\,Ori, Fields 1 \& 6): {\it Field 1:} Observed with BAb and UBr. The available data show ellipsoidal variability with a period 2.526~d. {\it Field 6:} Observed with BTr, BHr, BLb and BAb. The data confirm the main period of 2.526~d. \cite{2017sbcs.conf..120P} found in the BRITE data several additional frequencies attributable to $p$ modes.

HD\,36267 (32 Ori, Field 6): Observed with BTr, BHr and BLb. The data show variability with a period of 3.772~d.

HD\,36485/6 ($\delta$\,Ori, Fields 1, 6 \& 13): Massive eclipsing and spectroscopic binary. {\it Field 1:} Observed with BAb and UBr. The data clearly show eclipses with an orbital period of $5.733$~d and some additional variability. {\it Field 6:} Observed with BTr, BHr, BLb and BAb. The data confirm the binary period. {\it Field 13:} Observed with UBr and BAb. The data confirm the binary period.

HD\,36512 ($\upsilon$\,Ori, Field 6): No variability was detected in the BTr, BHr or BLb data.

HD\,36822 ($\phi^1$\,Ori, Field 6): No variability was detected in the BTr data.

HD\,36861/2 ($\lambda$\,Ori A/B, Fields 1, 6 \& 13): {\it Field 1:} No variability was detected in the BAb and UBr data. {\it Field 6:} No variability was detected in the BTr, BHr and BLb data. The BAb light cvurve is scarce and does not show any variability. {\it Field 13:} No variability was detected in the BAb data.

HD\,36959/60 (Field 6): No variability was detected in BHr, BTr, BLb or BAb data.

HD\,37018 (42 Ori, Field 6): No variability was detected in the BTr and BAb data.

HD\,37022/41 ($\theta^1$/$\theta^2$\,Ori, Fields 6 \& 13): {\it Field 6:} Observed with BTr and BAb. The FS of the BTr data shows some marginal variability with a frequency of 0.62~d$^{-1}$. No variability was detected in the BAb data. {\it Field 13:} No variability was detected in the UBr data.

HD\,37043 ($\iota$\,Ori, Fields 1, 6 \& 13): Known spectroscopic binary. {\it Field 1:} Observed with BAb and UBr. The FSa of the data show some variability in the low-frequency regime, but with no significant peaks. {\it Field 6:} Observed with BTr, BHr, BAb and BLb. The FSa of these data confirm variability in the low-frequency domain without a clear periodicity. {\it Field 13:} Observed with UBr and BAb. The FSa of the data confirm the variability in the low-frequency domain as seen from Field 1 and 6 data. BRITE data of this star were discussed by \citet{pablo2017}.

HD\,37128 ($\varepsilon$\,Ori, Fields 1, 6 \& 13): B-type supergiant. {\it Field 1:} Observed with BAb and UBr. Both data sets show similar long-term variability, but no clear periodicity. This is characteristic for $\alpha$~Cyg-type variability. {\it Field 6:} Observed with BAb, BTr, BHr and BLb. The BAb data set is scarce and does show variability. The other data sets are of excellent quality and their FSa show clear and strong variability at low frequencies. {\it Field 13:} Observed with UBr and BAb. The FSa of these data clearly show variability in the low-frequency regime, but without a clear periodicity.

HD\,37468 ($\sigma$\,Ori, Fields 1, 6 \& 13): {\it Field 1:} Observed with BAb and UBr. The FS of the BAb data shows peak at a frequency of 0.84~d$^{-1}$. No variability was detected in the UBr data. {\it Field 6:} Observed with BAb, BLb, BHr and BTr. In the FSa of the BHr and BTr data, the strongest peak is located at a frequency of 1.68~d$^{-1}$, twice the frequency detected in the BAb data. In the FS of the BLb data the strongest peak lies again at frequency of 0.84~d$^{-1}$. The BAb data set is shorter and scarce. {\it Field 13:} Observed with UBr and BAb. The FS of the UBr data shows two peaks at 0.84 and 2.52~d$^{-1}$. No variability was detected in the BAb data, which are of lower quality.

HD\,37490 ($\omega$\,Ori, Fields 6 \& 13): {\it Field 6:} Observed with BAb, BHr, BTr and BLb. The data show variability with the main period of about 0.95\,d. {\it Field 13:} No variability was detected in the UBr data.

HD\,37742/3 ($\zeta$\,Ori A/B, Fields 1, 6 \&13): O-type spectroscopic binary showing stochastic variability typical for massive stars \citep{Buysschaert2017}. {\it Field 1:} Observed with BAb and UBr. The BAb light curve shows a period of $3.275$~d. The corresponding frequency is not dominant in the FS of the UBr data, but a phase fold with the same period reveals similar variability. {\it Field 6:} Observed with BAb, BHr, BTr and BLb. The light curves show clear variability although the period of $3.275$~d is not dominant in their FSa. {\it Field 13:} Observed with BAb and UBr. The FS of the UBr data clearly shows low-frequency variability. The BAb data are scarcer, but its FS confirms the variability at low frequencies.

HD\,38771 ($\kappa$\,Ori, Fields 1, 6 \& 13): B-type supergiant. {\it Field 1:} Observed with UBr and BAb. The FSa of the data show low-frequency signals with a dominating peak at a frequency of $\sim$0.44~d$^{-1}$. {\it Field 6:} Observed with BHr, BLb, BTr and BAb. The FSa of the BHr and BLb data show clear variability with the strongest peak at a frequency of 0.9~d$^{-1}$. The BTr data set is shorter and its FS shows twice the frequency detected in the BHr and BLb data. The BAb data have a short time base and are scarce. {\it Field 13:} Observed with UBr and BAb. The FS of the UBr data shows variability in the low-frequency regime with a dominant peak at a frequency of 0.3016~d$^{-1}$. The BAb data are scarcer but its FS confirms the low-frequency variability.

HD\,39060 ($\beta$ Pic, Field 8): Young $\delta$ Scuti star, observed with BHr. The FS of the data reveals $\delta$ Scuti-type pulsations around 43\,d$^{-1}$. Studied in detail by \cite{zwintz2019}, \cite{mollous2018} and \cite{kenworthy2021}.

HD\,39801 ($\alpha$\,Ori, Betelgeuse, Fields 1, 6 \& 13): {\it Field 1:} Observed with BAb and UBr, but the acquired CCD subraster of this star was severely overexposed. {\it Field 6}: Observed with BAb, BTr, BHr and BLb. The data show clear variability, although part of this variability can be instrumental due to strong saturation \citep{weiss2021}. The BAb data set is short and scarce. {\it Field 13:} Observed with UBr and BAb. The UBr data show some variability. No variability was detected in the scarce BAb data set.

HD\,42933 ($\delta$\,Pic, Field 8): Eclipsing binary, observed with BHr. The light curve clearly shows eclipses with a period of 1.672~d.

HD\,44402 ($\zeta$\,CMa, Field 12): Observed with BLb. The FS of the data shows a strong peak at a frequency of 0.2569~d$^{-1}$. The presence of additional frequencies is also evident.

HD\,44743 ($\beta$\,CMa, Field 12): Known $\beta$~Cep star. Observed with BTr and BLb. The FSa of the data show a clear peak at a frequency of 3.9791~d$^{-1}$.

HD\,45348 ($\alpha$\,Car, Field 8): No variability was detected in the BHr data.

HD\,45871 (IY\,CMa, Field 12): Observed with BTr and BLb. The FS of the BTr data shows clear variability with the strongest peak at a frequency of 1.8503~d$^{-1}$, but additional frequencies can also be seen. The BLb data set is shorter, but confirms the variability detected from the BTr data.

HD\,46328 ($\xi^1$\,CMa, Field 12): Known magnetic $\beta$~Cep-type pulsator. Observed with BHr, BTr and BLb. The FSa of the BTr and BLb data show clear variability with the strongest peak at a frequency of 4.7717~d$^{-1}$. The BHr light curve has two large gaps, but the frequency detected in the BLb and BTr data is confirmed. This star has been studied in detail using BRITE data by \cite{2020MNRAS.492.2762W}.

HD\,47306 (N\,Car, Field 8): No variability was detected in the BHr data.

HD\,47670 ($\nu$ Pup, Field 8): Observed with BHr. The FS of the data displays a significant peak at a frequency of 0.6575~d$^{-1}$. The BRITE data were studied in detail by \cite{baade2018}.

HD\,48917 (10 CMa, Field 12): Known Be star, observed with BTr and BLb. The FS of the BTr data shows strong variability with the strongest peak at a frequency of 1.3364~d$^{-1}$. It is evident that several more frequencies are present. The BLb data have two large gaps but confirm the variability found from the BTr data.

HD\,49131 (HP CMa, Field 12): No variability was detected in the BLb or BTr data.

HD\,50013 ($\kappa$ CMa, Field 12): Observed with BLb and BHr. FSa of both data sets show strong variability with a main peak at a frequency of 0.056~d$^{-1}$.

HD\,50123 (HZ\,CMa, Field 12): Observed with BTr and BLb. The FS of the BTr data shows a frequency of 0.07~d$^{-1}$, which corresponds to twice the frequency of the known ellipsoidal variability \citep{1994A&A...291..473S}. The BLb data set is shorter, but confirms this variability.

HD\,50337 (V415\,Car, Field 8): No variability was detected in the BHr data.

HD\,50707 (15\,CMa, Field 12): Observed with BTr, BHr and BLb. The FSa of the data show clear $\beta$\,Cep-type pulsations with the strongest peak at a frequency of 5.418~d$^{-1}$.

HD\,50877 ($o^1$\,CMa, Field 12): Observed with BHr. The light curve shows strong long-term variability.

HD\,50896 (EZ\,CMa, Field 12): Observed with BHr. The data show rotational modulation with a period of 3.656~d. They are discussed by \citet{schmutz2019}.

HD\,51309 ($\iota$\,CMa, Field 12): Observed with BTr, BHr and BLb. The FSa of the data show variability with the strongest peak at a frequency of 0.077~d$^{-1}$, but additional frequencies that can be attributed to $\beta$~Cep pulsations are evident.

HD\,52089 ($\varepsilon$\,CMa, Field 12): No variability was detected in the BHr or BLb data.

HD\,52670 (LS\,CMa, Field 12): Observed with BTr. The data show three eclipses, two primary and one secondary.

HD\,52877 ($\sigma$\,CMa, Field 12): Observed with BHr. The data show strong long-term variability with a period of about 34 d.

HD\,53138 ($o^2$\,CMa, Field 12): Blue supergiant, observed with BHr and BLb. The FSa of the data reveal strong variability with low frequencies.

HD\,53244 ($\gamma$\,CMa, Field 12): No variability was detected in the BTr data.

HD\,54309 (FV\,CMa, Field 12): No variability was detected in the BLb data.

HD\,54605 ($\delta$\,CMa, Field 12): No variability was detected in the BHr or BLb data.

HD\,55892 (QW\,Pup, Field 8): Observed with BHr. The FS of the data shows multi-periodic $\gamma$~Dor-type pulsations with a dominant frequency of 0.975~d$^{-1}$.

HD\,56014 (27\,CMa, Field 12): Observed with BLb, BHr and BTr. The FSa of the data reveal multi-periodic variability with the strongest peak at a frequency of 0.7925~d$^{-1}$ for this Be star. The BLb light curve is of lower quality, but confirms the variability seen in the other light curves.

HD\,56022 (OU Pup, Field 8): No variability was detected in the BHr data.

HD\,56139 ($\omega$\,CMa, Field 12): Observed with BHr, BTr and BLb. The data show strong variability, typical for Be stars (Fig.\,\ref{fig:sampling}).

HD\,56455 (PR\,Pup, Field 8): Observed with BHr. The data are scarce, but its FS shows a significant frequency at 0.484~d$^{-1}$.

HD\,56855 ($\pi$\,Pup, Field 8): No variability was detected in the BHr data.

HD\,57060 (29\,CMa, Field 12): Observed with BHr, BLb and BTr. All data sets show clear eclipses with a period of about 4.4~d.

HD\,57061 ($\tau$\,CMa, Field 12): Observed with BTr, BHr and BLb. The FSa of the data are dominated by a frequency of 1.56~d$^{-1}$, corresponding to twice the orbital frequency of the known variability due to eclipses \citep{1997A&A...327.1070V}.

HD\,58155 (NO CMa, Field 12): Observed with BTr and BLb. The BTr data reveal variability with a main period of 0.4774~d. The BLb data set is shorter and scarce.

HD\,58286 (Field 12): No variability was detected in the BLb data.

HD\,58343 (FW\,CMa, Field 12): Observed with BTr and BLb. The FS of the BTr data shows variability with the strongest peak at a frequency of 0.0772~d$^{-1}$. The BLb data set is shorter and scarce.

HD\,58350 ($\eta$\,CMa, Field 12): Blue supergiant, observed with BHr and BLb. The FS of the data shows strong low-frequency variability with the strongest peak at 0.1222~d$^{-1}$.

HD\,61068 (PT\,Pup, Field 12): Observed with BLb and BTr. The FSa of the data show $\beta$~Cep-type variability with the strongest peak at a frequency of 5.9977~d$^{-1}$, see \cite{2017sbcs.conf..151H}.

HD\,61715 (MY\,Pup, Field 8): Known classical Cepheid. Observed with BHr. The data show clear variability with a period of about 5.7~d.

HD\,62623 (3\,Pup, Field 12): Observed with BHr and BLb. The FSa of the data show low-frequency variability typical for $\alpha$~Cyg-type stars.

HD\,62747 (V390 Pup, Field 12): Known binary system, observed with BTr. The data show shallow eclipses with an orbital period of about 3.93~d.

HD\,63462 ($o$\,Pup, Field 12): Observed with BTr, BHr and BLb. The FSa of all data sets show some low-frequency variability around 0.05~d$^{-1}$.

HD\,63744 (Q\,Pup, Field 7): No variability was detected in the BTr data.

HD\,63922 (P\,Pup, Fields 7 \& 8): {\it Field 7:} No variability was detected in the BAb or BTr data. {\it Field 8:} No variability was detected in the scarce BHr data set.

HD\,63949 (QS\,Pup, Field 8): No variability was detected in the BHr data.

HD\,64440 (a\,Pup, Fields 7 \& 8): {\it Field 7:} Observed with BAb and BTr. The FS of the BTr data shows a significant peak at $\sim$1.9~d$^{-1}$. No variability was detected in the scarce BAb data set. {\it Field 8:} No variability was detected in the BHr data.

HD\,64740 (Fields 7, 8 \& 14): {\it Field 7:} Observed with BAb and BTr. The FSa of the data show a clear peak at a frequency of 0.7519~d$^{-1}$. In addition, the FS of the BTr data shows the presence of additional frequencies up to 2~d$^{-1}$. {\it Field 8:} No variability was detected in the scarce BHr data set. {\it Field 14:} No variability was detected in the short and scarce BAb data set.

HD\,64760 (J Pup, Fields 7, 8 \& 14): {\it Field 7:} Observed with BAb and BTr. The data show some indications for low-frequency variability in the FS. {\it Field 8:} No variability was detected in the scarce BHr data. {\it Field 14:} No variability was detected in the short and scarce BAb data.

HD\,65575 ($\chi$\,Car, Fields 7, 8 \& 14): {\it Field 7:} Observed with BTr and BAb. The FS of the BTr and BAb data show multi-periodic variability with the strongest peaks at frequencies of 0.6438~d$^{-1}$ and 0.6126~d$^{-1}$, respectively. {\it Field 8:} Observed with BHr. The light curve is scarce, but its FS shows variability with a frequency of 0.66~d$^{-1}$. {\it Field 14:} No variability was detected in the short and scarce BAb data set.

HD\,65818 (V\,Pup, Fields 7, 8 \& 14): Known eclipsing binary consisting of two B-type stars. {\it Field 7:} Observed with BTr and BAb.  The data clearly show eclipses with the orbital period of about 1.45~d. {\it Field 8}: Observed with BHr. The data set is relatively short, but the variability is clearly detectable. {\it Field 14:} Observed with BAb. Again, the data set is relatively short, but confirms the period found from the other BRITE observations.

HD\,66811 ($\zeta$\,Pup, Fields 7, 8 \& 14): O-type supergiant. {\it Field 7:} Observed with BTr and BAb. The FSa of the data show clear variability with a frequency of 0.562~d$^{-1}$. {\it Field 8:} Observed with BHr. The FS of the data shows the same frequency as detected in the Field 7 data set. {\it Field 14:} No variability was detected in the short and scarce BAb data. The star was studied in detail by \citet{ramiaramanantsoa2018b}.

HD\,67523 ($\rho$\,Pup, Field 12): Known $\delta$ Sct star observed with BLb, BHr and BTr. The FSa of the data show the variability with the strongest peak at a frequency of 7.098~d$^{-1}$.

HD\,68243/73 ($\gamma^1$/$\gamma^2$\,Vel, Fields 7, 8 \& 14): Wolf-Rayet star. {\it Field 7:} Observed with BTr and BAb. The FSa of the data clearly show low-frequency variability. {\it Field 8:} Observed with BHr. The FS of the data confirms the low-frequency variability. {\it Field 14:} No variability was detected in the short and scarce BAb data set. Using BRITE-Constellation data, $\gamma$\,Vel was studied in detail by \citet{Richardson2017}.

HD\,68553 (h$^1$\,Pup, Field 7): Observed with BTr. The data show clear variability.

HD\,69142 (h$^2$\,Pup, Field 7): No variability was detected in the short and scarce BTr data set.

HD\,71129 ($\varepsilon$\,Car, Field 7): Observed with BAb and BTr. The BAb data set is short, and its FS shows variability with the strongest frequency at 1.1289~d$^{-1}$. The BTr light curve confirms this variability.

HD\,72127 (Field 7): No variability was detected in the short and scarce BAb data set.

HD\,73634 (e\,Vel, Field 7): No variability was detected in the BTr and short and scarce BAb data sets.

HD\,74006 ($\beta$\,Pyx, Field 7): No variability was detected in the BTr data and in the short and scarce BAb data set. The star was included in the study by \citet{kallinger2019}.

HD\,74180 (b\,Vel, Field 7): No variability was detected in the BAb and BTr data.

HD\,74195 ($o$\,Vel, Fields 7 \& 14): {\it Field 7:} Observed with BAb and BTr. The FSa of the data show pulsations with a main frequency of 0.3574~d$^{-1}$. {\it Field 14:} Observed with BAb. The data set is relatively short and scarce, but its FS reveals variability with a frequency of 0.3597~d$^{-1}$.

HD\,74375 (d\,Car, Field 7): Observed with BAb and BTr. The FSa of the data show clear pulsational variability with the strongest peak at a frequency of 0.4202~d$^{-1}$.

HD\,74560 (HY Vel, Field 7): Observed with BAb. The data set is short and scarce, but its FS shows variability with a frequency of 0.63~d$^{-1}$.

HD\,74575 ($\alpha$ Pyx, Field 7): Observed with BAb and BTr. The FSa of the data clearly show multi-periodic variability with the strongest peak at a frequency of 0.1876~d$^{-1}$.

HD\,74772 (d\,Vel, Field 7): No variability was detected in the BTr data and short and scarce BAb data set.

HD\,74956 ($\delta$\,Vel, Field 7): Known eclipsing binary, observed with BAb and BTr. The light curves show eclipses with the orbital period of 45.15~d.

HD\,75063 (a Vel, Field 7): Observed with BTr and BAb. The BTr data set is short and shows weak variability. No variability was detected in the short and scarce BAb data set.

HD\,75311 (f\,Car, Fields 7 \& 14): Known Be star. {\it Field 7:} Observed with BTr. The FS of the data reveals complex variability, with multiple intrinsic frequencies. {\it Field 14:} No variability was detected in the short and scarce BAb data set.

HD\,75821 (f\,Vel, Field 7): No variability was detected in the short and scarce BAb data set.

HD\,76728 (c\,Car, Fields 7 \& 14): {\it Field 7:} Observed with BTr and BAb. The FS of the BTr data shows variability with a frequency of 0.293~d$^{-1}$. No variability was detected in the BAb data. {\it Field 14:} No variability was detected in the short and scarce BAb data set.

HD\,77002 (b$^1$\,Car, Field 7): No variability was detected in the short and scarce BTr data set.

HD\,78004 (c\,Vel, Field 7): No variability was detected in the BTr data set. The BAb data set consists only of a few data points.

HD\,78647 ($\lambda$\,Vel, Field 7): Observed with BTr and BAb. The BTr light curve shows complex long-term variability. The BAb data set is short and scarce.

HD\,79351 (a\,Car, Fields 7 \& 14): {\it Field 7:} Observed with BTr and BAb. The FS of the BTr data shows multi-frequency variability with the strongest peak at a frequency of 0.3725~d$^{-1}$. The FS of the BAb data confirms the multi-frequency variations, but the data are noisier. {\it Field 14:} No variability was detected in the short and scarce BAb data set.

HD\,79940 (k\,Vel, Field 7): No variability was detected in the short and scarce BTr data.

HD\,80230 (g\,Car, Field 7): Observed with BTr. The light curve shows a dominant period of 12.17~d.

HD\,80404 ($\iota$\,Car, Fields 7 \& 14): {\it Field 7:} No variability was detected in the BTr and BAb data. {\it Field 14:} Observed with BAb. The data set is short and scarce, but its FS shows variability with a frequency of 1.1509~d$^{-1}$.

HD\,81188 ($\kappa$\,Vel, Fields 7 \& 14): {\it Field 7:} Observed with BTr and BAb. The FS of the data shows multi-frequency variability with the strongest peak at a frequency of 0.2431~d$^{-1}$. {\it Field 14:} No variability was detected in the short and scarce BAb data set.

HD\,82434 ($\psi$\,Vel, Field 7): No variability was detected in the BTr data and the short and scarce BAb data set.

HD\,82668 (N\,Vel, Field 7): Observed with BTr and BAb. The FS of the BTr data shows complex multi-frequency variability, while the BAb data set is short and scarce.

HD\,83183 (h\,Car, Field 7): No variability was detected in the BTr data and in the short and scarce BAb data set.

HD\,83446 (M\,Vel, Field 7): Observed with BTr and BAb. The FS of the BTr data show $\delta$~Sct-type variability with the strongest peak at a frequency of 31.08~d$^{-1}$. The BAb data set is shorter and of inferior quality.

HD\,86440 ($\phi$\,Vel, Field 7): No variability was detected in the BAb and BTr data.

HD\,118716 ($\varepsilon$\,Cen, Field 2): Known $\beta$~Cep star, observed with BTr, BLb, UBr and BAb. The FSa of the UBr and BAb data show clear pulsational variability with many frequencies and the strongest peak at a frequency of 5.8955~d$^{-1}$. The BTr and BLb data sets are scarce, but the main pulsation frequency can still be identified in the FSa of both data sets.

HD\,120307 ($\nu$\,Cen, Field 2): Observed by BTr, BLb, BAb and UBr. The BAb and UBr data are of excellent quality and show clear variability with a period of about 2.62~d. The BTr and BLb data sets are significantly shorter, but confirm the variability. The variability is due to the reflection effect in a binary system \citep{2021MNRAS.503.5554J}.

HD\,120324 ($\mu$\,Cen, Field 2): Known Be star, observed by BLb, BTr, BAb and UBr. The BAb and UBr data show significant, possibly irregular variations. The BLb and BTr data sets are shorter, but confirm the variability. BRITE data of this star were studied in detail by \citet{Baade2016}.

HD\,121263 ($\zeta$\,Cen, Field 2): Heartbeat binary star observed with BAb, UBr, BLb and BTr. In the FSa of the BAb and UBr data the strongest peaks lie at a frequency of 0.1244~d$^{-1}$, corresponding to a period of $\sim$8.026\,d. The data sets from BLb and BTr are shorter, but confirm the detected period.

HD\,121743 ($\phi$\,Cen, Field 2): Observed with UBr, BTr, BAb and BLb. The FSa of the BAb and BLb data show the strongest peak at a frequency of 0.88~d$^{-1}$. This signal is weaker in the FSa of the UBr and BTr data, however.

HD\,121790 ($\upsilon^2$\,Cen, Field 2): No variability was detected in the BAb data or UBr data sets. The BAb observations have a large gap in the middle of the run. The BTr data set is short and scarce. 

HD\,122451 ($\beta$\,Cen, Field 2): Observed with BAb, UBr, BTr and BLb. As this star is very bright, the acquired CCD subraster was overexposed and parts of the light curves have high scatter. The FS of the BAb data shows multi-frequency pulsations with the strongest peak at a frequency of 0.8689~d$^{-1}$. The UBr data are of similar quality. The BTr and BLb data sets are shorter.

HD\,122980 ($\chi$\,Cen, Field 2): Observed with BAb, UBr and BTr. The BAb data set has a gap in the centre and does not show any significant variability. The FS of the UBr data shows the strongest peak at $\sim$1.002~d$^{-1}$, which is of instrumental origin. The BTr data set is significantly shorter and does not show any variability.

HD\,125238 ($\iota$\,Lup, Field 2): Observed with UBr, BAb and BTr. The UBr data set has the longest time base and best quality and its FS shows several peaks in the low frequency domain; the strongest peak is located at a frequency of 0.2974~d$^{-1}$. The BAb data set has a gap in the centre and the BTr data set has a significantly shorter time base. The FSa of both data sets confirm the low-frequency variability detected in the UBr data.

HD\,125823 (a\,Cen, Field 2): Observed with UBr, BTr and BAb. The UBr light curve reveals rotational variability with a period of $\sim$8.85\,d. The BAb light curve has a gap in the centre, but confirms the period obtained with the UBr data. The BTr light curve is significantly shorter, but the variability is consistent with that seen in the UBr and BAb light curves.

HD\,126341 ($\tau^1$\,Lup, Field 2): Known $\beta$ Cep-type pulsator, observed with BAb, UBr, BTr and BLb. The FSa of the BAb and UBr data show the strongest peak at frequency of $\sim$5.6~d$^{-1}$. The BTr and BLb data have shorter time bases, but confirm this variability.

HD\,126354 ($\tau^2$\,Lup, Field 2): Observed with UBr, BAb and BTr. The FS of the UBr data shows an instrumental frequency of about 1~d$^{-1}$ and a few lower-amplitude peaks. The BAb data set has a gap in the middle and the BTr data set has shorter time base, but their FSa confirm the frequencies found from the UBr data set.

HD\,127381 ($\sigma$\,Lup, Field 2): Observed with UBr, BAb and BTr. The FS of the UBr data shows a clear peak at a frequency of 0.33~d$^{-1}$. The BAb data set has a gap in the middle of the run. The BTr data set is significantly shorter, but both data sets confirm the variability detected from the UBr observations.

HD\,127972/3 ($\eta$\,Cen, Field 2): Known Be star, observed with BTr, BLb, UBr and BAb. The FSa of the UBr and BAb data show clear variability with the strongest peak at a frequency of 1.55~d$^{-1}$. The BTr and BLb data sets are shorter, but their analysis confirmed the main frequency found from the UBr and BAb data. BRITE data of this star were studied in detail by \citet{Baade2016}.

HD\,128345 ($\rho$\,Lup, Field 2): Observed with UBr, BAb and BTr. The FS of the UBr data shows clear variability with a frequency of 2.245~d$^{-1}$. The BAb data set has a gap in the middle of the run, but its analysis confirms this variability. The BTr data set is shorter and of poorer quality.

HD\,128620/1 ($\alpha$ Cen, Field 2): Observed with BAb, BTr and BLb. The BAb light curve shows a period of 29.85~d. The BTr and BLb data sets are shorter and of poorer quality. Due to the brightness of the star, the acquired CCD subraster is severely overexposed and parts of the light curves have increased scatter.

HD\,128898 ($\alpha$\,Cir, Field 2): Known roAp star. Observed with BAb, UBr, BLb and BTr. The FS of the BAb data has the strongest peak at 0.225~d$^{-1}$. The strongest peak in the FS of the UBr data is located at the harmonic of this frequency. The roAp pulsations are also clearly detectable in the UBr and BAb data sets. The BLb and BTr data sets are shorter, but theis FSa confirm the strongest peak found from the BAb data and the roAp pulsations. All available BRITE data of this star were analysed by \cite{weiss2016} and \cite{weiss2020}.

HD\,129056 ($\alpha$\,Lup, Field 2): Known $\beta$~Cep-type star. Observed with BAb, UBr, BTr and BLb. The $\beta$~Cep variability is seen in the FSa of the BAb and UBr data with the main peak at a frequency of 3.8487~d$^{-1}$ and several smaller peaks at other frequencies. The BTr and BLb data sets are shorter, but their FSa confirm the strongest peak found in the analysis of the BAb and UBr data sets.

HD\,129116 (b\,Cen, Field 2): No variability was detected in the UBr, BAb, and BTr data.

HD\,130807 ($o$\,Lup, Field 2): Observed with UBr, BAb and BTr. The FS of the UBr data shows several significant frequencies with the strongest peak at a frequency of 1.106~d$^{-1}$. The BAb data set has a gap in the middle of the run and the BTr data set is significantly shorter, but FSa of both confirm the strongest frequency found in the FS of the UBr data. The variability of this magnetic star was described by \citet{buysschaert2019}.

HD\,132058 ($\beta$\,Lup, Field 2): Observed with UBr, BAb, BTr and BLb. The UBr and BAb data show variability with a main period of 3.632\,d. The BTr and BLb data have significantly shorter time bases, but confirm the main period obtained from the UBr and BAb data.

HD\,132200 ($\kappa$\,Cen, Field 2): Observed with UBr, BAb, BTr and BLb. The FSa of the UBr and BAb data show multiple significant frequencies around $\sim$0.698~d$^{-1}$. The BTr and BLb data sets are significantly shorter, but their analysis confirms this variability.

HD\,133242/3 ($\pi$\,Lup, Field 2): No variability was detected in the BTr data.

HD\,134481/2 ($\kappa$\,Lup, Field 2): No variability was detected in the UBr, BAb and BTr data. The BAb data set has a gap in the middle of the run and the BTr data set is rather short.

HD\,134505 ($\zeta$\,Lup, Field 2): No variability was detected in the UBr, BAb, BTr and BLb data. The BTr and BLb data sets are short.

HD\,135379 ($\beta$\,Cir, Field 2): Observed with UBr, BAb and BTr. The FSa of the UBr and BAb data show low-frequency variability. The BAb data set has a gap in the middle of the run and the BTr data have a relatively short time base.

HD\,135734 ($\mu$\,Lup, Field 2): No variability was detected in the UBr, BAb and BTr data. The BTr data set is short.

HD\,136298 ($\delta$\,Lup, Fields 2 \& 9): {\it Field 2:} Observed with UBr, BAb, BTr and BLb. The FSa of the UBr and BAb data show several significant frequencies in the domain of SPB stars, with the strongest peak at $\sim$1.27~d$^{-1}$. The BTr and BLb data sets are significantly shorter, but their FSa confirm the strongest peak discovered from the UBr and BAb observations. {\it Field 9:} Observed with BAb, BLb, BHr and UBr. The FSa of the data show a clear signal at 5.902~d$^{-1}$ as well as many other frequencies. The BRITE data for this star were discussed by \citet{2017sbcs.conf..186C}.

HD\,136415/6 ($\gamma$\,Cir, Field 2): Observed with UBr, BAb and BTr. The FS of the UBr data shows a few significant frequencies, the strongest signal at 2.46~d$^{-1}$ and the second strongest at 4.49~d$^{-1}$. The BAb data set has a large gap in the middle of the run and does not show any variability. The BTr data set is significantly shorter, but its FS confirms the two strongest peaks found in the FS of the UBr data.

HD\,136504 ($\varepsilon$\,Lup, Fields 2 \& 9): {\it Field 2}: Observed with UBr, BAb, and BTr. The UBr light curve shows a clear period of $\sim$4.55~d; some smaller peaks in the FS of the UBr data might also be intrinsic. The BAb data set has a gap in the middle of the run, but confirms the main period found from the UBr observations. The BTr data set is too short for any conclusive analysis. {\it Field 9}: Observed by BLb. The BLb light curve was analysed by \citet{2019MNRAS.488...64P}, who identified a weak intrinsic signal at the orbital frequency. The light curve is otherwise free of any obvious variability.

HD\,136664 ($\phi^2$\,Lup, Field 9): No variability was detected in the noisy BHr data set.

HD\,138690 ($\gamma$\,Lup, Fields 2 \& 9): Close visual double, one component of which is a single-lined spectroscopic system with an orbital period of 2.8~d. {\it Field 2:} Observed with UBr, BAb, BTr and BLb. The UBr and BAb data show clear variability with the orbital period. The BTr and BLb data sets are significantly shorter, but the same variability can also be identified. {\it Field 9:} Observed with BLb, UBr, BHr and BAb. The FSa of the BLb, UBr and BHr data show a clear peak at 0.351~d$^{-1}$, corresponding to the orbital period. The BAb data set is short and scarce. A detailed analysis of the BRITE data of this star was presented by \citet{2021MNRAS.503.5554J}.

HD\,139127 ($\omega$\,Lup, Field 2): No variability was detected in the short (8 days long) UBr data set.

HD\,139365 ($\tau$\,Lib, Field 9): Early B-type spectroscopic binary exhibiting a heartbeat. Observed with BHr, UBr and BLb. The data show variability with the orbital period of 3.45~d. These results were first reported by \citet{2018pas8.conf..115P}.

HD\,141556 ($\chi$\,Lup, Field 9): No variability was detected in the BLb, BHr or UBr data.

HD\,142669 ($\rho$\,Sco, Field 9): Observed with BLb, UBr and BHr. The BLb light curve exhibits small scatter and its FS reveals a highly significant signal at 0.3697~d$^{-1}$. This frequency is also found in the FS of the BHr data. The UBr data are of inferior quality and show no significant signal. When phased according to the corresponding period, the BLb and BHr data yield a light curve that is consistent with ellipsoidal variability.

HD\,143018 ($\pi$\,Sco, Field 9): Observed with BAb, BHr, BLb and UBr. The FSa of the data show clear variability with the most prominent frequency at 1.2735~d$^{-1}$ which is twice the orbital frequency. The data were analysed in detail by \cite{2017sbcs.conf..120P}, who found also pulsation with a frequency corresponding to a {\sl p} mode.

HD\,143118 ($\eta$\,Lup, Field 9): Observed with BLb, UBr and BHr. The FS of the BLb data shows a weak signal at 0.206~d$^{-1}$. No signal was detected in the noisier UBr and the shorter BHr data sets.

HD\,143275 ($\delta$\,Sco, Field 9): This is an early B-type Be star in an extraordinarily eccentric, 10.8-yr spectroscopic binary. Observed by BLb, UBr and BAb. The FSa of the data show a weak low-frequency variation.

HD\,144217/8 ($\beta^1/\beta^2$\,Sco, Field 9): Observed with UBr, BLb and BAb. The FSa of the UBr and BLb data show a signal at 0.448~d$^{-1}$ with a possible triplet structure. The BAb data set is sparse.

HD\,144294 ($\theta$\,Lup, Field 9): No variability was detected in the BLb data.

HD\,144470 ($\omega^1$\,Sco, Field 9): No variability was detected in the BLb, BHr, UBr and BAb data.

HD\,145482 (c$^2$\,Sco, Field 9): Observed with BHr. The FS of the data shows clear signals at 0.173~d$^{-1}$ and 1.177~d$^{-1}$.

HD\,145502/1 ($\nu$\,Sco, Field 9): Observed with UBr and BLb. The FS of the UBr data shows a clear signal at $0.172$~d$^{-1}$ and possible additional peaks at 0.824~d$^{-1}$ and 1.174~d$^{-1}$ with a possible triplet centred at 1.574~d$^{-1}$. The BLb data set is of insufficient length to reveal the signal.

HD\,147165 ($\sigma$\,Sco, Field 9): This is a binary system consisting of two early B-type main-sequence components, with the primary star exhibiting $\beta$~Cep-type pulsations. Observed by BHr, BLb, UBr and BAb. The FSa of the data show complex variability with a principal peak at 4.0509~d$^{-1}$, but with several other peaks of comparable amplitude. The BRITE observations were first discussed and interpreted by \citet{2018pas8.conf...77P}.

HD\,148478/9 ($\alpha$\,Sco, Antares, Field 9): M-type supergiant, observed by UBr, BLb and BAb. The light curves show long-term variability.

HD\,148688 (V1058\,Sco, Field 9): Early B-type supergiant, observed by BHr. The FS of the data shows stochastic low-frequency variability typical for evolved B-type stars.

HD\,148703 (N\,Sco, Field 9): Eccentric eclipsing binary system observed by BHr, UBr and BLb. The BHr and UBr light curves reveal eclipses. The orbital period of the system is long and the BLb light curve does not cover any eclipse.

HD\,149038 ($\mu$\,Nor, Field 9): Observed with BHr. The FS of the data shows the typical stochastic low-frequency signal characteristic of evolved O and B-type stars.

HD\,149404 (V918\,Sco, Field 9): Evolved, post-Roche-lobe overflow non-eclipsing O-type binary, observed with BHr. The FS appears as  typical for an O-type supergiant  with a strong peak at 0.205~d$^{-1}$ corresponding to half the 9.84-d orbital period. A detailed study of the BHr data (in combination with the SMEI data) was presented by \citet{rauw2019}.

HD\,149438 ($\tau$\,Sco, Field 9): No variability was detected in the UBr, BHr and BLb data. The BLb measurements show some minor contamination due to scattered moonlight near dates of lunar conjunction.

HD\,151680 ($\varepsilon$\,Sco, Field 9): No variability was detected in the BLb and UBr data.

HD\,151804 (V973\,Sco, Field 9): Observed with BHr. The FS of the data exhibits a typical stochastic low-frequency signal of evolved O-type stars.

HD\,151890 ($\mu^1$\,Sco, Field 9): Observed with UBr, BHr and BLb. The FSa of the data show strong variability with a frequency of $1.383$~d$^{-1}$, corresponding to half of the known 1.446~d orbital period.

HD\,151985 ($\mu^2$\,Sco, Field 9): Observed with BLb, BHr and UBr. The FSa of the BLb and BHr data show a weak signal at 1.383~d$^{-1}$, the same as detected in the nearby (5.8$\arcmin$ apart) $\mu^1$\,Sco. This means that the photometry pipeline did not separate well the contribution from both stars. The FS of the UBr data shows a somewhat weaker signal at this frequency, but also at twice this value which is 2.765~d$^{-1}$.

HD\,157056 ($\theta$\,Oph, Field 3): This well known $\beta$ Cephei star was observed by UBr. The data set is short and scarce, but its FS reveals the most prominent pulsation frequency of $7.118$~d$^{-1}$. This star was studied in detail using BRITE data by \citet{Walczak2019}.

HD\,157792 (b\,Oph, Field 3): Observed by UBr. No variability was detected in the short and scarce UBr data set.

HD\,157919 (d\,Oph, Field 3): Observed by UBr. No variability was detected in the short and scarce UBr data set.

HD\,158408 ($\upsilon$\,Sco, Field 3): Observed with UBr. The data set is short and scarce, but the light curve shows signal with a period of 2.47~d.

HD\,158926 ($\lambda$\,Sco, Field 3): The star is known to be an eclipsing $\beta$ Cep star. Observed with UBr. The data set is short and scarce and its FS shows a high amplitude peak at a frequency of $4.679$~d$^{-1}$.

HD\,159433 (Q\,Sco, Field 3): No variability was detected in the short and scarce UBr data set.

HD\,159532 ($\theta$\,Sco, Field 3): Observed with UBr. The data set is short and scarce and shows a period of $\sim$13\,d.

HD\,160578 ($\kappa$\,Sco, Field 3): Known $\beta$~Cep-type star. Observed with UBr. The data set is short and scarce and its FS shows a high amplitude peak at a frequency of $5.001$~d$^{-1}$.

HD\,161471 ($\iota^1$\,Sco, Field 3): No variability was detected in the short and scarce UBr data set.

HD\,161592 (X\,Sgr, Field 3): Classical Cepheid, observed with UBr. The data set is short and scarce but clearly shows the variability with the known period of 7.1~d.

HD\,161892 (G\,Sco, Field 3): No variability was detected in the short and scarce UBr data set. The star was included in the study of \cite{kallinger2019}.

HD\,164975 ($\gamma^1$\,Sgr, Field 3): Classical Cepheid, observed with UBr. The data set is short and scarce, but clearly shows variability with a period of 7.7~d.

HD\,165135 ($\gamma^2$\,Sgr, Field 3): No variability was detected in the short and scarce UBr data set. The star was included in the study of \cite{kallinger2019}.

HD\,166937 ($\mu$\,Sgr, Field 3): Observed with UBr. The data set is short and scarce with some indication of variability.

HD\,167618 ($\eta$~Sgr, Field 3): Observed with UBr. The light curve is short and scarce and shows an irregular long-term variability.

HD\,168454 ($\delta$\,Sgr, Field 3): No variability was detected in the short and scarce UBr data set.

HD\,169022 ($\varepsilon$\,Sgr, Field 3): No variability was detected in the short and scarce UBr data set.

HD\,169916 ($\lambda$\,Sgr, Field 3): No variability was detected in the short and scarce UBr data set.

HD\,173300 ($\phi$\,Sgr, Field 3): No variability was detected in the short and scarce UBr data set.

HD\,186882 ($\delta$\,Cyg, Fields 4 \& 10): {\it Field 4:} No variability was detected in the BLb, UBr and BTr data sets. {\it Field 10:} No variability was detected in the BAb, BLb and UBr data.

HD\,187849 (19\,Cyg, Field 4): Observed with BTr. The data set is relatively short and shows strong and clear variability with a period of $\sim$28.4~d.

HD\,188892 (22\,Cyg, Fields 4 \& 10): {\it Field 4:} No variability was detected in the short and scarce BLb data set. {\it Field 10:} No variability was detected in the BTr and UBr data.

HD\,188947 ($\eta$\,Cyg, Field 4): No variability was detected in the BTr, UBr and BLb data. The star was included in the study by \cite{kallinger2019}. BTr data are of good quality, but UBr and BLb data sets are short and scarce.

HD\,189178 (Field 10): No variability was detected in the short BTr data set.

HD\,189687 (25\,Cyg, Fields 4 \& 10): Known Be star. {\it Field 4:} Observed with BLb. The data set is short and scarce, but shows the evidence of weak variability. {\it Field 10:} Observed with BTr and UBr. Twelve days of BTr data show clear and complex variability. The 17 days long UBr light curve has inferior quality and does not show the variability. The strongest signal in the FS of these data is at 1.76\,d$^{-1}$.

HD\,189849 (15\,Vul, Field 10): Observed with BAb, BTr and UBr. No variability was detected in UBr and BTr data. The BAb light curve is scarce and of poor quality.

HD\,191610 (28\,Cyg, Fields 4 \& 10): This is a known Be star. {\it Field 4:} Observed with BTr and BLb. The data sets are short and scarce, but show some complex variability. {\it Field 10:} Observed with BLb, UBr and BTr. The FSa of the data reveal complex Be-type variability with the main peak at a frequency of 1.377~d$^{-1}$. The BRITE data for this star have been studied in detail by \citet{baade2018}.

HD\,192577/8 (31\,Cyg, Fields 4 \& 10): {\it Field 4:} Observed with BTr, UBr and BLb. The BTr data reveal complex variability. The UBr and BLb data sets are short and scarce. {\it Field 10:} No variability was detected in the scarce BLb and BAb data sets.

HD\,192640 (29\,Cyg, Fields 4 \& 10): Known $\delta$\,Sct-type variable. {\it Field 4:} Observed with BTr and BLb. The FSa of the data show many peaks at frequencies typical for a $\delta$\,Sct star, the highest lies at a frequency of 37.4\,d$^{-1}$. {\it Field 10:} Observed with BAb, BTr and UBr. A very similar to Field 4 $\delta$\,Sct-type variability can be seen in these data.

HD\,192685 (QR\,Vul, Field 10): Observed with BTr, BLb and UBr. The FS of the 156-d long BTr data shows strong and complex variability with the main peak at a frequency of 0.195~d$^{-1}$. The BLb data set is short and UBr data set is of poor quality.

HD\,192806 (23\,Vul, Field 4): Observed with BTr. The data do not show any prominent variability, but several low-frequency peaks can be seen in their FS.

HD\,192909/10 ($o$\,Cet, Field 4): Observed with BTr and UBr. The BTr light curve shows some strong and complex variability. No variability was detected in the short and scarce UBr data set.

HD\,193092 (Field 4): No variability was detected in the BTr data.

HD\,193237 (P\,Cyg, Fields 4 \& 10): Blue supergiant with known variability. {\it Field 4:} Observed with BTr, BLb and UBr. The BTr and BLb light curves show strong variability. The UBr data set is short and scarce. {\it Field 10:} Observed with BTr, UBr, BLb and BAb. The BTr data show clear variability. The UBr, BLb and BAb data sets are shorter. The BRITE data for this star were investigated in detail by \citet{elliott2022}.

HD\,194093 ($\gamma$\,Cyg, Fields 4 \& 10): {\it Field 4:} Observed with BTr, BLb and UBr. The BTr light curve shows some low-amplitude long-term variability, while the BLb data do not reveal any variability. The UBr data set is short and scarce. {\it Field 10:} No variability was detected in the UBr, BAb and BLb data.

HD\,194317 (39\,Cyg, Field 4): Observed with BTr and UBr. The FS of the BTr data reveals a significant peak at a frequency of 0.76~d$^{-1}$. The UBr data set is short and scarce.

HD\,194335 (Field 10): Known Be star. No variability was detected in the BLb data.

HD\,195068/9 (43\,Cyg, Field 10): Observed with BTr and BLb. The FS of the BTr data clearly shows $\gamma$\,Dor-type pulsations with the main frequency at 1.25~d$^{-1}$. No variability was detected in the short BLb data set. The star was studied in detail by \citet{zwintz2017}.

HD\,195295 (41\,Cyg, Fields 4 \& 10): {\it Field 4:} No variability was detected in the BTr, UBr and BLb data. {\it Field 10:} No variability was detected in the UBr data.

HD\,195556 ($\omega^1$\,Cyg, Field 10): Observed with BAb and BTr. The BAb data set is of poor quality. The BTr data set is only 12.4 days long and shows evidence of long-term variability.

HD\,196093/4 (47\,Cyg, Field 4): Observed with BTr and UBr. The BTr data show complex variability. The UBr data set is scarce.

HD\,197345 ($\alpha$\,Cyg, Fields 4 \& 10): {\it Field 4:} Observed with BTr, BLb and UBr. The FSa of the BTr and BLb data show clear variability at low frequencies; the UBr data set is scarce. {\it Field 10:} Observed with UBr, BAb and BLb. The FSa of the data show variability at low frequencies.

HD\,197912 (52\,Cyg, Field 4): No variability was detected in the short BTr and in the short and scarce BLb and UBr data sets.

HD\,197989 ($\varepsilon$\,Cyg, Fields 4 \& 10): {\it Field 4:} No variability was detected in the BLb and BTr data or in the short and scarce UBr data set. {\it Field 10:} No variability was detected in the BAb, BLb and UBr data. The star was included in the study by \cite{kallinger2019}.

HD\,198183 ($\lambda$\,Cyg, Fields 4 \& 10): {\it Field 4:} Observed with BLb, UBr and BTr. The BLb and UBr data sets are short. The BTr data show complex variability. {\it Field 10:} Observed with BTr, UBr, BLb and BAb. A complex variability is clearly present in the BTr data and less evident in the UBr data. The BAb data set is of poor quality, the BLb data set is short.

HD\,198478 (55\,Cyg, Fields 4 \& 10): {\it Field 4:} Observed with BTr and BLb. The FS of the BTr data shows complex low-frequency variability. The BLb data set is short and scarce. {\it Field 10:} Observed with BTr, BAb, UBr and BLb. The BTr light curve shows complex variability which is less evident in the BAb light curve due to the sparsity and low quality of the data. The FS of the UBr data shows low-frequency variability. The BLb data set is short.

HD\,198639 (56\,Cyg, Field 10): No variability was detected in the BTr or BAb data.

HD\,198726 (T\,Vul, Fields 4 \& 10): Classical Cepheid. {\it Field 4:} Observed with BTr and UBr. The BTr light curve shows variability with a period of $\sim$4.44~d. The UBr data set is shorter, but the variability is clearly detectable despite the sparsity of the data. {\it Field 10:} Observed with BTr. The light curve shows variability with the same period as in the Field 4 observations.

HD\,198809 (31\,Vul, Field 4): The UBr data set is short and scarce. No variability was detected.

HD\,199081 (57\,Cyg, Fields 4 \& 10): {\it Field 4:} Observed with BLb and BTr. The FSa of the data show strong frequency at 0.934~d$^{-1}$. {\it Field 10:} Observed with BTr, BAb, UBr and BLb. The FS of the BTr data shows multiple frequencies below $\sim$1.1~d$^{-1}$; the strongest peak is at a frequency of 0.934~d$^{-1}$. The BAb and UBr data sets are shorter, but the low-frequency variability is confirmed in their FSa. The BLb data set is of poor quality.

HD\,199629 ($\nu$\,Cyg, Fields 4 \& 10): {\it Field 4:} No variability was detected in the UBr data. {\it Field 10:} No variability was detected in the UBr data.

HD\,200120 (59\,Cyg, Fields 4 \& 10): Known Be star. {\it Field 4:} No variability was detected in the short and scarce BLb data set. {\it Field 10:} Observed with BTr, BAb and BLb. The FS of the BTr data shows clear variability with frequencies between $\sim$1.5 and 2.2~d$^{-1}$. The BAb and BLb data sets are short and scarce.

HD\,200310 (60\,Cyg, Field 10): Known Be star, observed with BTr, BAb and BLb. The FS of the BTr data shows clear variability with frequencies below 4~d$^{-1}$. The BAb data set is shorter and of poorer quality. The BLb data set is scarce.

HD\,200905 ($\xi$\,Cyg, Field 4): Observed with BTr and UBr. The BTr light curve shows variability with a period of $\sim$17.3\,d. No variability was detected in the short and scarce UBr data set.

HD\,201078 (DT\,Cyg, Fields 4 \& 10): Classical Cepheid. {\it Field 4:} Observed with BTr and UBr. The light curves show classical Cepheid variability with a period of $\sim$2.5\,d. {\it Field 10:} Observed with BTr and BLb. The BTr light curve shows variability with the same period as the Field 4 data. The BLb light curve is short, but the variability can also be seen.

HD\,201251 (63\,Cyg, Field 4): No variability was detected in the BTr data or in the short and scarce UBr data set.

HD\,201433 (V389 Cyg, Field 10): This is a triple system with SPB pulsations, observed with BTr and BLb. The BTr light curve shows SPB-type pulsations, confirmed also in the BLb data despite their scarcity. The BRITE data of this star were discussed in detail by \citet{kallinger2017}.

HD\,202109 ($\zeta$\,Cyg, Fields 4 \& 10): {\it Field 4:} No variability was detected in the BLb and BTr data or in the short and scarce UBr data sets. {\it Field 10:} No variability was detected in the UBr and BAb data or in the short and scarce BLb data set.

HD\,202444 ($\tau$\,Cyg, Fields 4 \& 10): {\it Field 4:} No variability was detected in the BLb and BTr data or in the short and scarce UBr data set. {\it Field 10:} No variability was detected in the BAb data or in the short and scarce BLb data set.

HD\,202850 ($\sigma$\,Cyg, Fields 4 \& 10): {\it Field 4:} Observed with BTr, BLb and UBr. The FSa of the BTr and BLb data show some low-frequency variability. The UBr data set is short and scarce. {\it Field 10:} Observed with BAb, BLb and BTr. The FS of the BTr data shows low-frequency variability. The BAb and BLb data sets are shorter and scarce.

HD\,202904 ($\upsilon$\,Cyg, Fields 4 \& 10): Known Be star. {\it Field 4:} Observed with BTr, BLb and UBr. The FS of the BTr data shows strong variability at very low frequencies. The FS of the BLb data reveals a frequency of $\sim$1.47~d$^{-1}$ on the top of the lower frequency variability. The UBr data set is short and scarce. {\it Field 10:} Observed with BTr, BAb and BLb. The FS of the BTr data shows clear variability with a frequency of 1.47~d$^{-1}$. The BAb and BLb data sets are shorter and scarce.

HD\,203064 (68\,Cyg, Fields 4 \& 10): {\it Field 4:} No variability was detected in the BTr and BLb data. {\it Field 10:} No variability was detected in the BTr data or in the short and scarce BAb and BLb data sets.

HD\,203156 (V1334\,Cyg, Fields 4 \& 10): Known classical Cepheid. {\it Field 4:} Observed with BTr and UBr. The light curves clearly show variability with a period of $\sim$3.3~d. {\it Field 10:} Observed with BTr and BLb. The variability seen in the Field 4 data is confirmed in the BTr data set. The BLb data set is short and scarce.

HD\,203280 ($\alpha$\,Cep, Field 11): No variability was detected in the short BLb data set.

HD\,205021 ($\beta$\,Cep, Field 11): The BHr and BLb data sets are very sparse and not useful for scientific analysis.

HD\,205435 ($\rho$\,Cyg, Field 10): No variability was detected in the good-quality BTr and UBr data or in the short and scarce BLb data set.

HD\,206570 (V460\,Cyg, Field 4): No variability was detected in the BTr data.

HD\,207260 ($\nu$\,Cep, Field 11): Observed with BHr and BLb. The FS of the BHr data shows weak low-frequency variability. The BLb data set is short and of lower precision.

HD\,209790 ($\xi$\,Cep, Field 11): No variability was detected in the BHr data and in the short BLb data set.

HD\,209975 (19\,Cep, Field 11): Observed with BHr and BLb. The FSa of the BHr data show some stochastic low-frequency variability. The BLb data set is short and scarce.

HD\,210745 ($\zeta$\,Cep, Field 11): No variability was detected in the BHr data.

HD\,211336 ($\varepsilon$\,Cep,Field 11): Known $\delta$ Sct-type pulsator, observed with BTr, BLb and BHr. The FSs of the BTr and BLb data show $\delta$~Sct-type variability. The BHr data set is short and not usable.

HD\,213306 ($\delta$\,Cep, Field 11): This is the prototype of classical Cepheids. Observed by BTr, BLb and BHr. The data are of excellent quality and show variability with a period of $\sim$5.4\,d.

HD\,216228 ($\iota$\,Cep, Field 11):  No variability was detected in the relatively short BHr data set.

HD\,218376 (1\,Cas, Field 11): No variability was detected in the short BLb and BHr data sets.

HD\,220652 (4\,Cas, Field 11): Observed with BHr. The data set is short and its FS shows some low-frequency variability.

HD\,221253 (AR\,Cas, Field 11): Known eclipsing and spectroscopic binary. Observed with BLb and BHr. The light curves show clear eclipses with an orbital period of $\sim$6.1\,d.

HD\,224572 ($\sigma$ Cas, Field 11): No variability was detected in the BAb data set.

HD\,225289 (V567~Cas, Field 11: No variability was detected in the short BHr data set.

\FloatBarrier

\section{Format of BRITE-Constellation data in the archives}
\label{sec:archives}

\subsection{Contents of the BRITE-Constellation data in the archives}

All decorrelated BRITE-Constellation data described and illustrated in this article are freely available for download from the BRITE Public Data Archive (PDA)\footnote{https://brite.camk.edu.pl/pub} and the Canadian Astronomy Data Centre (CADC)\footnote{https://www.cadc-ccda.hia-iha.nrc-cnrc.gc.ca/en/}. 

Both databases provide the data as ASCII files split into setups (explained in Appendix \ref{sect:setups}). For each setup, five different files are available. They contain the original time series (i.e., before the decorrelation procedure), the decorrelated time series, the time series containing averages per satellite orbit, the variability model used in the decorrelation, and the log file generated by the software during the decorrelation. All files have headers that describe their contents in detail. The files containing photometric time series (i.e., original, decorrelated and orbit averaged data) include columns with the Heliocentric or Barycentric Julian Date, instrumental magnitudes including errors and magnitudes with the mean instrumental magnitude of the given setup subtracted. Observations with the largest errors are filtered out during the decorrelation process. For further analyses, the provided errors in magnitudes can be used to filter the data if needed.

\subsection{BRITE-Constellation photometry subrasters and setups}
\label{sect:setups}

The following brief explanation of the subrasters and setups used for BRITE-Constellation observations provides essential guidance for a better understanding of BRITE-Constellation data. More details are given in \citet{popowicz2017}. 

The full frame CCD images of the BRITE-Constellation satellites cannot be transferred to ground at the cadence taken in space which is typically about 20~s.
Therefore, in the context of BRITE photometry, the expression ‘subrasters’ concerns the definition of which parts of the full frame CCD are rendered into images before transmitting to the ground. For that the $(x,y)$ starting position on the CCD is provided and the $x$ and $y$ size of each subraster is required. In stare mode, those regions of interest or subrasters are 24\,$\times$\,24 pixels in size, while in chopping mode they are rectangular and cover 24\,$\times$\,48 pixels. Typically 15 to 30 subrasters are set up for each satellite and observing campaign. With a few exceptions, each subraster contains the profile of a single target in its central part in the case of stare mode. In chopping mode, a star is located either on one or the other side of the rectangular subraster. In addition to the positions of the subrasters, the exposure time and the time between exposures has to be specified as well as the central right ascension, declination and the roll angle with respect to the optical axis. It is important that the stellar profile and a number of neighbouring background areas are fully rendered in each subraster even when the pointing changes slightly from exposure to exposure.

Data obtained by a given satellite are split into parts, called setups.
Each observing run starts with an initial setup number (1), which is the first attempt to acquire the selected stars properly in the defined subrasters. Very often this first setup needs to be changed or optimised, and, hence, a new setup is generated during an early phase of an observing campaign. Furthermore when stars (and new subrasters) are added or removed, a new setup is generated and uploaded to the respective satellite. 
During some observing runs, a slow drift on the orientation of a given satellite over time had been apparent and, therefore, a resetting of the subraster was needed in order to keep the stellar profiles well in range of the subraster borders.
Also, a change in exposure time or observing cadence always results in a new setup. Hence, during a full campaign for each satellite several setups might have been utilized. In some cases only two or three were used to collect data over up to 180 d but it can be up to seven or more if needed. 
Data can also be split into setups during the initial reductions, that is on the ground (see Appendix A of \citet{popowicz2017}).

New setups introduced during observations can have an effect on the integrity of a light curve (time, magnitude) after reduction and even decorrelation as described in this paper. When a new setup changes the position of the star in a subraster from (too) close to the border to centre (or accidentally in reverse), jumps in the signal levels can occur. 
Jumps can also occur as a result of the reduction, because each setup is processed independently and for each setup an independent aperture is defined. Therefore, if a setup covers only a short part of a long period variable star light curve, the reduction and decorrelation of such a small segment can result in serious offsets that do not reflect the real long-term variability of the object.
If setups are generated during the initial process of data reduction, no offsets or other changes in data quality are introduced. In this case, the light curve is split only into two or more parts.

In particular during this first phase of the BRITE-Constellation mission, many individual setups were required for technical reasons (such as the very first observations conducted with the individual satellites that needed testing or the change from stare mode to chopping mode). In later years of the BRITE-Constellation mission, the number of setups decreased and were introduced more often by the reduction process that required splitting the data into parts.

\subsection{Recommendations on combining individual setups and data sets from different BRITE satellites}

The decorrelated photometric time series contain the Barycentric Julian date, instrumental magnitudes and magnitudes with the mean instrumental magnitude of the given setup subtracted (see the headers of the files stored in the archives for more information). In many cases, the time series stored as individual setups can just simply be stitched together (e.g., if the setups were generated during the reduction process) without any issues. 

However, very often different setups can have different mean magnitudes which cause different offsets to be taken into account when combining them into a full light curve. A good example is shown in Figure \ref{fig:sampling}: individual setups are marked with vertical lines and the numbers on top of the light curve illustrate the offsets that were applied and the $\sigma_{\rm orbit}^{\rm med}$ parameter as a measure for the data quality.

There are some cases where combining different setups needs a bit more care. For any type of a long-period variable, for example, stitching together the time series with the mean magnitude subtracted will not yield a reliable result. Introducing individual offsets by adding constants to a given setup might be necessary to cover the intrinsic variability of the star. 

Full light curves obtained by different BRITE satellites using the same filter can be combined by subtracting the mean magnitude from each light curve individually. 
Data taken in different BRITE filters can be combined after subtracting the mean magnitudes only if the amplitudes are expected to be similar in both bands or if the observations in one filter are scaled to the amplitude of the other.

\FloatBarrier

\section{Data quality per satellite}
\label{sec:quality}

\begin{figure}[htb]
\includegraphics[width=0.48\textwidth]{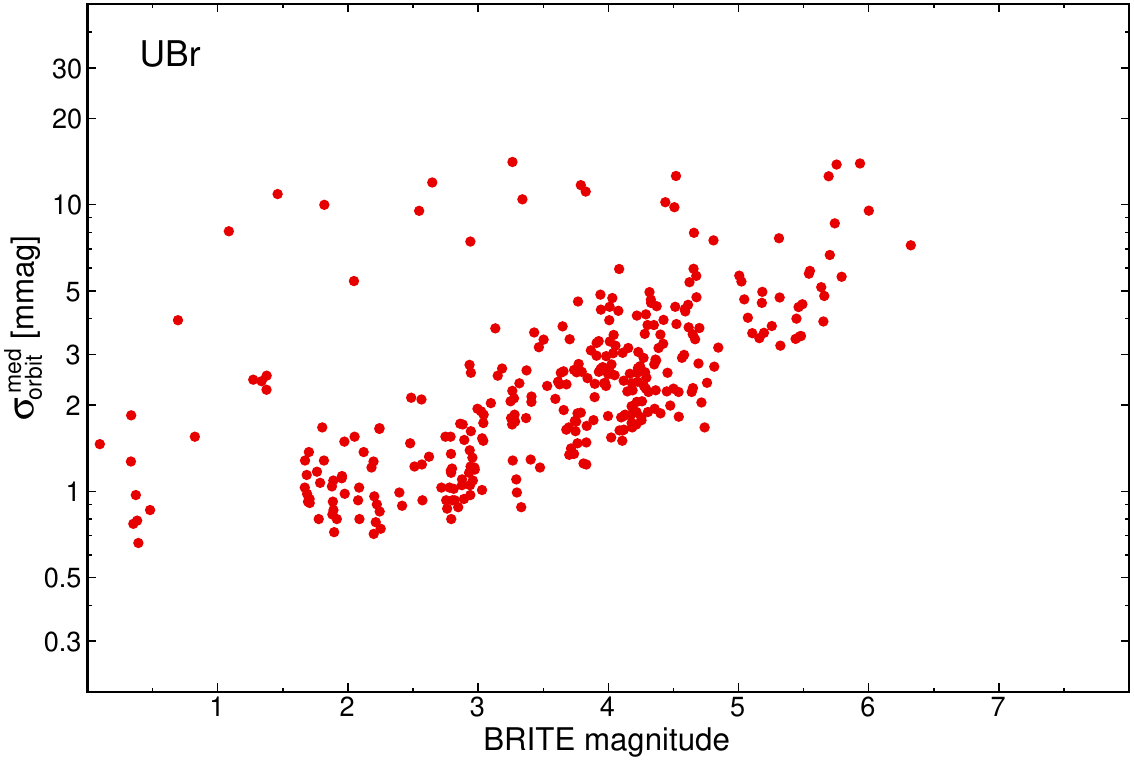}
\caption{
Values of $\sigma_{\rm orbit}^{\rm med}$ plotted as a function of instrumental BRITE magnitude for UBr setups.}
\label{fig:rms_UBr}
\end{figure}

\begin{figure}[htb]
\includegraphics[width=0.48\textwidth]{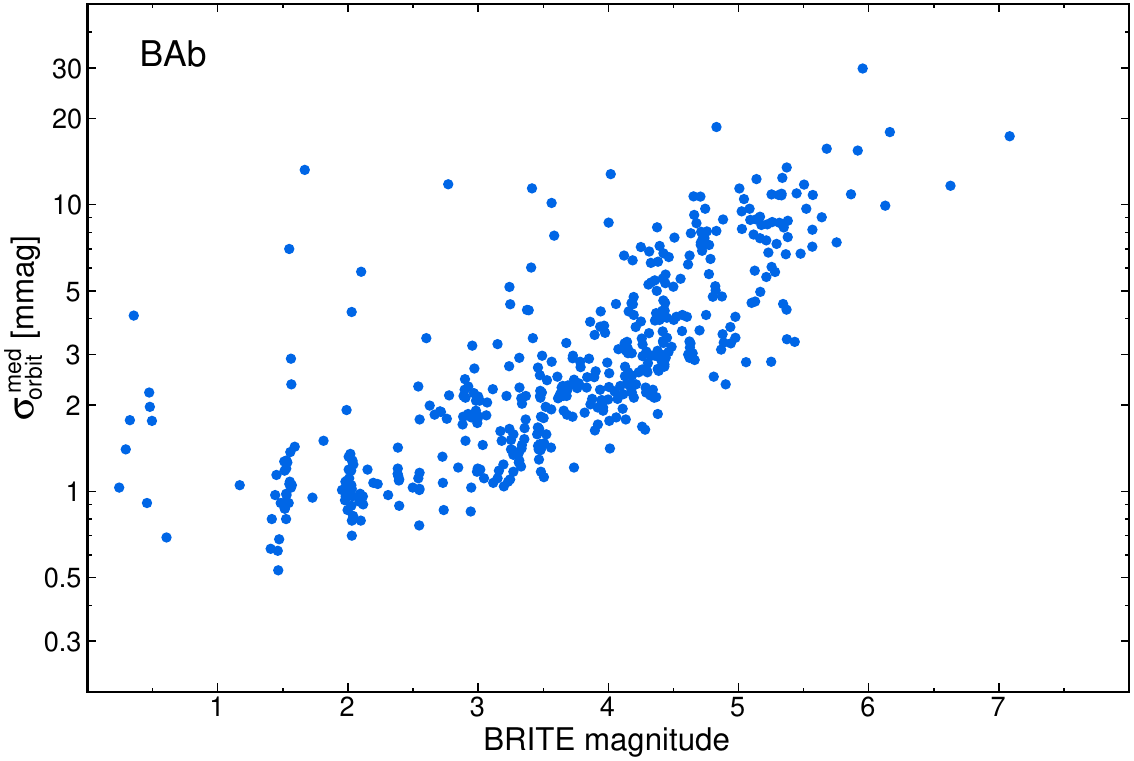}
\caption{
Values of $\sigma_{\rm orbit}^{\rm med}$ plotted as a function of instrumental BRITE magnitude for BAb setups.}
\label{fig:rms_BAb}
\end{figure}

\begin{figure}[htb]
\includegraphics[width=0.48\textwidth]{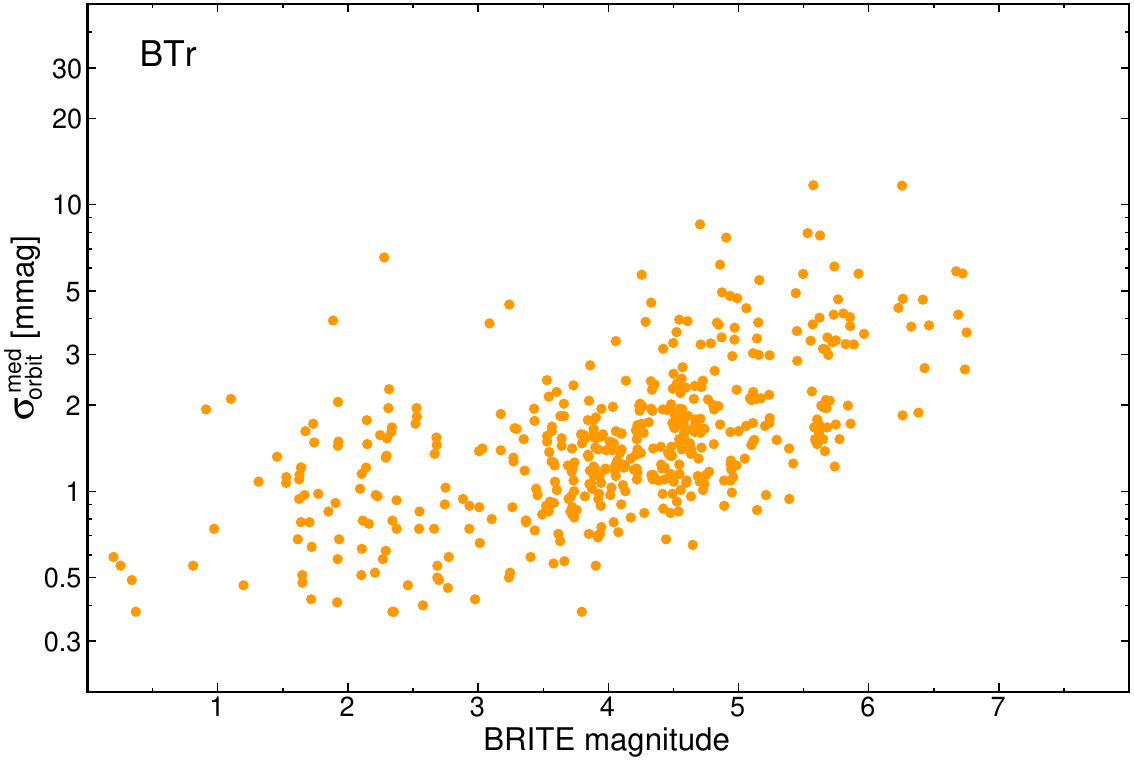}
\caption{
Values of $\sigma_{\rm orbit}^{\rm med}$ plotted as a function of instrumental BRITE magnitude for BTr setups.}
\label{fig:rms_BTr}
\end{figure}

\begin{figure}[htb]
\includegraphics[width=0.48\textwidth]{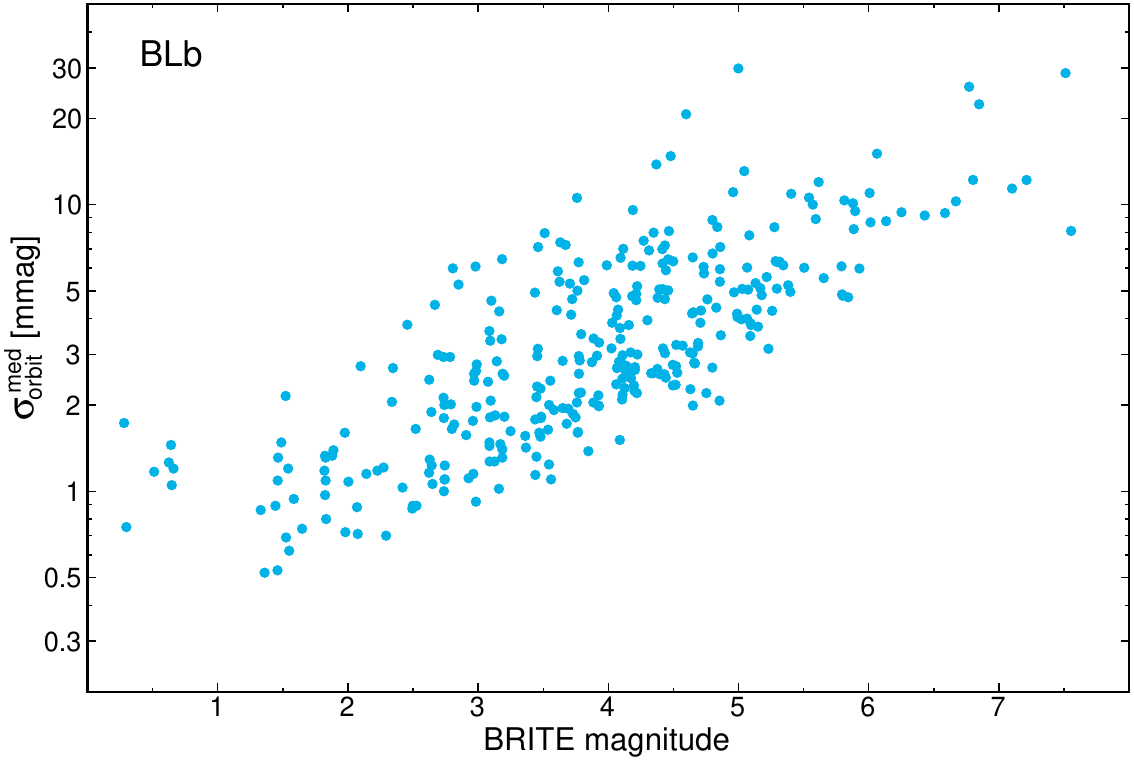}
\caption{
Values of $\sigma_{\rm orbit}^{\rm med}$ plotted as a function of instrumental BRITE magnitude for BLb setups.}
\label{fig:rms_BLb}
\end{figure}

\begin{figure}[htb]
\includegraphics[width=0.48\textwidth]{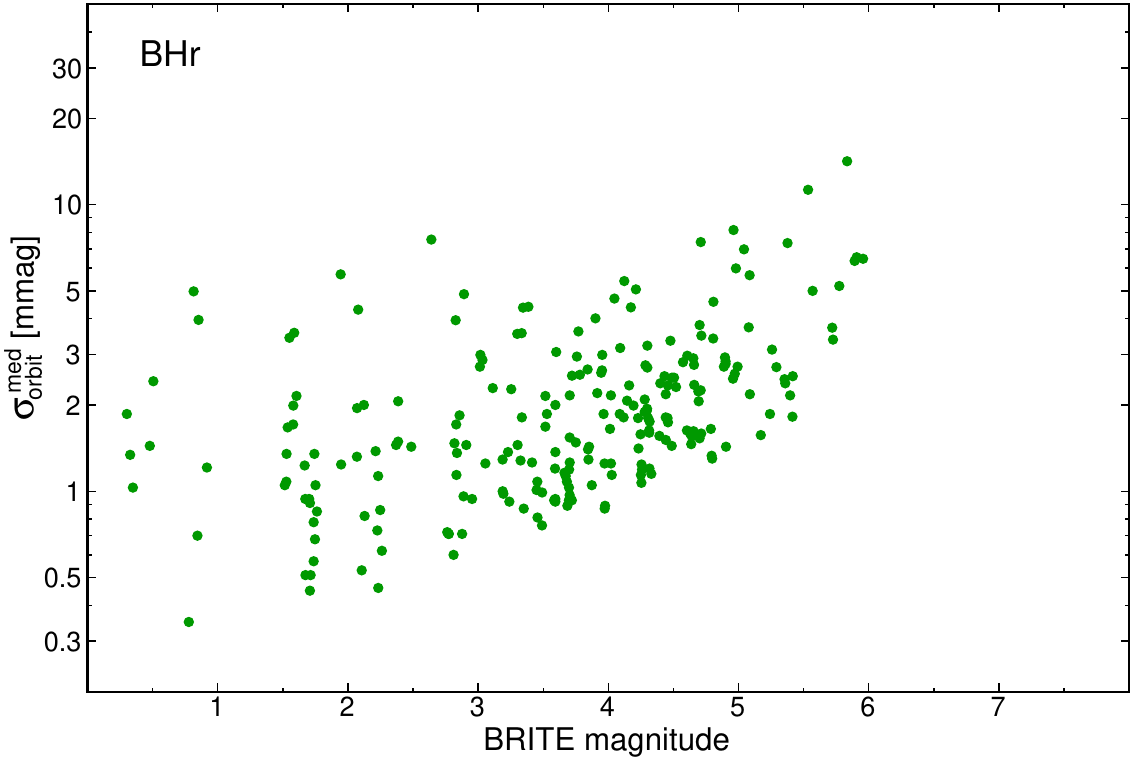}
\caption{
Values of $\sigma_{\rm orbit}^{\rm med}$ plotted as a function of instrumental BRITE magnitude for BHr setups.}
\label{fig:rms_BHr}
\end{figure}

Figures \ref{fig:rms_UBr} to \ref{fig:rms_BHr} illustrate the noise properties for each of the five BRITE-Constellation instruments individually.

\FloatBarrier
\clearpage

\section{Field maps}
\label{sec:fieldmaps}

\begin{figure*}[!htb]
    \centering
    \includegraphics[width=0.42\linewidth]{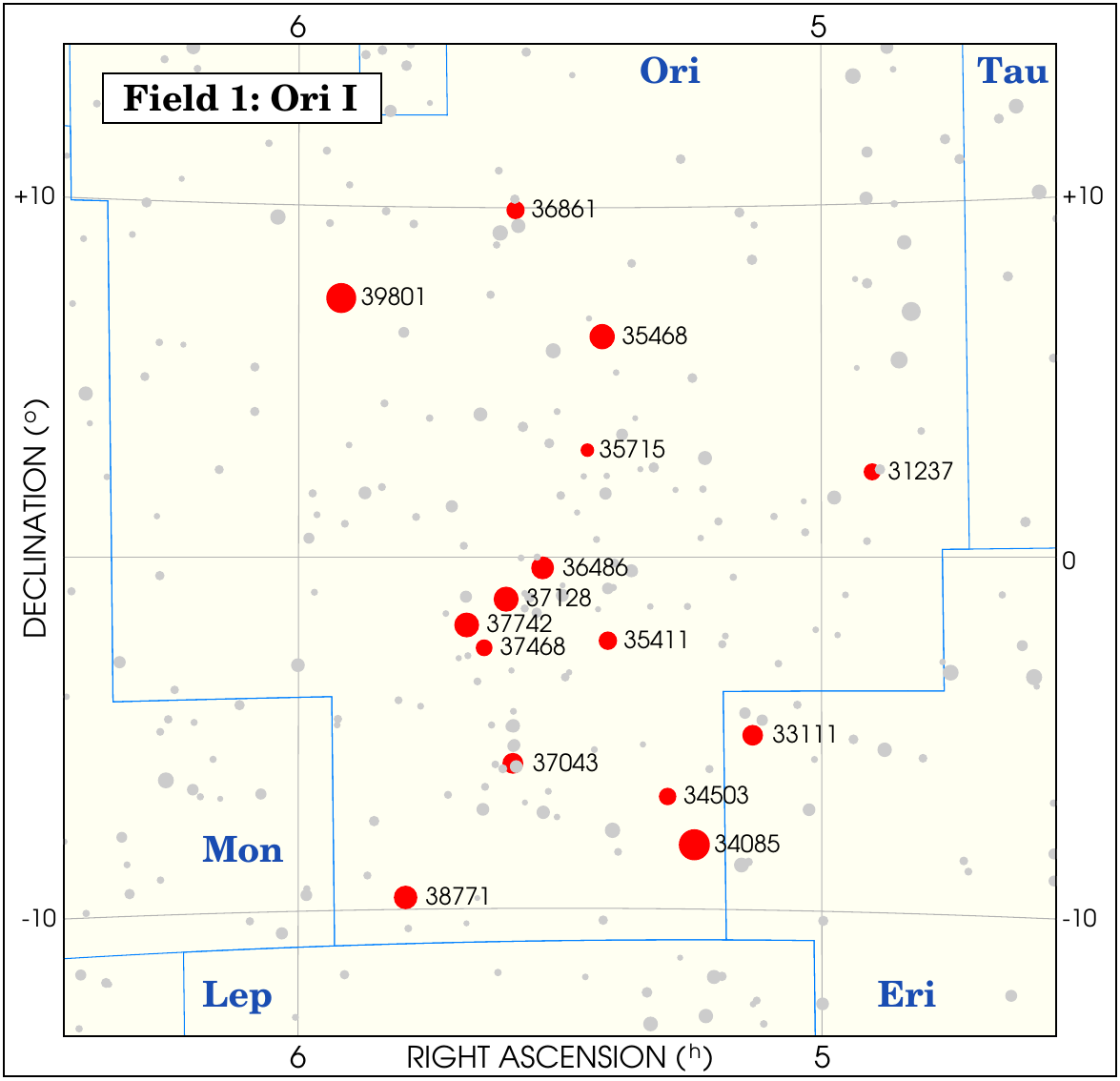}\hspace{0.5cm}\includegraphics[width=0.42\linewidth]{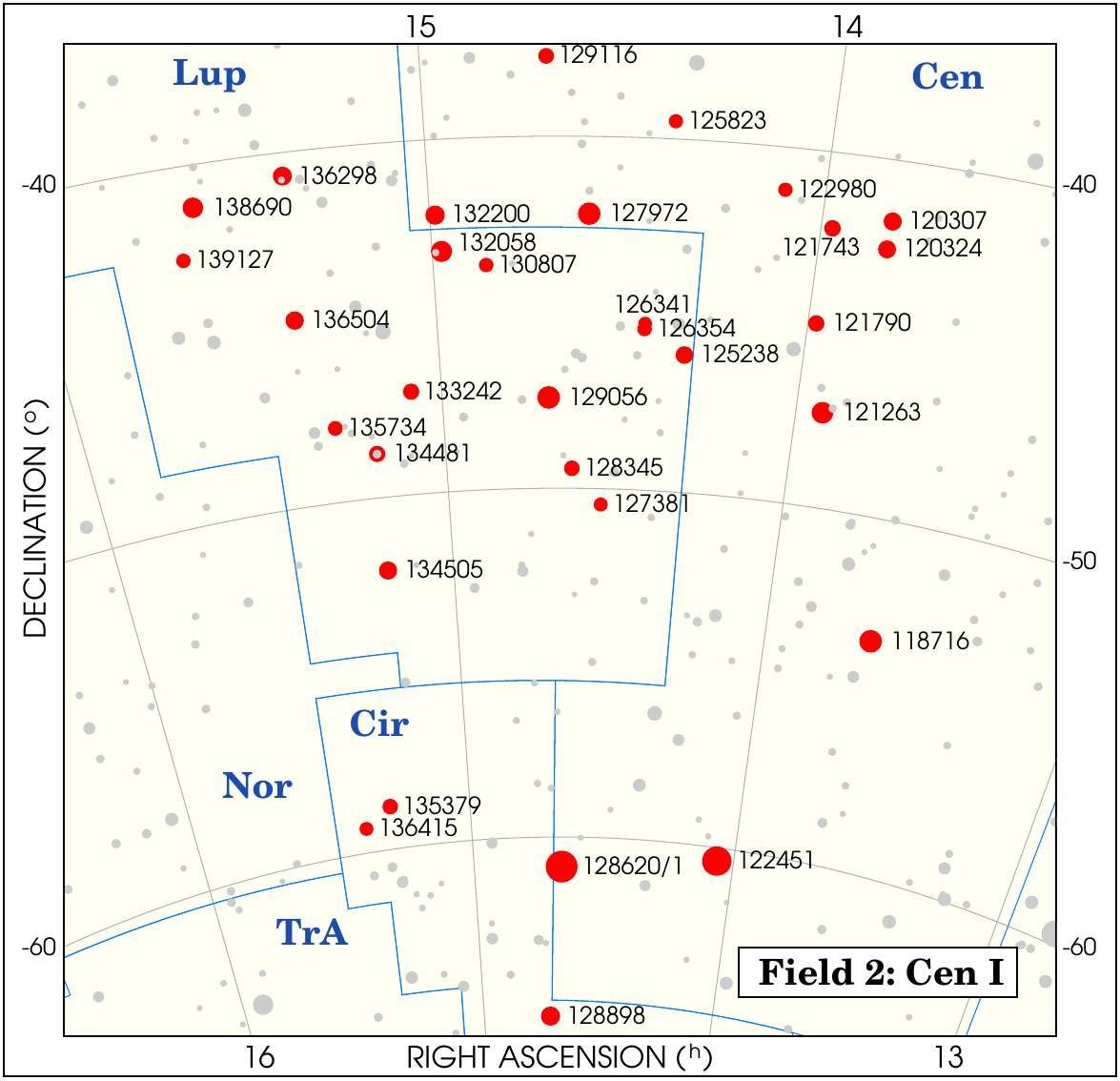}\\
    \vspace{0.25cm}
    \includegraphics[width=0.42\linewidth]{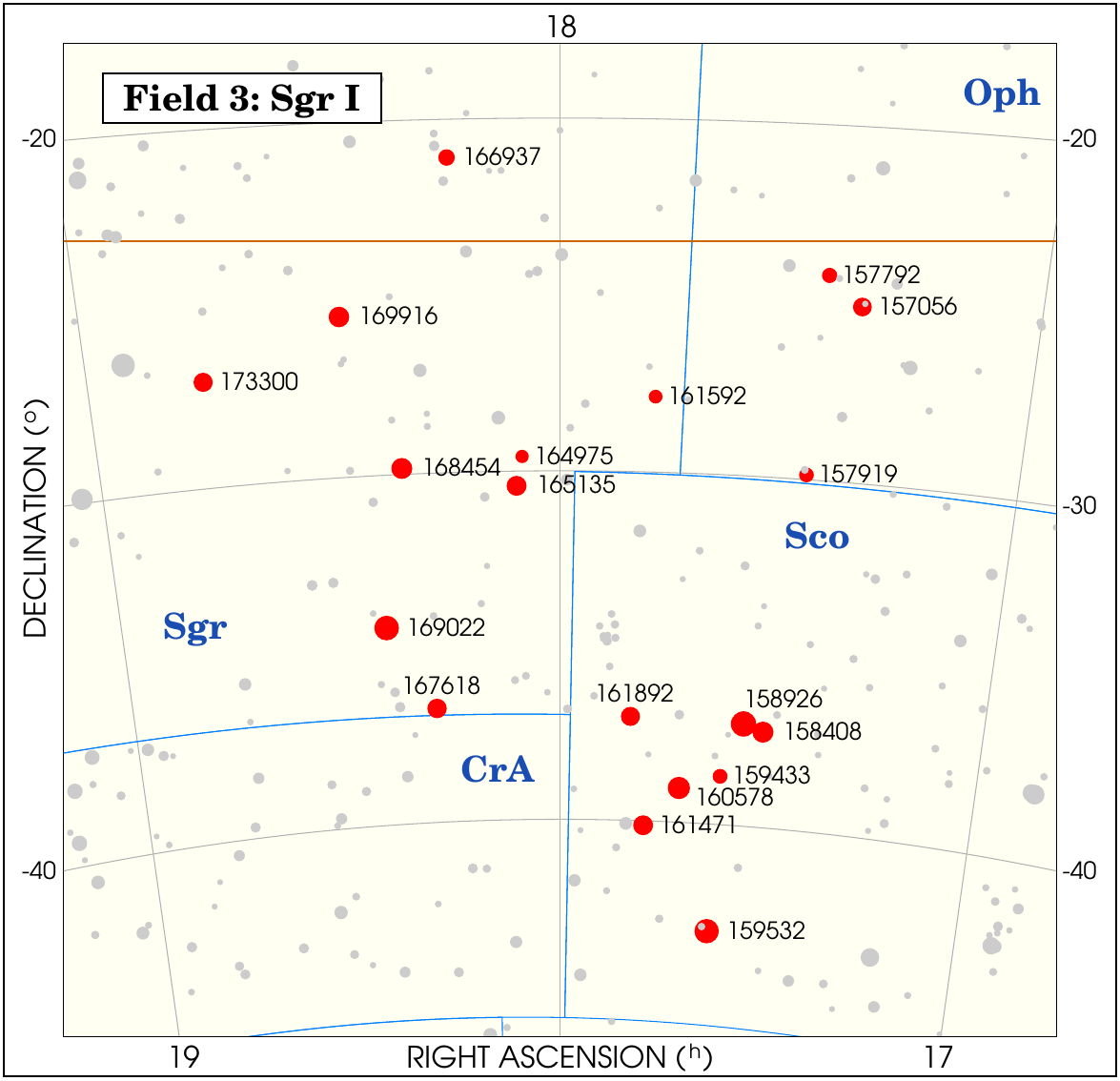}\hspace{0.5cm}\includegraphics[width=0.42\linewidth]{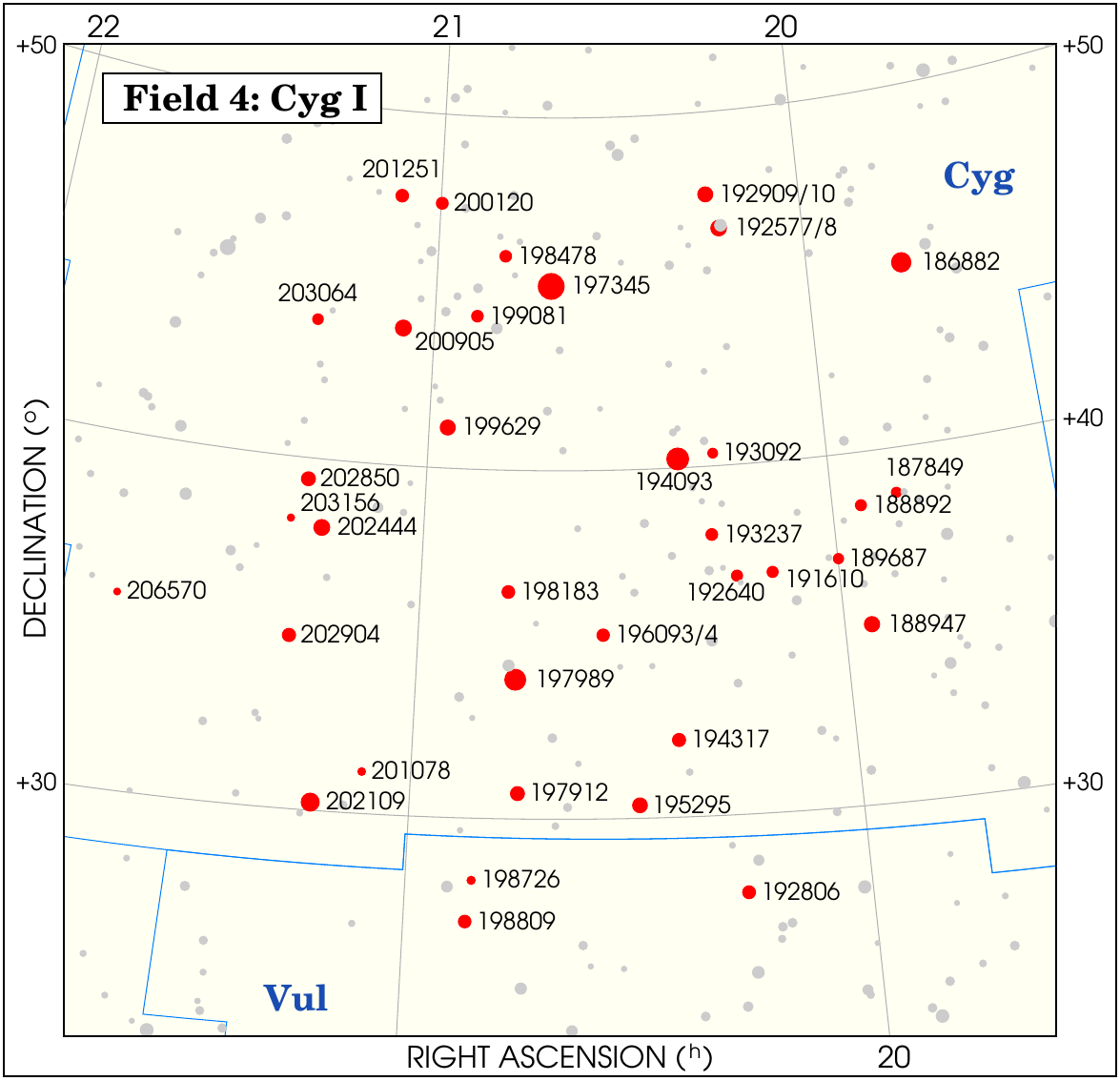}\\
    \vspace{0.25cm}
    \includegraphics[width=0.42\linewidth]{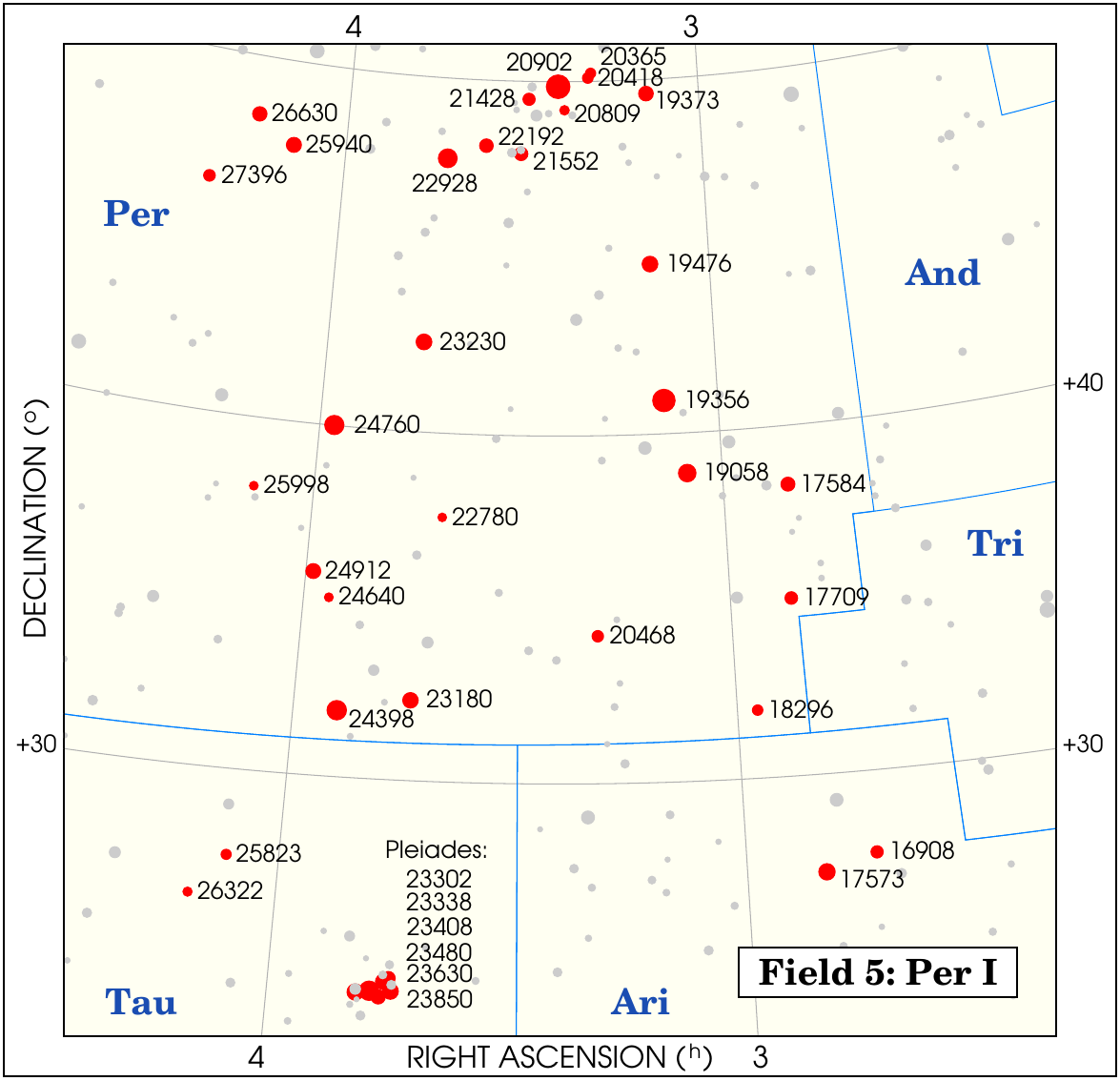}\hspace{0.5cm}\includegraphics[width=0.42\linewidth]{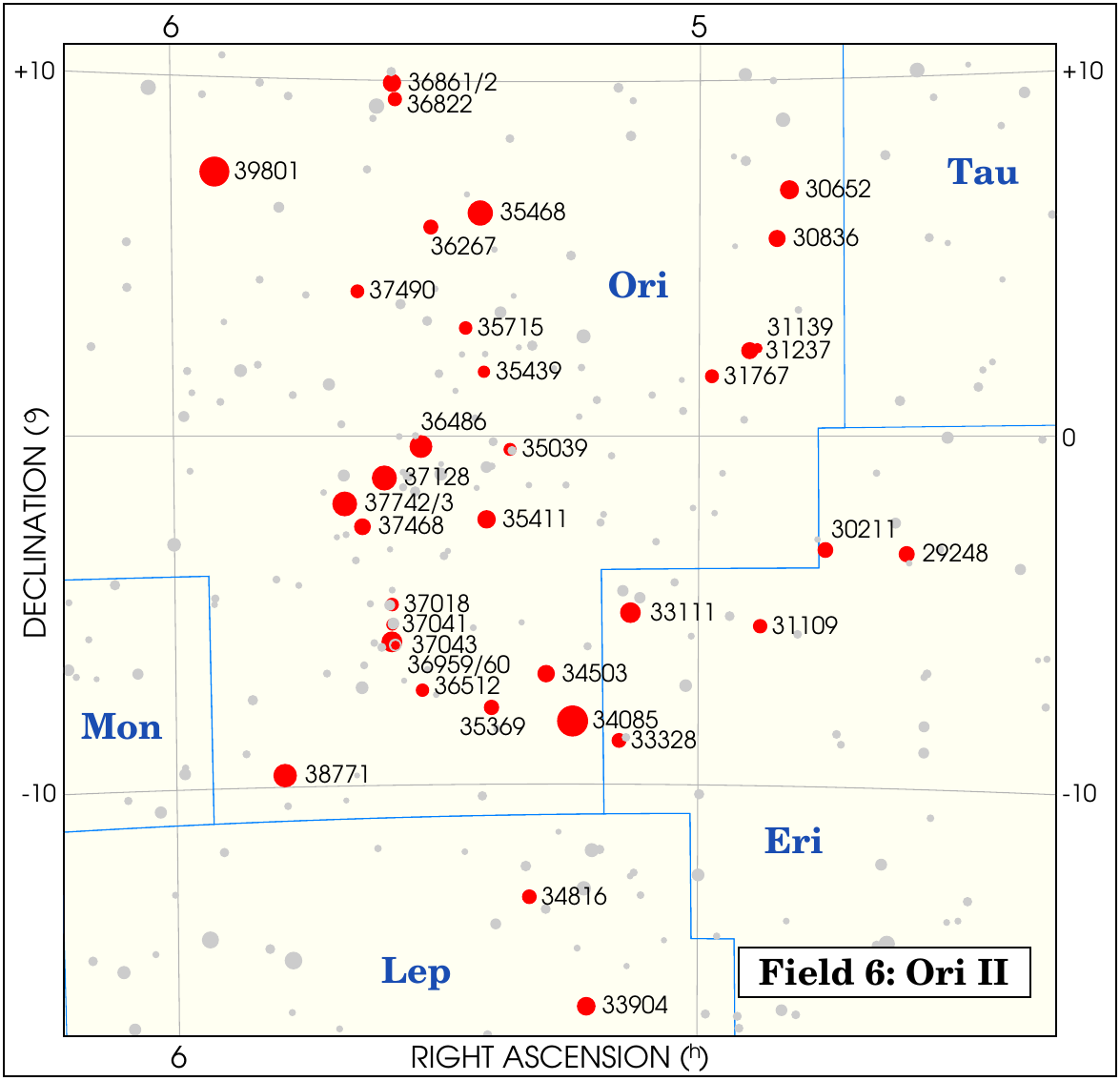}\\
  \caption{BRITE-Constellation field maps: Fields 1 to 6.}
    \label{fig:f1-6_map}
\end{figure*}

\begin{figure*}[!htb]
\centering
  \includegraphics[width=0.42\linewidth]{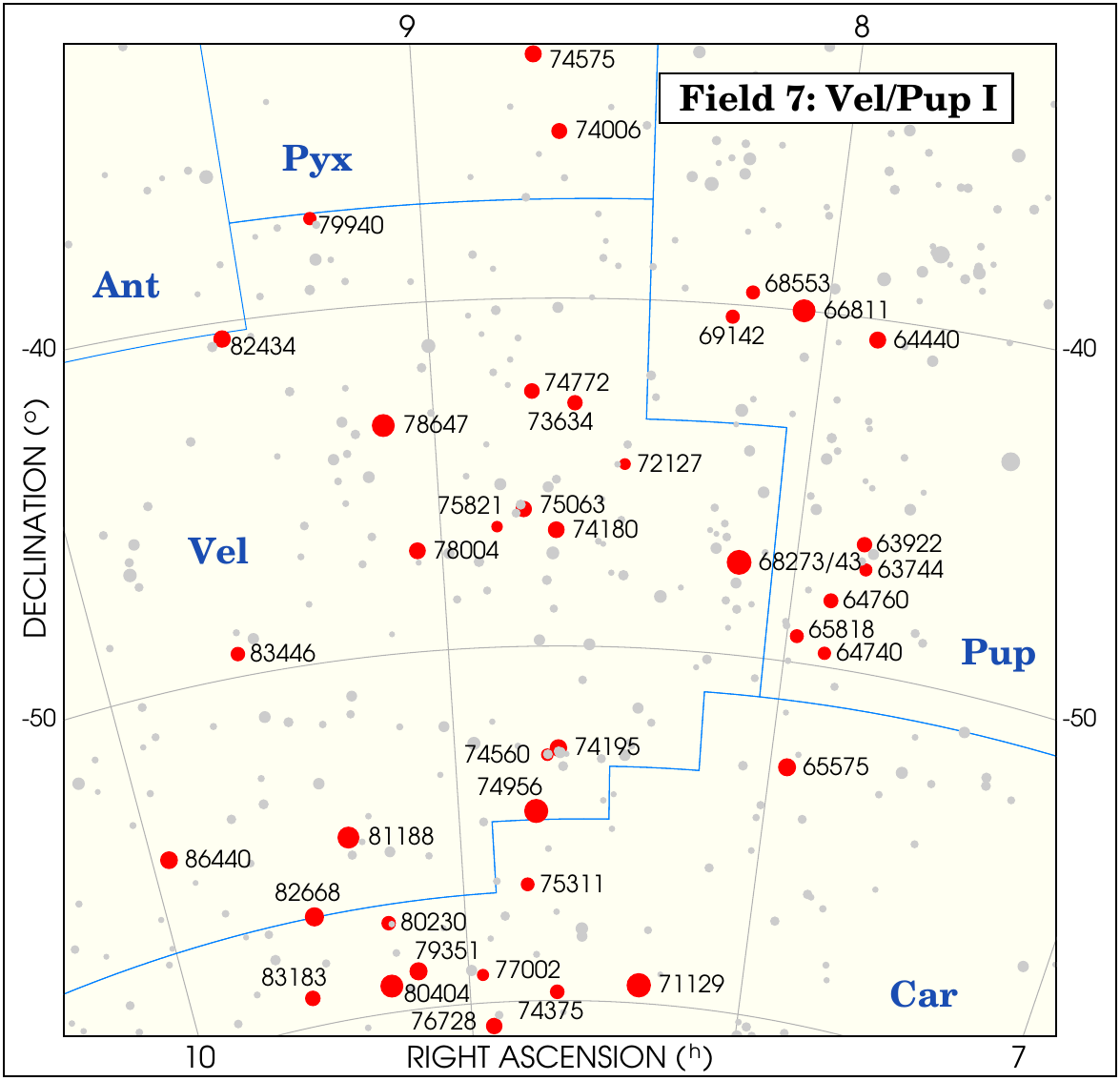}\hspace{0.5cm}\includegraphics[width=0.42\linewidth]{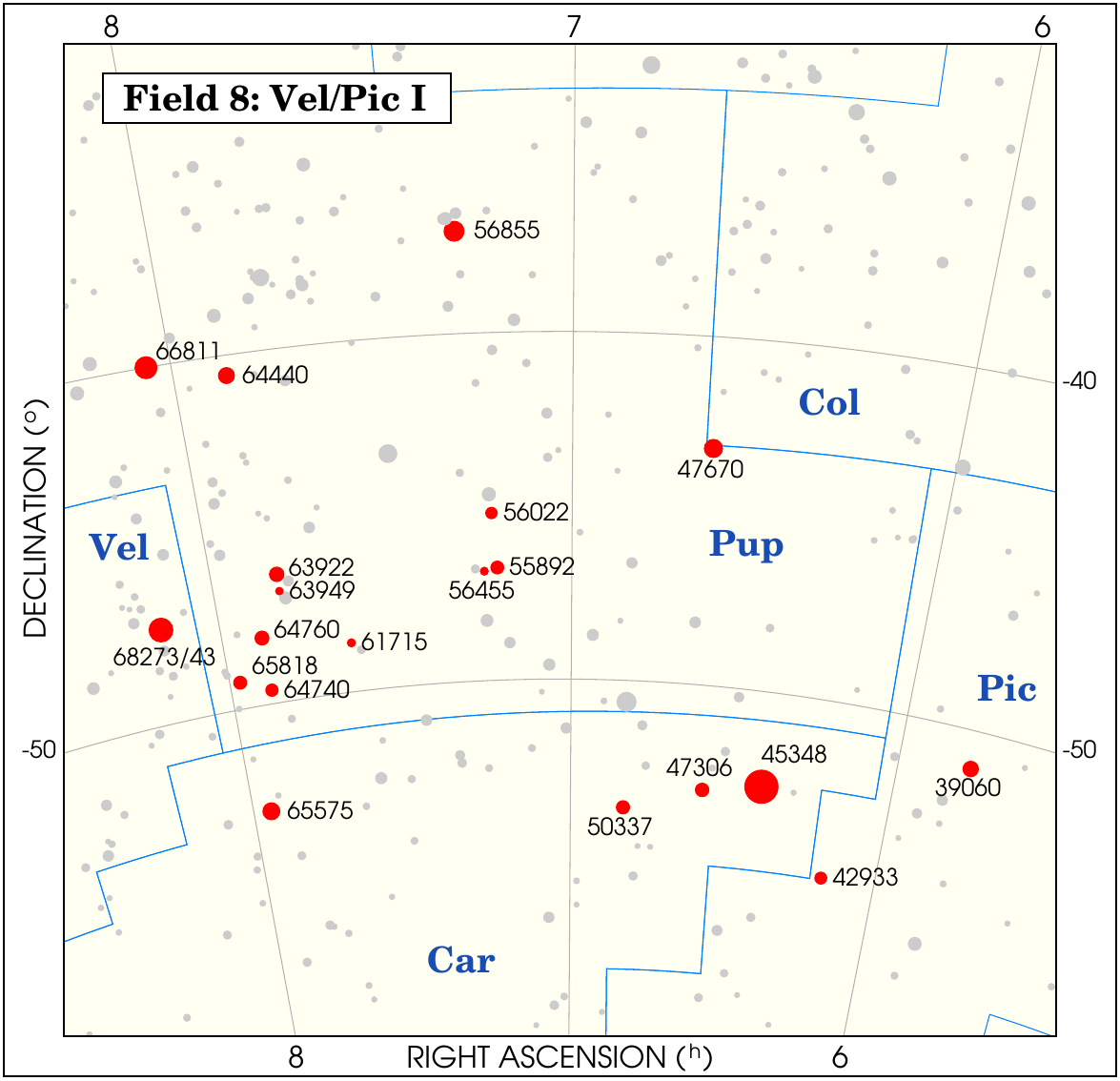}\\
   \vspace{0.25cm}
  \includegraphics[width=0.42\linewidth]{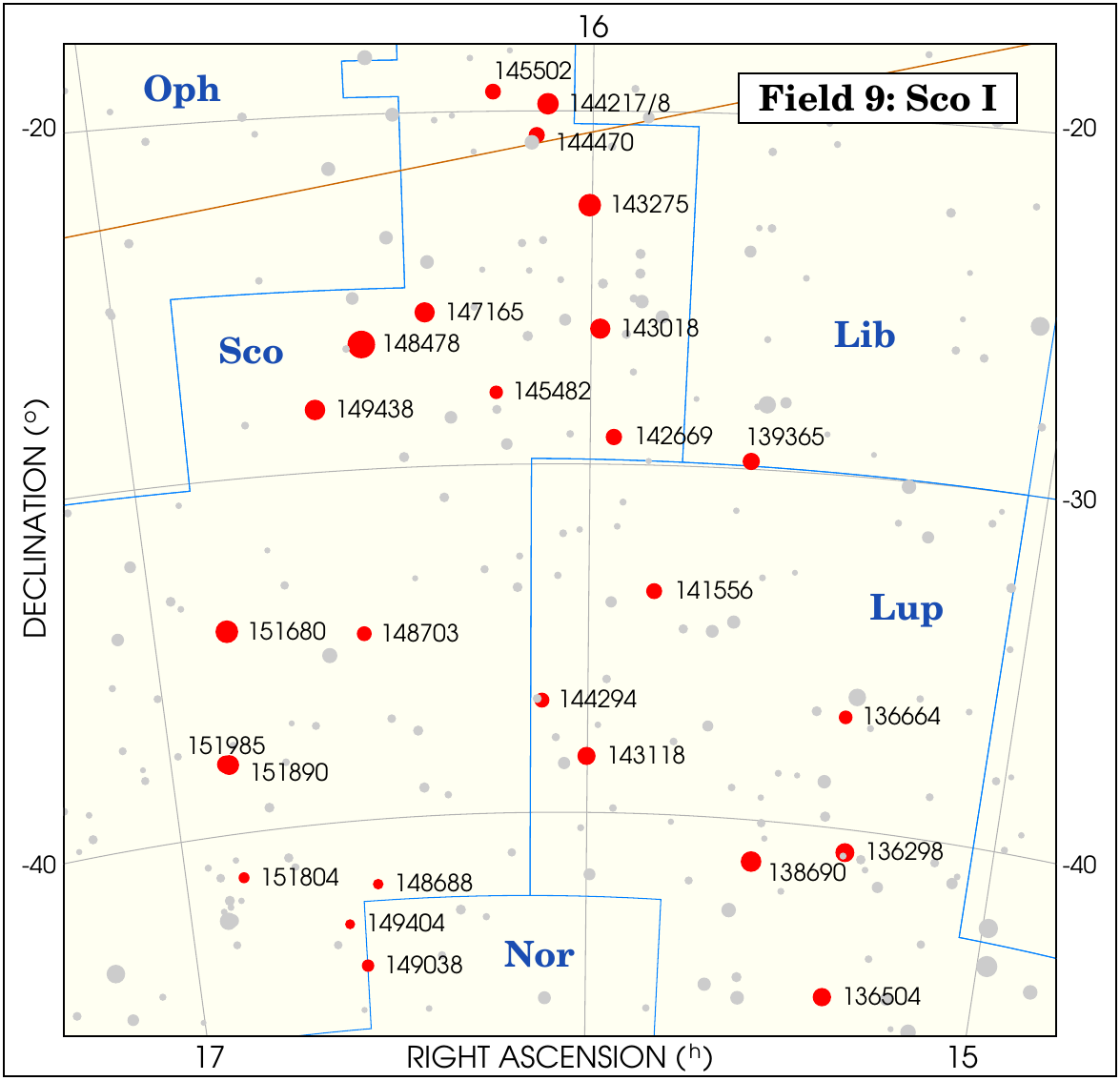}\hspace{0.5cm}\includegraphics[width=0.42\linewidth]{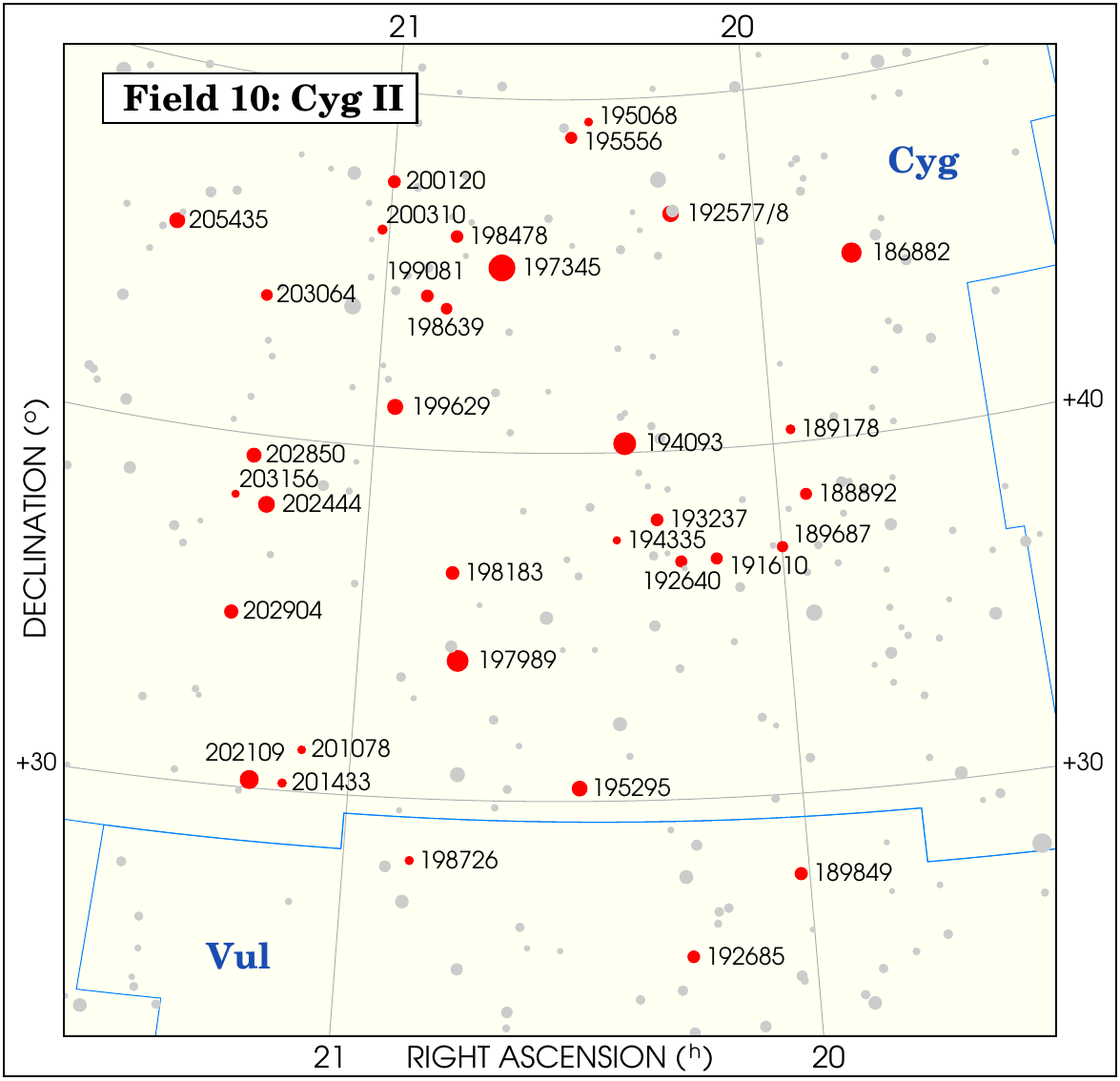}\\
  \vspace{0.25cm}    \includegraphics[width=0.42\linewidth]{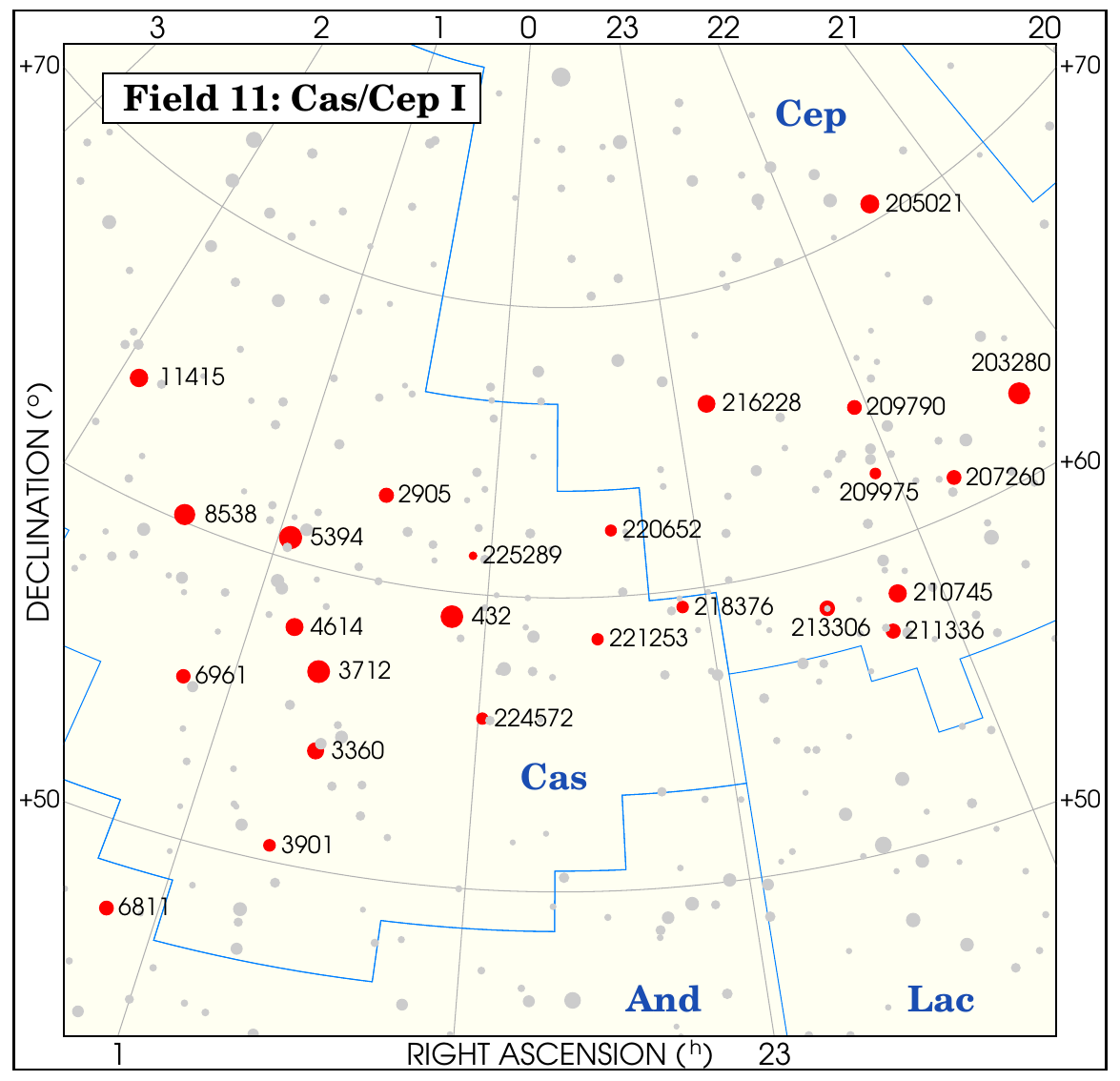}\hspace{0.5cm}\includegraphics[width=0.42\linewidth]{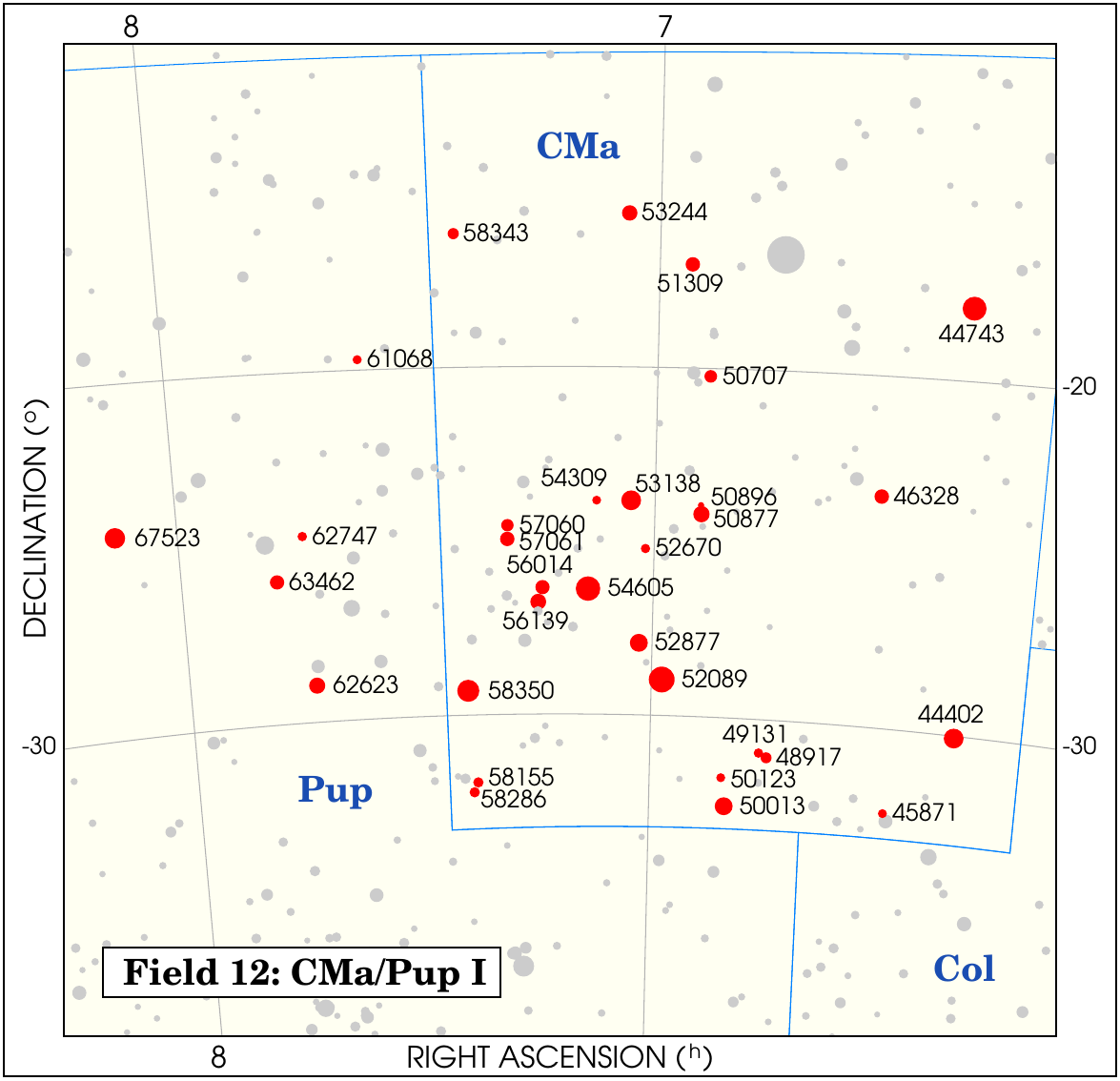}\\
\caption{BRITE-Constellation field maps: Fields 7 to 12.}
\label{fig:f7-12_map}
\end{figure*}

\begin{figure*}[!htb]
    \centering
    \includegraphics[width=0.42\linewidth]{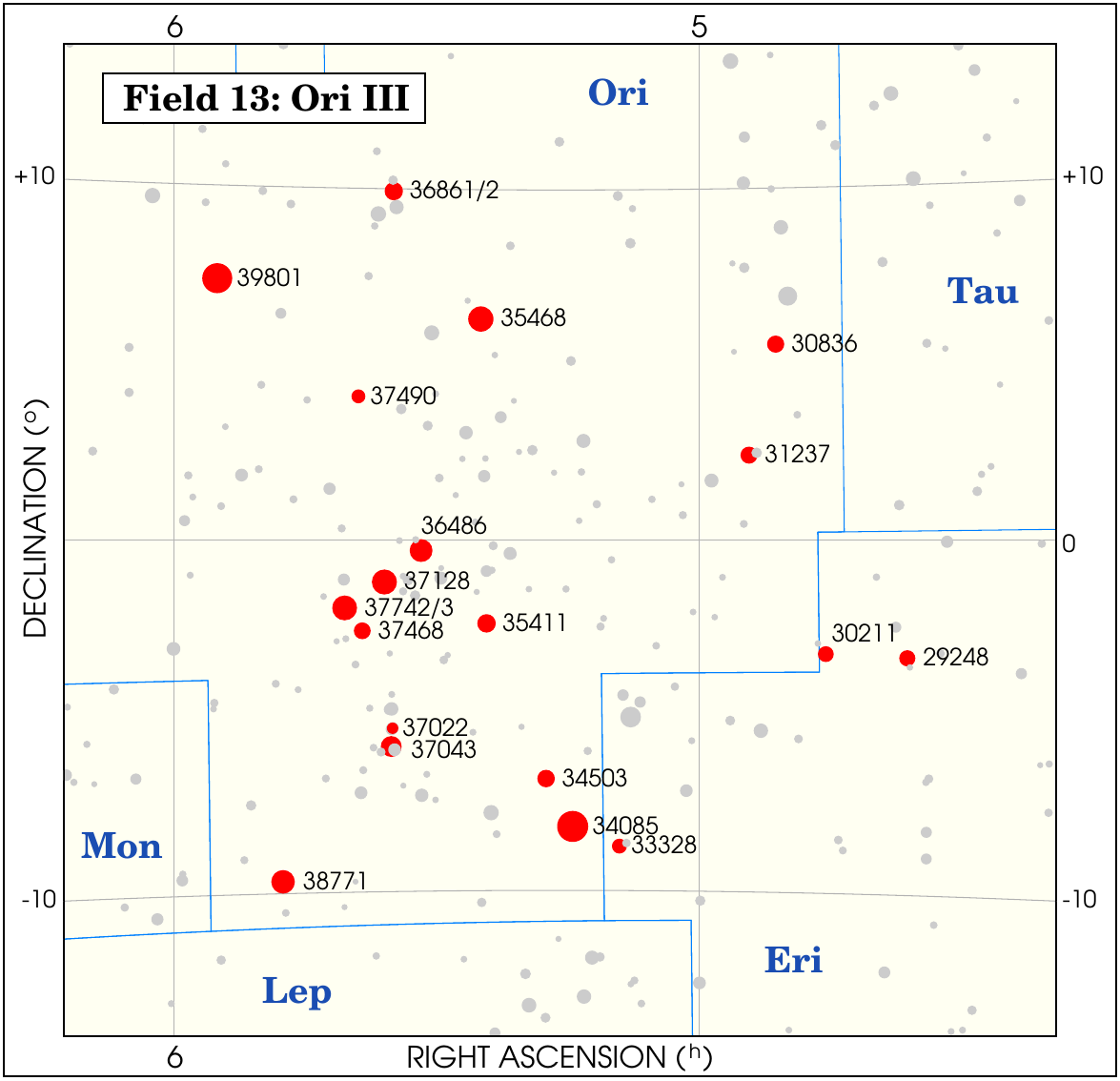}\hspace{0.5cm}\includegraphics[width=0.42\linewidth]{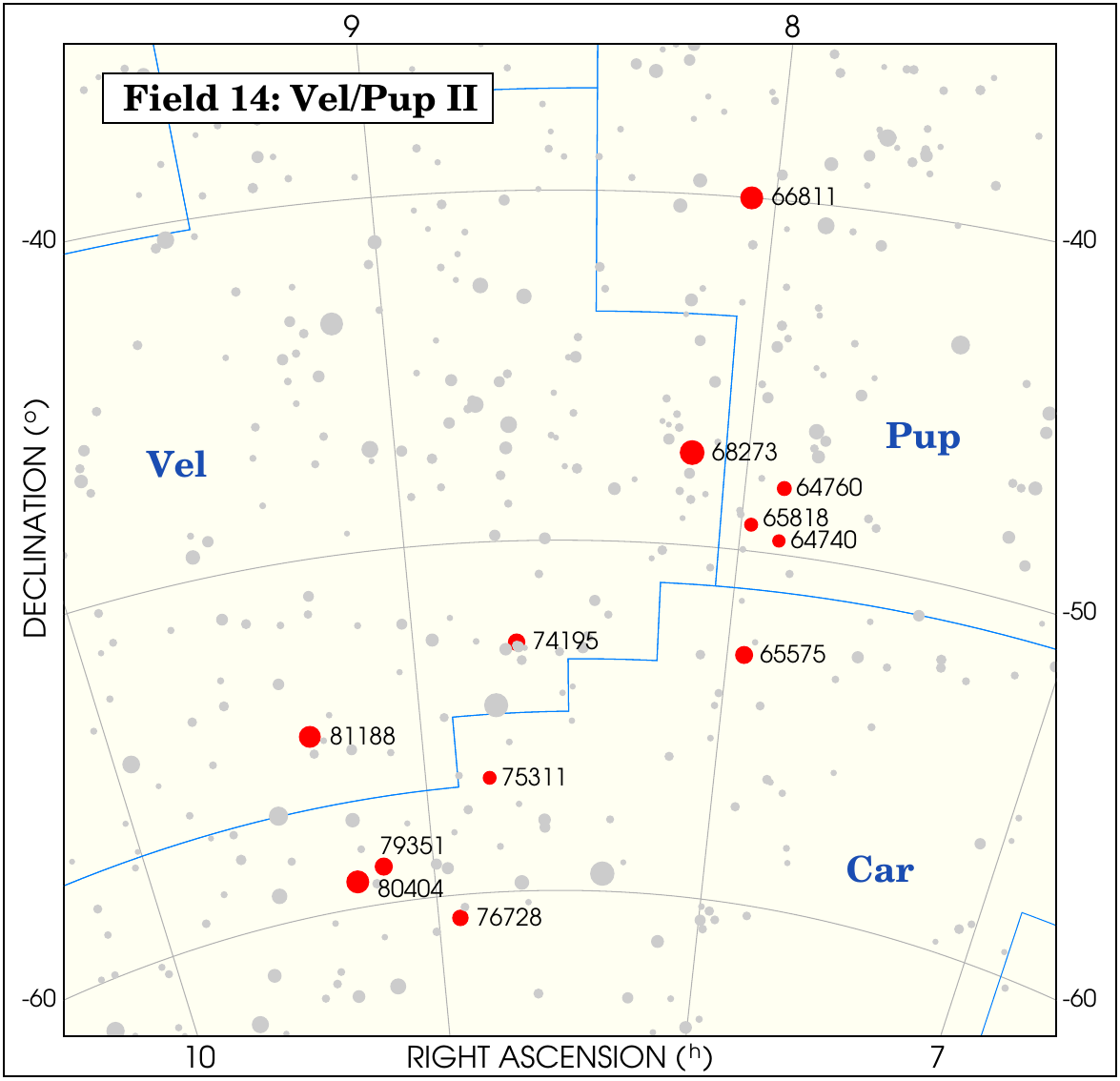}\\
    \caption{BRITE-Constellation field maps: Fields 13 and 14.}
    \label{fig:f1314_map}
\end{figure*}

Sky maps of the Fields 1 to 14 observed by BRITE-Constellation are shown in Figs. \ref{fig:f1-6_map}, \ref{fig:f7-12_map} and \ref{fig:f1314_map}: Observed stars are marked as red filled circles where the size of the symbol corresponds to the magnitude. The black numbers next to those symbols are HD identifiers (see Table \ref{stars300}). The names of the stellar constellations are given in blue. 


\end{appendix}


\end{document}